\documentclass[pre,twocolumn,superscriptaddress,aps,amsmath,amssymb]{revtex4-1}

\usepackage{color}
\usepackage{amsmath}    
\usepackage{amssymb} 
\usepackage{amsfonts}   
\usepackage{graphicx}
\graphicspath{{FIGS/}}
\usepackage{pgfplots}
\usepackage{mathtools}
\usepackage{bm}
\usepackage{hyperref}

\DeclareMathOperator*{\argmax}{arg\,max}

\definecolor{RawSienna}{cmyk}{0,0.72,1,0.45}
\definecolor{Dandelion}{RGB}{255,212,100}
\definecolor{Emerald}{RGB}{4,99,7}
\definecolor{Grey}{RGB}{50,50,50}
\definecolor{LightGrey}{RGB}{ 150,150,150}
\definecolor{DarkRed}{RGB}{230,130,130}

\newcommand{\s}{\sigma}	
\newcommand{\bs}{\bm{\s}}
\newcommand{\bx}{\bm{x}}
\newcommand{\bh}{\bm{h}}

\renewcommand{\>}{\rangle}
\newcommand{\<}{\langle}
\newcommand{\beq}{\begin{equation}}
\newcommand{\eeq}{\end{equation}}
\newcommand{\bea}{\begin{eqnarray}}
\newcommand{\eea}{\end{eqnarray}}
\newcommand{\av}[1]{\langle{#1}\rangle{}}
\newcommand{\abs}[1]{|{#1}|}

\newcommand{\eg}{\textit{e.g. }}

\newcommand{\rev}{}

\begin{document}

\title{Quantitative Immunology for Physicists}

\author{Gr\'egoire Altan-Bonnet}
\thanks{Authors are listed alphabetically.}
\affiliation{Immunodynamics Section, Cancer \& Inflammation Program,
  National Cancer Institute, Bethesda MD 20892, USA}
\author{Thierry Mora}
\thanks{Authors are listed alphabetically.}
\affiliation{Laboratoire de physique de l'\'Ecole normale sup\'erieure
  (PSL University), CNRS, Sorbonne  Universit\'e, Universit\'e de
  Paris, 75005 Paris, France}
\author{Aleksandra M. Walczak}
\thanks{Authors are listed alphabetically.}
\affiliation{Laboratoire de physique de l'\'Ecole normale sup\'erieure
  (PSL University), CNRS, Sorbonne  Universit\'e, Universit\'e de
  Paris, 75005 Paris, France}

\begin{abstract}
The adaptive immune system is a dynamical, self-organized multiscale system that protects vertebrates from both pathogens and internal irregularities, such as tumours. For these reason it fascinates physicists, yet the multitude of different cells, molecules and sub-systems is often also petrifying. Despite this complexity, as experiments on different scales of the adaptive immune system become more quantitative, many physicists have made both theoretical and experimental contributions that help predict the behaviour of ensembles of cells and molecules that participate in an immune response. Here we review some recent contributions with an emphasis on quantitative questions and methodologies. We also provide a more general methods section that presents some of the wide array of theoretical tools used in the field.

\end{abstract}

\maketitle

\tableofcontents{}

\section{Introduction}

The role of the immune system is to detect potential pathogens, confirm they really are undesirable pathogens, and destroy them. The goal is in principle well defined. However recognizing molecular friends from foes is not easy, and organisms have evolved many complementary ways of dealing with this problem. Immunologists separate the molecularly non-specific response of the ``innate'' immune system, which includes everything from scratching to the recognition of protein motifs characteristic of bacteria, and the molecularly specific ``adaptive'' response, by which specialized cells recognize evolving features of never encountered before pathogens. From another angle we can consider different ways of destroying a pathogen: either swallowing pathogens whole, which is done by cells of the innate immune system called neutrophils and macrophages; or killing our own cells that have been infected by a pathogen or are tumourous --- as performed by representatives of the adaptive immune system called killer T-cells. Alternatively, the adaptive immune system produces specialized molecules called antibodies that smother the invader: they attach to pathogens to prevent them from entering cells and multiplying; they bind to bacterial toxins, thereby disarming them; or they bind directly to bacterial cells, flagging them for consumption by macrophages. 

\begin{figure}
\includegraphics[width=\linewidth]{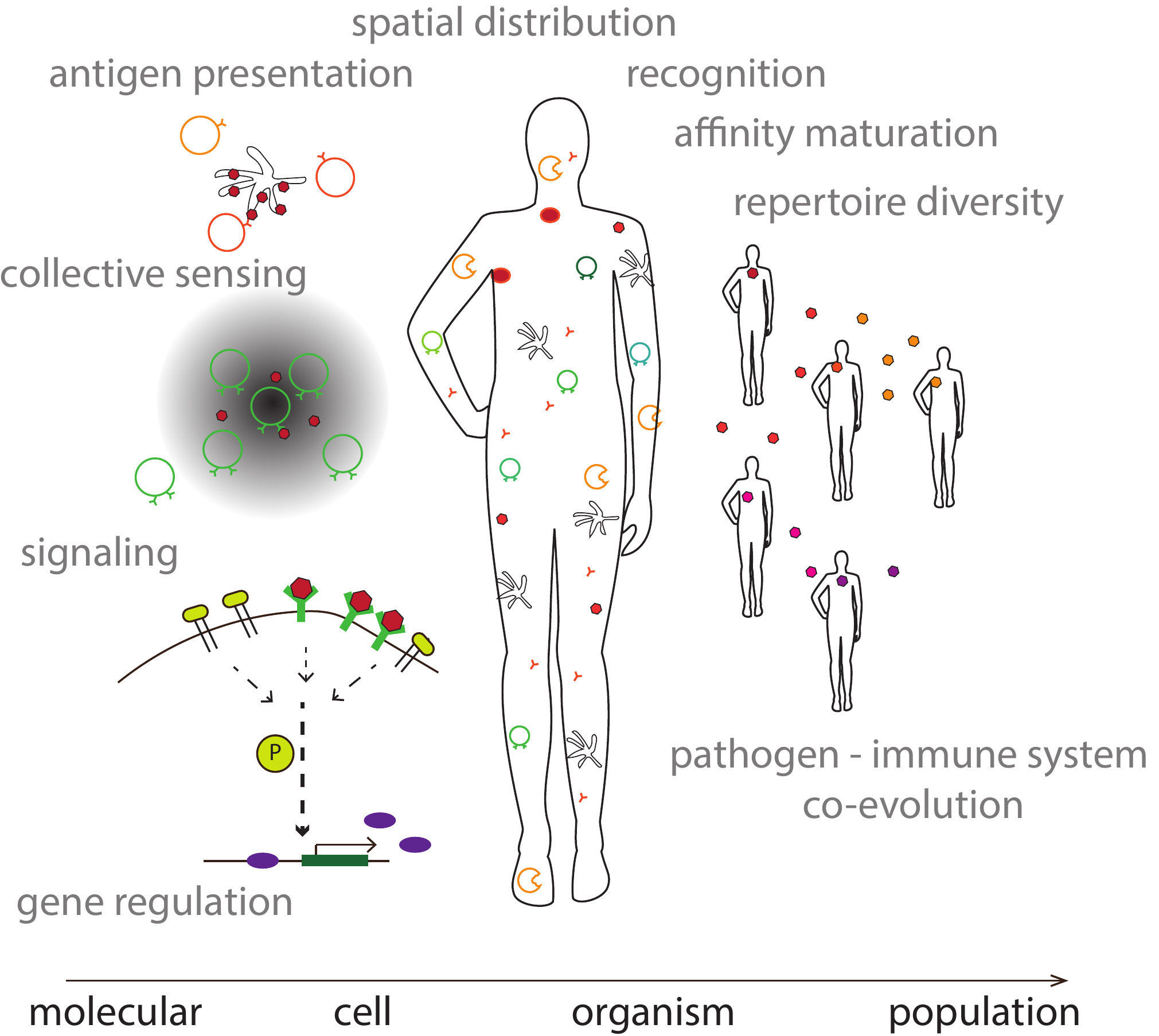}
\caption{\textbf{The many scales of the immune system}. The immune system works at many scales from the molecular of receptor-protein interactions, gene regulation, activation of biochemical pathways, to the cellular of cell-to-cell communication directly and through signalling molecules, and organismal level responses using cells of the innate and adaptive immune systems, to the population level where global viral evolution drives the co-evolution of immune systems of different individuals.  }
\label{fig:intro}
\end{figure}

\begin{figure}
\includegraphics[width=\linewidth]{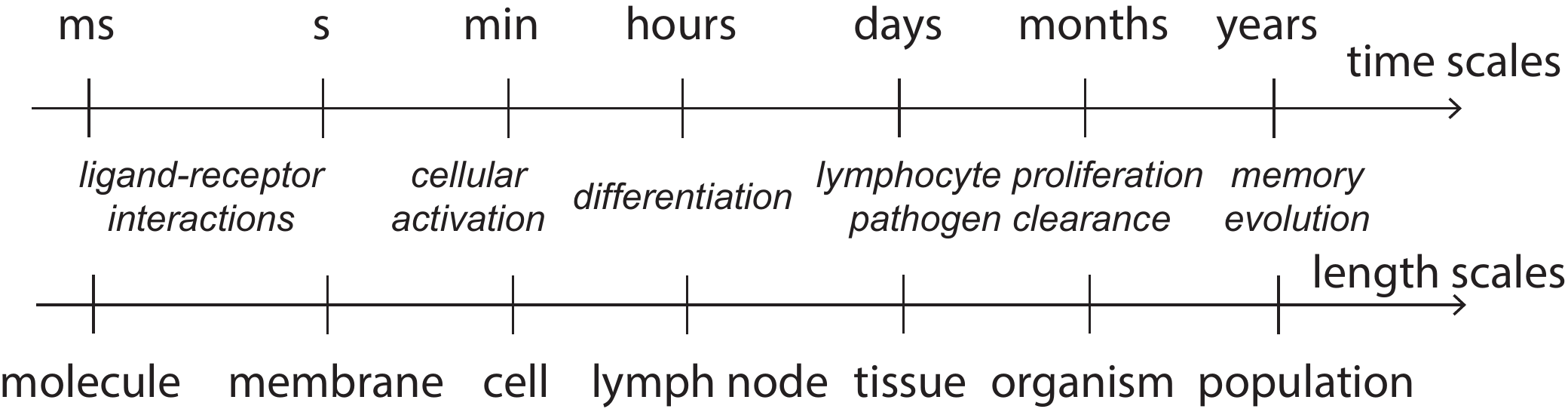}
\caption{\textbf{Placing immune processes in length and timescales}. The timescales of immune interactions cary from seconds to years, and the scales from $10^{-10}$m for molecular interactions to thousand of kilometers for pathogen evolution  }
\label{fig:scales}
\end{figure}

That short overview gives us an idea of the many strategies that both pathogens and the host organism have at their disposal (Fig.~\ref{fig:intro}). Pathogenic cells (a bacteria, virus, or a tumour cell) are programmed to divide. The immune system is there to prevent this. To a large extend, its main challenge is to recognize the unknown. Pathogens are constantly evolving to escape recognition by the immune system. Although the host organism does have a certain list of ``warning sign features,'' most of which are taken care of by the innate immune system, it is up against a large set of constantly moving targets --- as illustrated in our everyday experience by the evolving influenza virus, which requires a new vaccination every year. For this reason, the strategies developed by the immune system are mostly statistical, and require multiple interactions between different types of cells, a lot of checks and balances that leads to a multiplicity of time and length scales (Fig.~\ref{fig:scales}). Despite this complexity, the immune system works remarkably reliably. How do these interactions on many scales dynamically come together in a self-organized way to build a complex sensory system against a high dimensional moving target? This review breaks this high level question into smaller problems and presents some results and concepts contributed by physicists. It also presents a summary of the current methods, experimental and theoretical, used in the field. We do not shy away from the biology of the immune system, but to help the physicist navigate the complexity of immunology, biological details will be introduced as we go along. For an introduction of the immune system for the non-specialist, we refer the reader to the short but excellent book by Lauren Sompayrac \cite{Sompayrac1999}.
This review does not aim to be exhaustive, but rather focuses on the important physical concepts of immune function, reducing biological complexity to a minimum whenever appropriate.

From an evolutionary perspective, all organisms have some form of protection. Bacteria protect themselves from viruses through specific CRISPR (Clustered Regularly Interspaced Short Palindromic Repeats)-Cas, or unspecific restriction modification systems. The innate immune system is shared by many animals, and is largely similar between us and flies. The adaptive immune system evolved in jawed vertebrates and also has changed little between fish and mammals. Plants also have a well-studied innate system, and have recently been shown to have elements of adaptive immunity. Immunity is therefore a basic element of life. However the details of its implementation and spatial organisation scale with the organism. 

Burnet's clonal selection theory~\cite{Burnet1957}, built on previous observations and ideas~\cite{Ehrlich1900,Jerne1955,Hodgkin2007}, provides a theory of the adaptive immune system in the same sense that Darwin's theory provides a theory or framework for evolution. Burnet's theory states that molecules of the pathogens stimulate specific B and T lymphocytes (cells of the adaptive immune system) among a pre-existing ensemble of possible cells, which leads to proliferation of this specifically selected clone. Molecules thus recognized are called antigens.
It explains the diversity and specificity of the adaptive immune system, also highlighting its {\it adaptive nature}. The framework is often summarized in four assumptions: (i) each T and B lymphocyte cell has one type of receptor; (ii) receptor-antigen binding is required for cell or receptor proliferation; (iii) offspring of the stimulated cell have the same receptor as their parents; (iv) cells that have receptors that recognize the organism's self molecules (self-antigens) are removed early in their development. The theory was validated by showing that B-cells always produce one receptor~\cite{NossalLederberg} and later by experiments showing immunological tolerance to factors introduced in the embryonic period or immediately after \cite{Medawar1953,Medawar1956}. Burnet's theory provides an incredibly successful framework to understand the adaptive immune system, yet it is not a quantifiable theory that can be tested against concrete measurements. While what remains to be ``filled in'' may seem like details, these details still provide a lot of puzzles and uncertainties, also at the conceptual scale. In this review, we present certain examples of quantified ideas 
 that have recently emerged. 

We also present some (albeit not all) of the quantitative puzzles, starting from the smallest molecular (sections~\ref{biophys} and~\ref{discrimination}) and cellular (sections~\ref{cytokines} and~\ref{cellfate}) scales and moving to the organismal (sections ~\ref{repertoires},~\ref{dynamics} and ~\ref{affinity_maturation})  and population wide (section~\ref{popleveldyn}) scales. How do cells discriminate between self and non-self in such an exquisite way integrating processes taking part on many timescales (section~\ref{discrimination} on antigen discrimination)? How do cells communicate and coordinate to orchestrate the immune response as a collective phenomenon (section ~\ref{cytokines} on cytokines)? How do cells adopt specialized phenotypes through cell differentiation to span the range of functions they must fulfil (section~\ref{cellfate} on cell fate)? How does the system cover the space of possible specificities allowing for complete protection against the unknown (section ~\ref{repertoires} on repertoires)? How does the immune system as a whole adapt to the changing environment (section~\ref{affinity_maturation}) and how does it influence virus evolution (section~\ref{popleveldyn})?

This review covers a lot of topics,  many of them connected. We try to mention these connections, however we have tried to make the sections stand alone and the reader does not have to (and probably should not attempt to) read them all at once, or in the presented order. At the end we include a glossary of the biological terms (section~\ref{glossary}) to help navigate the immunological terminology. Detailed appendices (Sec.~\ref{sec:methods}) present general methods that are used in physical immunology as well as other fields.

\section{Physical chemistry of ligand-receptor interaction: specificity, sensitivity, kinetics.}~\label{biophys}
Our exploration of the immune system starts at the molecular scale, through the binding of ligands and receptors expressed at the surface of immune cells. Ligands carry information about the pathogenic threat and can be of two types: antigens, i.e. bits of proteins that are recognized by the immune system; and cytokines, which are small molecules secreted by immune cells to communicate with each other about their current state and experience.
Binding events between these ligands and their cognate receptors provide the raw signals that cells must interpret to adapt their individual and collective behavior, eventually mounting  an immune response in the case of recognition.

All immune decisions start with the classical physical chemistry of ligand--receptor interactions:
\begin{equation}
\label{eq:L+R}
{\rm Ligand} + {\rm Receptor} 
\begin{array}{c}
k_{\rm on}\\
\rightleftharpoons \\
k_{\rm off}
\end{array} 
{\rm Complex} \longrightarrow {\rm Activation}
\end{equation}
This extremely simple reaction leads to strong limits on how fast signals can spread in the immune system (and in biological systems in general), which in turn allows us to discriminate possible regulatory scenarios. This constraint stems from physical-chemical limits on the parameters of these reactions, typically in the picomolar to millimolar range for the equilibrium dissociation constant $K_D= k_{\rm off}/k_{\rm on}$ (we recall that $1nM= 10^{-9}{\rm mol}/L = 10^{-24}\mathcal{N}_{\rm Avogadro} \mu m^{-3}  \approx 0.6$ molecules per $\mu m^3$, with $\mathcal{N}_{\rm Avogadro}\approx 6\cdot 10^{23} {\rm mol}^{-1}$). In this section, we discuss the physical considerations leading to estimates of the biochemical rates driving in these reactions. Such a discussion of the physical chemistry of ligand-receptor interactions is necessary to understand key quantitative aspects of immune activation, as well as discrimination between self and non-self antigens by B \& T cells (as we will discuss in Sec.~\ref{discrimination}).

\subsection{Diffusion-limited reaction rates}
The basic laws of physical kinetics~\cite{Landau_Physkin} can be used to obtain an estimate of the diffusion-limited rate of molecular association  (see Sec.~\ref{diff_lim} for a derivation): $k_{\rm on}\leq k_{\rm diffusion}$, with
\beq
k_{\rm diffusion}=4\pi\left(R_{\rm Ligand}+R_{\rm Receptor}\right)\left(D_{\rm Ligand}+D_{\rm Receptor}\right),
\label{eq:association_rate}
\eeq
where $R_{\rm Ligand}$ and $R_{\rm Receptor}$ are the radii of the ligand and of receptor binding pockets (both assumed spherical), while $D_{\rm Ligand}$ and $D_{\rm Receptor}$ are their diffusion coefficients.
Since the receptor in Eq.~\ref{eq:L+R} is usually embedded in the lipid bilayer of an immune cell, we can assume it is relatively immobile compared to its ligand $L$, which diffuses more rapidly in the extracellular medium or in the cytoplasm, $D_{\rm Receptor}=D_{\rm membrane}\ll D_{\rm Ligand}=D_{\rm solution}$.

The ligand's diffusion coefficient can be estimated from Stokes-Einstein's formula:
\beq
D_{\rm Ligand}=D_{\rm solution}=\frac{k_B T}{6\pi \eta R_{\rm Ligand}}\approx 300\,\mu m^2s^{-1}
\eeq

for a small ligand of size $R_{\rm Ligand} =1nm$, diffusing in extracellular medium (whose viscosity is given by that of water, $\eta=0.7\,mPa \cdot s$), at body temperature $T=310 K$ with Boltzmann's constant $k_B =1.38\cdot 10^{-23}$J/K). Then, from Eq.~\ref{eq:association_rate}, assuming a small target ($R_{\rm Receptor}\ll R_{\rm Ligand}$), we obtain: 
\beq \label{k_diff_num}
k_{\rm  diffusion}=4\pi R_{\rm Ligand} D_{\rm Ligand}=\frac{2k_B T}{3 \eta}\approx 2\cdot 10^9 M^{-1}s^{-1}.
\eeq
Eq.~\ref{k_diff_num} gives a general upper bound for any ligand-receptor association rate, whether it is immunological or not. This limit is conceptually equivalent to the speed of light (although clearly not as fundamental). 

Yet, there are examples when ligand-receptor associations between two small proteins seem to ``beat'' this diffusion limit: $k_{\rm on} > k_{\rm  diffusion}$~\cite{Hager2009}. Such an apparent paradox can be resolved when considering the limiting step for this association. If the ligand interacts weakly with a large object,  (e.g. the entire plasma membrane of a cell, DNA coils), inducing directed diffusion in a constrained space before reaching its specific targets~\cite{Halford2004,Gorman2008}, this pre-equilibration step implies that the cross-subsection of the object our ligand needs to hit can be much larger. The diffusion molecules then can ``hop'' around these non-specific binding sites to accelerate their search for the specific binding sites. 
In these cases, the effective $R_{\rm Receptor}$ of collision is the macroscopic scale associated with the large object of weak/non-specific interactions ($5\mu m$ for cells, $>0.1\mu m$ for DNA coils), hence, an apparent increase in the rate of collision according to Eq.~\ref{eq:association_rate}. A careful calculation (see e.g. \cite{Slutsky2004}) yields the appropriate bound.

The diffusion coefficient will  be slowed down for larger macromolecular complexes, and for molecules diffusing inside the cells (there, the apparent viscosity can be increased by 10-fold). Additionally, when considering biomolecules embedded within cell membranes, one must take into consideration the dramatically increased viscosity that leads to slowed-down diffusion for receptor proteins on the surface of immune cells ($D_{\rm solution} \approx200 \mu m^2/s$ vs $D_{\rm membrane} \approx 5\mu m^2/s$ for a typical protein of radius 2nm). For this reason, when considering a soluble protein interacting with proteins embedded within the plasma membrane of immune cells (\eg extracellular cytokines being captured by a cytokine receptor, intracellular enzymes --such as kinases or phosphatases, that catalyze the phosphorylation/dephosphorylation of proteins-- interacting with an activated receptor, etc.), the diffusion of the membrane proteins is so small ($D_{\rm membrane} \ll D_{\rm solution}$), than it can be neglected and
\begin{equation}
k_{\rm  diffusion}=4\pi\left(R_{\rm Ligand}+R_{\rm Receptor}\right)\left(D_{\rm solution}\right).
\label{eq:association_rate_immobilized}
\end{equation}

\subsection{Extrapolating collision rates in solution to association rates in the physiological context}
While the rate of collision calculated above is an upper bound to the association rate,
a very important limitation must be taken into account: not every molecular collision will lead to their association, and we must estimate the probability of a successful association event. This probability is given by Arrhenius's law:
\beq
k_{\rm on}=k_{\rm diffusion} e^{-{\Delta G_{\rm association}}/{\left(k_B T\right)}},
\label{eq:association_full}
\eeq
where $\Delta G_{\rm association}$ is the free energy barrier (entropic and enthalpic) that molecules must overcome to associate.

The estimation of $\Delta G_{\rm association}$ is tricky as it requires a deep structural understanding at the atomic level of the entropic, energetic and conformational changes associated with bond formation between two large biomolecules. For the entropic contribution, one rule of thumb is to estimate the numbers of degrees of freedom constrained by the association between biomolecules: simply aligning two biomolecules (in rotation or in translation) incurs an entropic cost of at least $6 k_BT$, reducing the probability of association of each collision by $e^6\sim400$. Hence, the basal association rate (before taking into account more subtle molecular constraints) is $\sim 5 \cdot10^6 M^{-1}s^{-1}$ rather than the $\sim 2 \cdot 10^9 M^{-1}s^{-1}$ estimated in Eq.~\ref{k_diff_num}. Additional corrections should be made for each pair of biomolecules (as we will discuss in Sec.~\ref{sec:rates_numbers}).

\subsection{Rates and numbers in the physiological context}

\subsubsection{Association rates}

The formula in Eq.~\ref{eq:association_rate} is particularly useful in the context of quantitative immunology, when one must estimate the kinetics of molecular interactions in different context (solution, intracellular cytoplasm, surface plasma membrane etc.). Most kinetic parameters for biomolecular interactions are measured in solution, as experimentalists routinely purify ligand--receptor pairs and test their interactions in their soluble form (e.g. by calorimetry or by surface plasmon resonance -- in the latter case, one molecule must be immobilized). Such measurements can be used to estimate the hard-to-predict activation barrier $\Delta G_{\rm association}$ for molecular association by inverting Eq.~\ref{eq:association_full}, using the measured {\em in vitro} $k_{\rm on}$, and the estimated $k_{\rm diffusion}$.
Eq.~\ref{eq:association_rate}  can be used to rescale the viscosity $\eta$ and cross-section $R_{\rm Ligand}+R_{\rm Receptor}$ of molecular interactions in such a way that one can translate soluble measurements into physiologically relevant parameters. 

\subsubsection{Dissociation rates}

As discussed below, immunological interactions span the range of very short lived interactions, $k_{\rm off}>10\,s^{-1}$ or $\tau_{\rm off}=(k_{\rm off})^{-1}<0.1\,s$ for antibody binding to its target in the initial phase of an immune response, to extremely long-lived interactions, $k_{\rm off}<10^{-4} s^{-1}$ or $\tau_{\rm off}>3$ hours for cytokines interacting with  their receptors. These estimates set a huge range of time scales the immune system must deal with, even before taking into consideration delays in cellular responses and how these affect the ligand environment on time scales ranging from hours to days.  These considerations form the crux of the matter for quantitative immunology at the cellular scale: while  the physical chemistry of ligand-receptor interactions is straightforward, the immune system builds a response of devilish complexity from such elementary interactions. 

\subsubsection{Numbers of receptors per cell}
In some situations, such as in T cell antigen recognition, where both the T cell receptor and the antigen exist in a membrane-bound form.
All dynamics of interactions must be estimated taking into account the surface concentrations of molecules, with proper adjustment for slower diffusion in the association rates.

One often measures the number of receptors per cell $\#R_{\rm cell}$, e.g. by assessing the number of cytokine receptors on the surface of cells using quantitative flow cytometry, while the ligand is provided in soluble forms, as is the case with cytokines. In this case, one must estimate the {\it solution}-level concentrations of available receptors $[R]_{\rm total}$ in the reaction volume $V$:
\beq
[R]_{\rm total}=\frac{\#R_{\rm cell} N_{\rm cell}}{V \mathcal{N}_{\rm Avogadro}},
\eeq
where $N_{\rm cell}$ is the number of cells in volume $V$. For example, the effective ``concentration'' of receptors binding the IL-7 cytokine within a lymph node can be estimated, knowing that each of $10^7$ T cells within this 50 $\mu$L volume express $10^3$ receptors:
\beq
[{\rm IL-7R}]_{\rm total}\approx 3 \cdot 10^{-10}M = 300pM,
\eeq
while $K_D =10^{-11}M = 10pM$ for IL-7 binding to its receptor. Hence  $[{\rm IL-7R}]_{\rm total} \gg K_D$ and any secreted IL-7 will rapidly be captured by the cells within a lymph node. The high density of receptors and cells within lymphoid tissues makes for an interesting regime of competition for soluble ligands, as discussed below (Sec.~\ref{cytokines}). Such normalization by the Avogadro number, while straightforward, is of critical significance in many immunological configurations, as one must bridge the molecular scale of immune agents (cytokine secretion and consumption) with the functional scale of the system, where competition between cells for soluble ligands takes place. 
 
\subsubsection{Typical binding rates and some biology}
\label{sec:rates_numbers}

We summed up above the basic and general physical chemistry that drives the molecular association and dissociation of molecules in order to delineate key quantitative parameters in immunological regulation. The precise values of these rates is in fact crucial for the immune system to recognized pathogens.

At its core, the immune system can be considered as a collection of cells, called leukocytes, whose activation registers the presence of ``new'' molecules --- lipid and nucleic acid signatures of viral and bacterial infection for the innate system, or unknown proteins and peptides for the adaptive immune system --- and translates into a defensive response --- secretion of neutralizing antibodies, killing or phagocytosing of infected targets, etc. 

In the innate immune system, non-self recognition is encoded in the structure of the ligand and the receptor. Monocytes, macrophages, dendritic cells, and Mast cells are endowed with genome-encoded receptors called Toll-Like Receptors whose ligands are non-mutable elements of pathogens --- peptidoglycans and liposaccharides from the membrane of bacteria, CpG unmethylated dinucleotides derived from viruses. Recognition of non-self in that case is a ``simple'' lock-and-key proposition whereby pathogenic ligands engage the receptors and trigger a signaling response that activate innate responses. Hence, for the innate immune system, self vs. non-self discrimination is hard-wired at the molecular level and gets triggered in the early moments of an immune response. 

The adaptive immune system offers a more challenging issue. Each B and T cell clone expresses its own unique receptor (and one only), whose assembly is driven by random events (described in detail in Sec.~\ref{repertoires} and Sec.~\ref{affinity_maturation}). The ligands of these random receptors, which are called antigens, are not pre-determined --- in fact, they may not even exist at the time of birth of the organism, if for instance they are derived from fast evolving strains of viruses. 
Understanding how the binding and unbinding rates of such ligand-receptor pairs contribute to self vs. non-self discrimination is one of the core issues in quantitative immunology. 
Here we give some orders of magnitude of the binding rates to be discriminated.

B and T cell differ fundamentally by the type of antigenic ligands they interact with.
T-cell receptors (TCR) interact with their antigen on the surface of antigen-presenting cells, which they scan for anomalies. These antigens are complexes which include a short peptide (around 10 amino acid-long) produced within the antigen-presenting cells by chopping up larger proteins expressed by the cell, either from the genome or from foreign pathogens. Each peptide is loaded onto a large protein called the Major Histocompatibility Complex (MHC), forming a peptide-MHC (pMHC) complex. 
By contrast, B cell receptors (BCR) bind directly to proteins. The exact position where the binding occurs is called an epitope. Antibodies are soluble versions of the BCR with the same antigenic specificity, and also bind directly to the pathogen proteins to neutralize them.

{The affinity of antibodies (produced by B-cells) are highly variable. At the onset of an immune responses, B-cells express and secrete Immunoglobulin M (IgM), which is a pentamerized version of antibodies whose affinity for the target is weak ($K_D>10 \mu M$): the ability of secreted IgM to bind to pathogens and guide them towards eradication 
is then driven by the multimeric interaction of IgM with its target. Upon engagement of the adaptive immune response, the interplay between T-cell help and B-cell somatic hypermutation drives the maturation of antibody affinity (see Sec.~\ref{affinity_maturation}). At this point, B cells increase the affinity of the antibody they express (down to $K_D<1nM$) and switch the class of antibodies from IgM to IgG i.e. from a pentameric version to a monomeric version that does not require as much multimerization of the antibody to bind to their target. 
Such decrease in $K_D$ and increase in affinity is not driven by better association rates: IgM binds to their target with typical $k_{\rm on}$ around $10^6 M^{-1} s^{-1}$ -- similar to IgG.  $k_{\rm on}$ is essentially driven by the collision rate and the contribution of the energetic barrier in the association rate is very limited. Instead, the improvement afforded by affinity maturation is driven by a better fit between antibody and antigen, resulting in a lower $k_{\rm off}$.

In the case of T-cell antigen discrimination, the difference between ligands that will or not activate the immune response is extremely sensitive. Some peptides can elicit a robust activation with a single pMHC molecule, while mutated versions of this peptide differing by just a single amino-acid are unable to trigger T cells even in large quantities ($>10^6$). 
The striking feature in T-cell biology is that a single mutation in the peptide of the pMHC only has a marginal effect on $k_{\rm off}$, while greatly impacting its functional capacity to activate T-cells \cite{Altan-Bonnet2005,Feinerman2008b}). 
Experimentalists can measure the biophysical characteristics of pMHC--TCR interactions in solution, either by surface--plasmon resonance on soluble pMHC interacting with surface--immobilized TCR, by directly observing the surface of cells using single-molecule imaging and fluorescence energy transfer, or by measuring mechanical forces (see Sec.~\ref{sec:mechanics}). Typically, an agonist ligand, defined as a pMHC that will elicit an immune response, binds to their TCR with $\tau_{\rm off}\sim 1 - 10$s, while non-agonist ligands bind with timescales that are at least 3-fold shorter (from 0.3 to 3 s), although this varies according to the particular TCR, and also depends on the MHC class (of which there are two, as we will see later).
This relatively small difference in binding results in a large fold change ($\geq 10^5$) in response, regardless of the concentrations.
Understanding how cells can perform such sensitive discrimination is a major challenge of quantitative immunology, which we will discuss in detail in Sec.~\ref{discrimination}.

\subsection{Receptor-antigen specificity}\label{rec_antigen_spec}
T- and B-cell receptors do not interact as set of locks and keys matched to each other. Instead, each receptor can bind many different possible ligands, and vice-versa. Here we discuss numbers, data, and models that characterize this many-to-many mapping.

\subsubsection{Cross-reactivity}
Despite the huge diversity of BCR and TCR (reviewed in Sec.~\ref{repertoires}), the diversity of antigens may be even greater. While such an estimate is difficult for BCR epitopes, a rough estimate of the number of antigens can be obtained for peptide-MHC complexes. For each of the two MHC classes that exist, only a few ($\sim 5\%$) percents of peptides can be presented on any of the 6 MHC genes expressed in an individual. For peptides of length 12, this amounts to $0.05\times 20^{12}\approx 2\cdot 10^{14}$ antigens. While this might be a manageable number for humans, who harbor $\sim 10^{12}$ B and T cells, this is not by mice, who have $\sim 10^8$ such cells, and this could be an issue even for humans for longer peptides. This argument  led Mason \cite{Mason1998} to conclude that each TCR and BCR must be able to recognize more than one antigen, a phenomenon called cross-reactivity or polyspecificity. Cross-reactivity has a less discussed counterpart, which is that each antigen must be recognized by a large variety of receptors. If there are $N$ antigens, each of which can be bound by $k$ receptors, and $R$ distinct antigen receptors, each of which can recognize $\ell$ antigens, then one must have $Nk=R\ell$. To fix ideas, the ratio $p=\ell/N=k/R$, the probability that a random antigen and receptor bind together, is thought to be around $10^{-5}$ \cite{Boer1993}.

It should be emphasized that finding the sequences of binding pairs of antigens and lymphocyte receptors remains an essentially experimental question, which has received renewed attention lately thanks to the reduced costs of sequencing allowing for high-throughput binding or functional assays \cite{Birnbaum2014,Adams2016,Glanville2017,Dash2017}. However, despite increasing amounts of data on these binding pairs compiled in useful databases \cite{Shugay2018,Tickotsky2017}, and current attempts to leverage these data to make prediction using modern machine learning techniques \cite{Jurtz2018,Sidhom2018,Jokinen2019}, there exists no good predictive model of antigen-receptor specificity. Therefore, the binding models presented in the next paragraph should be viewed as useful toy models for investigating broad properties of cross-reactivity.

\subsubsection{Models of receptor-antigen binding} \label{receptor_antigen_models}
The recognition process of antigens by immune receptors is based on molecular interactions between the two proteins: the receptor protein and the antigen, which can be summarized by $K_D=c_0e^{
\Delta G(a,r)/k_BT}$, where $\Delta G(a,r)$ is the interaction free energy of binding between antigen $a$ and receptor $r$, and $c_0$ a constant.
This energy may be modeled as a string matching problem~\cite{Chao2005}. In such toy models, each interaction partner is decribed by a string of length $N$ representing the interacting amino acids. The binding energy between these two specific interaction partners is assumed to depend additively on the interaction between pairs of facing amino-acids:
\beq\label{eq:specificity}
\Delta G(a, r) = \sum_{k=1}^N {J_k}(a_k, r_k),
\eeq
where the interaction matrix, $ {J_k}(a_k, r_k)$ is an $q\times q$, where $q$ is the size of the space that defines the variabilty of residues at each position. If we describe each residue as one of the $q=20$ amino acids, we need to define an amino acid interaction matrix. The Miyazawa-Jernigan matrix \cite{Miyazawa1996} was used in such a model to suggest that thymic selection favors moderately interacting residues in TCR \cite{Kosmrlj2008}, or more recently to study the effect of thymic selection on tumor antigenic peptides \cite{George2017}. The maximum affinity of a receptor to a large set of antigens, which plays a key role in thymic selection (Sec.~\ref{sec:thymic}), is amenable to a statistical mechanis analysis of extreme value statistics \cite{Kosmrlj2009,Butler2013a}, leading to universal features that are robust to the details of the models.

Alternatively, reduced models have been considered, where each residue is defined using a projection that attempts to capture the main biophysical and biochemical properties in a generalized ``shape space'' \cite{Perelson1979}. In the initial string model \cite{Detours1999a}, $a_k$ and $r_k$ were bounded natural integers, and $J_k=a_k \oplus r_k$, where $\oplus$ is the exlusive OR operator acting on each digit of the binary representations of $a_k$ and $r_k$. This choice was motivated by algorithmic ease rather than biophysical realism. These models were used by Perelson and collaborators to investigate the effect of thymic selection \cite{Detours1999a,Detours2000} or the immune response \cite{Chao2005}.

A more drastic reduction is to binarize the antigen: for each epitope position, the amino-acid is defined either as the one present in the viral wild type ($a_k=1$) or another one (a mutant, $a_k=-1$) \cite{Wang2015a}. The selective pressure exerted by each receptor, which acts as a ``field'' on $a_k$,  was drawn as a number $r_{k}$ from a random continuous distribution, which reduces the model to
$J_k(a_k,r_k)=r_k a_k$.
In \cite{Wang2015a}, it was additionally assumed that certain positions of the viral epitope were constrained to take the wildtype value, $a_k=1$ for $k>M$, because of strong conservation at these sites.
A similar description in terms of binary strings \cite{Nourmohammad2015} assumes both the viral epitope and BCR to be binary strings ($a_k=\pm 1,\ r_k=\pm 1$), with a fixed interaction strength,
$J_k(a_k,r_k)=\kappa_k r_k a_k$,
with again conserved viral positions $k>M$ for which $a_k=1$ is imposed.
Within this description a mismatch between the receptor and the viral ``spins'' induces an energy penalty. The convention is such that smaller energies imply stronger binding and better recognition. Similar models have been used recently for co-evolution of BCR and HIV \cite{Wang2015a, Nourmohammad2015}, the results of which we describe in detail in subsection~\ref{affinity_maturation}.

\subsubsection{Data-driven receptor-antigen binding models}
To go beyond the toy models described above, one must experimentally map out the binding energies between specific pairs of antigens and receptors. This can be done in a massively parallel way using high-throughput experiments assaying the binding affinity of many pairs in single experiments. Such an approach was applied to a deep mutational scan experiment reporting the dissociation constant, $K_D$, of a large number of antibody variants against a fixed antigenic target, fluorescin \cite{Adams2016}.

The simplest model assumes an additive contribution of each residue to the binding free energy $E(a,r)=\ln (K_D/c_0)$ as in \eqref{eq:specificity}, but with $J_k(a_k,r_k)=h_k(r_k)$ fixed to a single value of the antigen $a$. However, statistical analysis shows that such an additive model is not able to capture all the variability in the binding energy, accounting for less than $2/3$ of the variability in double and triple mutants. Epistasis, defined as non-additive effects, accounts for $25-35\%$ of variability in the binding energy between antibodies and the antigen and a large fraction of the epistasis was found to be beneficial \cite{Adams2019}.

Non-additivity in the binding energy is likely to be a general feature of both TCR and BCR. More sophisticated models including intra-protein or higher-order interactions will be needed to accurately predict receptor-antigen affinity. 
Ultimately, it would be interesting to reconcile the useful picture of an effective binding shape space with affinity landscapes inferred from data.
A major roadblock is that most current experimental techniques only allow for varying one element of the receptor-antigen pair, either testing a library of receptors against a fixed antigen, or a library of antigens against a fixed receptor \cite{Birnbaum2014}. Full characterization of the binding landscape would require new methods to test double libraries of antigens and receptors in an ultra-high-throughput manner. 

\subsubsection{Modeling immunogeneticity}\label{sec:immunogeneticity}
Rather than modeling the full details of the TCR-antigen interaction, an alternative strategy is to focus on the immunogenic potential of particular antigens, without explicitly modeling the TCR. Such an approach was followed to predict the response of cancer patients to immunotherapy. The prediction is based on the knowledge of the pMHC antigens presented by the tumor cells, called neo-antigens \cite{Luksza2017}. To estimate the ability of a neoantigen $a$ to be recognized by the TCR repertoire, a score was calculated to evaluate its similarity to a database $D$ of antigens known to elicit an immune response:
\beq
S(a)=\sum_{a'\in D} e^{ks_{a,a'}},
\eeq
where $s_{a,a'}$ is the alignment score between the neoantigen $a$ and antigens $a'$ from the database, and $k$ an adjustable parameter. The immunogenicity of $a$ is then predicted to be:
\beq
I(a)=\frac{S(a)}{S_0+S(a)}.
\eeq
This quantity was combined with the likelihood $A(a)$ that neoantigen $a$ is presented by class-I MHC (predicted by a neural network model \cite{Andreatta2015}) to form a global score $-I(a)A(a)$. With its minus sign, this score reflects the fitness of the tumor cells carrying antigen $a$. These scores were found to be predictive of patient survival after immunotherapy.

While this similarity-based approach holds great promise for generally predicting the immunogeneticity of antigens, further tests are needed to validate the method for broader use. The link between survival and TCR recognition is indirect, and is complicated by the fact that many neoantigens and tumor clones are involved. Direct functional tests of the immune response against libraries of antigens would help consolidate the foundations of this approach.

\section{Antigen discrimination}\label{discrimination}
We now move to the cellular scale, and describe how the signal propagates in minutes from antigen-receptor binding to biochemical networks, allowing for fine antigen discrimination. We leave to a later section the discussion how this signal is later integrated to make decisions about cell fate and response on timescales spanning hours to days (Sec.~\ref{cellfate}).

\subsection{T cells}\label{Tcelldiscrimination}

\subsubsection{Kinetic proofreading for ligand discrimination}~\label{proofreading}

To model T-cell antigen discrimination, we remain within the quantitative parameters of the ``lifetime dogma,'' in which the lifetime $\tau_{\rm off}$ of TCR-pMHC is the sole determinant of the discrimination process. While this dogma was derived from the biophysical measurements on TCR/pMHC interactions \cite{Feinerman2008b},  as any dogma it must be revisited regularly for exceptions and refinements. For example, recent studies in the mechanics of TCR triggering have highlighted a new mode of signal initiation, as we shall discuss in Sec.~\ref{sec:mechanics}.

Upon engaging their pMHC ligands, T-cells trigger a cascade of phosphorylation events, which consist of adding phosphoryl groups to conserved residues of TCR-associated chains. The phosphorylations in turn trigger other reactions, ultimately activating transcription factors that modulate gene regulatory responses.

Decision making relative to antigen discrimination can thus be modelled at the phenomenological level by the state of phosphorylation.
We assume that, when the TCR complex has accumulated a certain number of phosphoryl groups $N_{\rm P}$, it flips a digital switch (to be specific, the activation of the NFAT or ERK molecules \cite{Altan-Bonnet2005}) which defines the onset of T-cell activition.

One of the first quantitative models that tackled the exquisite specificity of antigen discrimination during T cell activation was proposed by McKeithan in 1995~\cite{McKeithan1995}, building upon the classical kinetic proofreading (KPR) scheme proposed by Hopfield and Ninio in the context of protein translation~\cite{Hopfield1974,Ninio1975}.
Upon binding its ligand, TCR progresses through the different steps of the cascade, controlled by a phosphorylation rate $k_{\rm P}$ and a de-phosphorylation rate $k_{\rm deP}$ (Figure~\ref{fig:AKPR}: A). A key aspect of the kinetic proofreading is that, at each step, it is assumed that the complex is de-phosphorylated into unoccupied receptors upon ligand unbinding: this is a reasonable assumption because the CD45 phosphatase ---\,the enzyme that removes phosphoryl groups from the TCR complex\,--- prevents rebinding of TCR by the antigen due to its large size, thus ensuring complete de-phosphorylation before the next binding event.
A consequence of this resetting upon unbinding is that antigens forming short-lived complexes with the TCR will typically only accumulate a small number of phosphoryl groups, below the threshold $N_{\rm P}$ necessary to activate the immune response, while long-lived complexes can progress more deeply through the phosphorylation cascade. 

\begin{figure*}
\includegraphics[width=\linewidth]{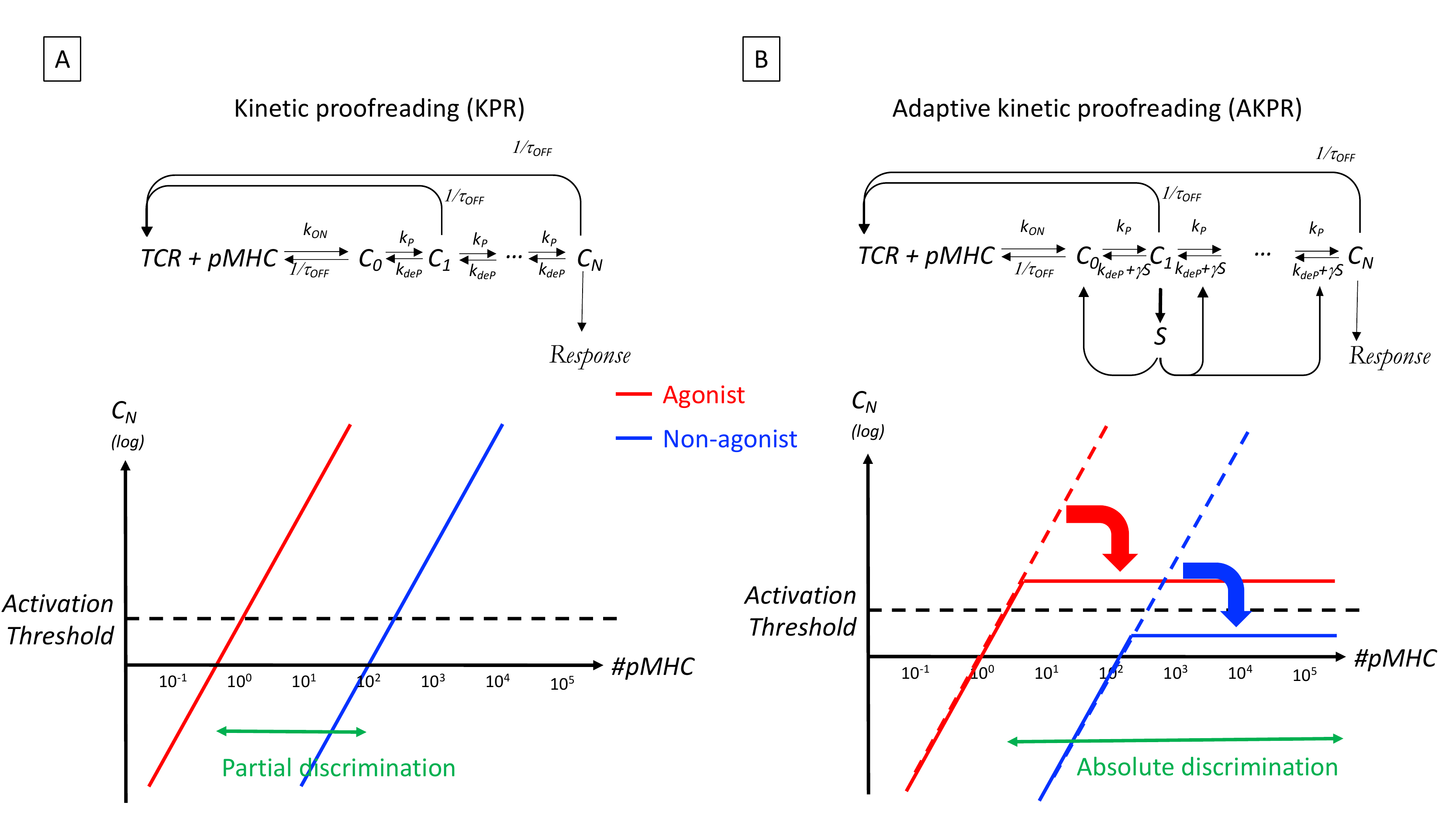}
\caption{\textbf{Biochemical scheme to reconcile specificity, sensitivity and speed in antigen discrimination.} A) a classical kinetic proofreading (KPR) scheme amplifies differences in TCR phosphorylation ($C_N$) based on the lifetime $\tau_{\rm OFF}$ of the pMHC--TCR complex but gets overwhelmed by a large quantity of pMHC and achieves only partial discrimination of antigens. B) the adaptive kinetic proofreading (AKPR) scheme relies on the activation of a phosphatase ($S$) to limit spurious activation by a large quantity of non--agonist ligands (see text for details). In the KPR scheme (A) a non-agonist (blue line) can activate a T-cell if it is present in high concentrations. The agonist (red line) activates T-cells also when it is present at lower concentrations. In the AKPR scheme (B) even high non-agonist concentrations cannot lead to T-cell activation, while an agonist still manages to activate T-cells.}
\label{fig:AKPR}
\end{figure*}

Neglecting dephosphorylation ($k_{\rm deP}\ll k_{\rm P},k_{\rm off}$), the number of pMHC--TCR complexes $C_i$ that accumulate $i$ phosphorylations scales with $i$ as \cite{McKeithan1995}:
\beq
C_{i} = C_{0} \left( \frac{k_{\rm P}}{k_{\rm P}+k_{\rm off}} \right)^{i}.
\eeq
At steady state, the total number of complexes, $C_{\rm total} = C_0\sum_{\rm i=1}^{N_p} C_i $, 
is given by a second-order equation describing the balance between the rate of binding  between free antigens and free receptors, $k_{\rm on}/(V.\mathcal{N}_A) (L-C_{\rm total}) (T-C_{\rm total})$, where $L$ is the total number of antigens, $T$ the total number of TCR, $V$ the volume, $\mathcal{N}_A$ --the Avogadro number, and the rate of unbinding events, $k_{\rm off}C_{\rm total}$.
In the limit when TCR are not limiting, $C_{\rm total} \ll T$, this balance yields the total number of complexes \cite{Francois2013}:
\beq
C_{\rm total} \sim \frac{ (k_{\rm on}/(V \mathcal{N}_A)) T L}{k_{\rm off} + (k_{\rm on}/(V \mathcal{N}_A)) T}.
\eeq

Further assuming a relatively slow phosphorylation rate, $k_{\rm P}\ll k_{\rm off}$, the number of fully phosphorylated complexes $C_{N_{\rm P}}$ scales as $\propto k_{\rm on} Lk_{\rm off}^{-N_{\rm P}}=L k_{\rm on}\tau_{\rm off}^{N_{\rm P}}$ as a function of the antigen characteristics --- concentration $L$ and binding affinity $\tau_{\rm off}$.
This scaling shows how alterations in the pMHC ligand impacting $\tau_{\rm off}$ will get amplified into large differences in the amount of phosphorylation accumulating on the TCR:
\beq
\frac{C_{N_{\rm p}}({\rm agonist})}{C_{N_{\rm p}}({\rm non-agonist})}=\left(\frac{\tau_{\rm off}({\rm agonist})}{\tau_{\rm off}({\rm non-agonist})}\right)^{N_p}.
\eeq

In other words, kinetic proofreading ``reads off'' the lifetime of the pMHC--TCR complex and amplifies differences in the output. As with the original KPR scheme of Hopfield and Ninio, this amplification implies energy expenditures caused by the phosphorylation steps.
Structurally speaking, the TCR complex contains 20 phosphorylation sites (specifically, tyrosine residues), which can potentially participate in the kinetic proofreading cascade. This means that $N_{\rm P}$ can be as large as 20, and a two-fold change in pMHC--TCR lifetime could be amplified into a $2^{20} \approx 10^6$ fold difference in the number of phosphorylated receptors $C_{\rm N_p}$. This mechanism can account for the fact that TCR ligands with minute differences in affinity may elicit very different signals. In addition, it is consistent with the lifetime dogma, in that $\tau_{\rm off}$ has a much more determining impact on activation than $k_{\rm on}$.

However, KPR is insufficient to capture all aspects of T cells' ability to discriminate between structurally-related ligands. As pointed out by Altan-Bonnet \& Germain~\cite{AltanBonnet2005}, T cells not only achieve high specificity in ligand discrimination, but they also maintain high sensitivity, as they are be able to trigger activation from a single agonist ligand~\cite{Sykulev1996,Irvine2002};. In addition, they respond very fast, typically within minutes of encountering an antigen presenting cell. These requirements for speed, sensitivity and specificity are hard to achieve all at once. For example, a large number of phosphorylation steps in the KPR scheme allows for high discriminability, but also implies a slower response, as each step must be slow enough to discriminate between agonists and non-agonists. It also affects sensitivity, as more steps imply a lower chance of making it the activation step. These considerations demonstrate how quantitative modeling invalidates a bare KPR scheme, and calls for additional mechanisms.

\subsubsection{Adaptive kinetic proofreading}

To fulfill the conflicting requirements of specificity, sensitivity, and speed, one must expand on the simple kinetic proofreading scheme by adding a mechanism of adaptation to modulate the proofreading steps.

The resulting
adaptive kinetic proofreading (AKPR) model \cite{Altan-Bonnet2005,Francois2013}, which is based on experimental evidence presented by \v{S}tefanova  {\it et al.}~\cite{Stefanova2003}, quantitatively accounts for key features of ligand discrimination by the TCR.

The adaptation module is introduced through a negative feedback mediated by a phosphatase $S$ (which removes phosphoryl groups), which is itself activated by the engaged TCR in state $C_1$, so that its steady-state concentration is $S=C_1/(C_1+K_S)$, where $K_S$ is a model parameter. The dephosphorylation rate is then enhanced in proportion to the phosphatase concentration, to $k_{\rm deP}+\gamma S$. The equations for steady state have a simple closed form, whose solution is given by the root of a fourth-order polynomial.

Fig.~\ref{fig:AKPR} presents a simple graphical argument illustrating how KPR and AKPR schemes perform in their task of discriminating ligands.
In the adaptive kinetic proofreading scheme, the negative feedback mediated by $S$ enforces a higher selectivity of antigen discrimination without affecting the high sensitivity of the response.

To better understand how this is possible, consider a large concentration $L$ of non-agonists. Their unspecific engagement with the TCR will cause the first phosphorylation of the complex into $C_1$, in proportion to their concentration $L$. This $C_1$ state activates the phosphatase $S$, which in turn increases the specificity of the TCR by accelerating its dephosphorylation. When $L$ is large, the amount of phosphatases, and hence desphosphorylation rate, will exactly balance out the amount of non-agonist-bound TCR complexes, preventing these complexes from reaching the activation state $C_{\rm N_P}$ (see Figure~\ref{fig:AKPR}~B).
As a result, for non-agonists $C_N$ is essentially flat and stays below the threshold for T cell activation, regardless of $L$.

The activation threshold for $C_{N_{\rm P}}$ must to be set at very low values to account for the extreme sensitivity of the TCR signalling cascade (a single ligand is sufficient to trigger activation). Hence, one must introduce stochastic equations to appropriately tackle the low number of molecules involved in the TCR decision making, although this does not affect the qualitative features of this particular model.

To conclude, the adaptive kinetic proofreading scheme satisfies the joint requirements of ligand discrimination and sensitivity.  It can explain how a single agonist ligand can trigger T cell activation, while a large number ($>10^5$) of non-agonist ligands cannot. Additionally, it can account for the speed of ligand discrimination by T leukocytes, as only two steps can be sufficient to sort ligands~\cite{Lalanne2013}.
It is important to emphasize again the importance of a quantitative approach and of modeling to appreciate why achieving speed, sensitivity and specificity in ligand discrimination is indeed such an amazing feat of the immune system.

\subsubsection{Coupling mechanics and biochemistry: the significance of forces for ligand discrimination}~\label{sec:mechanics}

\begin{figure*}
\includegraphics[width=\linewidth]{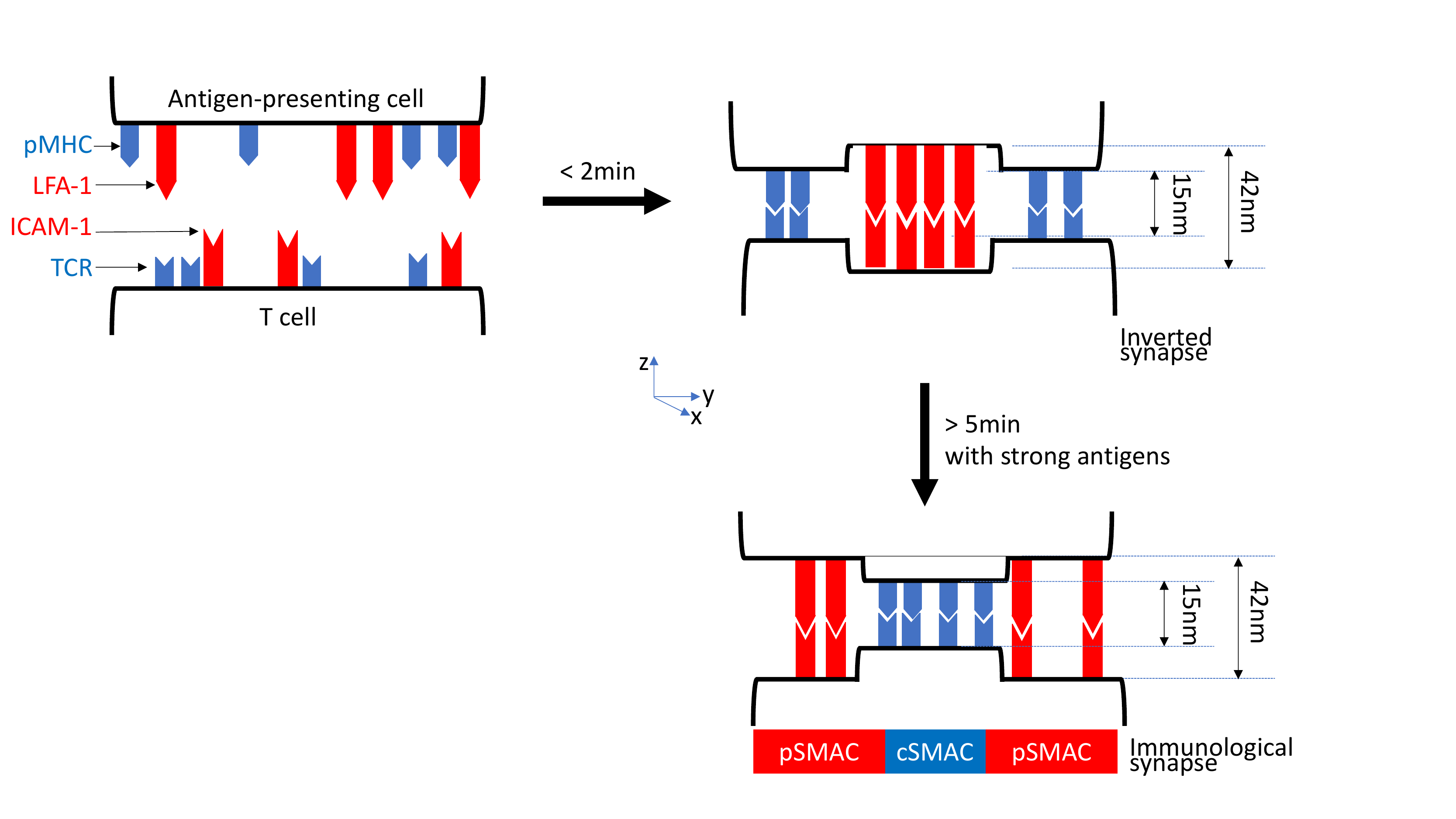}
\caption{\textbf{Molecular sorting in the immunological synapse.} Upon activation by antigen-presenting cells, T-cells reorganize their membrane proteins to form a bullseye structure (so--called immunological synapse) that stabilizes cell-cell interactions and drives cellular functions. Coupling of mechanical forces derived from the curvature of the membrane and differences in the sizes of pMHC/TCR and integrin (LFA-1/ICAM-1) complexes drives the formation of two ''Supramolecular Assembly Complexes" (central: cSMAC; peripheral: pSMAC). At early time points (< 2min) or for weaker antigens, the systems form an inverted synapse with the tall, rapidly-formed integrin bonds concentrating in the center (red bonds), and pushing the slowly-formed pMHC-TCR bonds to the periphery (blue bonds). For longer timescales and when strong antigen ligands are present, these molecular complexes get sorted into a classical immunological synapse to alleviate the cost of negative curvature in the inverted synapse. A Ginzburg--Landau model (as introduced by Qi~et al.~\cite{Qi2001}) demonstrates that such self-organized sorting could help in antigen discrimination. }
\label{fig:immunological_synapse}
\end{figure*}

One of the most striking events in the T-cell activation process is the dynamic membrane reorganization into what has been called an immunological synapse. Within minutes of signal initiation, the antigen receptors on the surface of lymphocytes congregate at the center of the interface with their antigen-presenting cells (this has been dubbed a central Supramolecular Assembly Complex or c-SMAC), while transmembrane receptors and phosphatases needed for signal propagation accumulate at the periphery of the contact area (p-SMAC and d-SMAC respectively, standing for peripheral and distal). Such a bullseye pattern has been studied quantitatively using high-resolution and single-cell time-lapsed microscopy. The functional significance of immunological synapses has been identified for the long-term activation of T cells, such as directed killing of antigen-presenting cells or directed secretion of cytokines \cite{Huse2006}.
As the cell biology of immunological synapse formation was being studied, the question arose as to whether such cellular structure could contribute to self vs non--self discrimination. Qi {\it et al.}~\cite{Qi2001} introduced a Ginzburg-Landau model for the mechanical deformation of the T-cell membrane during synapse formation. The model assumes the presence of several receptor-ligand pairs, indexed by $i$ (e.g. the TCR-pMHC pairs is one these pairs).
The free energy $\mathcal{F}$ for the membrane system is estimated as:
\begin{equation}
\label{eq:GLmodel}
\begin{split}
\mathcal{F}&=  \sum_i \frac{\lambda_i}{2}\int {dx dy \,C_i(x,y)\left[z(x,y)-z_i^{(0)}\right]^2} \\
&+\frac{1}{2}\int {dx dy\, \left[ \gamma (\nabla z(x,y))^2 + \kappa (\nabla^2 z(x,y))^2 \right]},
\end{split}
\end{equation}
and the system evolves according to:
\begin{eqnarray}
 \frac{\partial C_{i}}{\partial t}&=&  D_i\left[\nabla^2 C_{i}+\frac{1}{k_B T}\nabla \left(C_{i}\nabla\frac{\delta {\mathcal F}}{\delta C_{i}}\right)\right] \\ \nonumber
 &&+k_i^{\rm on}(z)L_iR_{i}-k_i^{\rm off}C_{i}+\zeta_i,
 \end{eqnarray}
and  
\beq
\frac{\partial z}{\partial t}=  -M \frac{\delta \mathcal{F}}{\delta z} +\zeta,
\eeq
where $z(x,y)$ is the membrane coordinate for the T-cell membrane, $R_i$, $L_i$, and $C_i$, are the surface densities of receptors, ligands, and receptor-ligand complexes of the $i^{\rm th}$ type,  $z_i^{(0)}$ is the length of the $C_i=R_i$-$L_i$ complex bond at rest; and the $\zeta$'s are thermal noises for each process (their distributions are not explicitely stated in the original publications~\cite{Qi2001}). The free energy $\mathcal F$ sums up the elastic energies of each type of complex $i$ (with stiffness $\lambda_i$), and the elastic energy of the membrane. $k_i^{\rm on}$ and $k_i^{\rm off}$ are the binding and unbinding rates of ligand-receptor complex formation.

This model is purely mechanical and passive, in the sense that no energy is injected nor is there any biochemical feedback. Yet it captures salient features of the spontaneous formation of immunological synapses. Importantly, synapse formation strongly depends on the lifetime of ligand-receptor complexes, and thus allows for fine discrimination between agonist and non-agonist ligands. The model also emphasizes the functional relevance of size differences in ligand-receptor pairs. For instance, TCR-pMHC complexes have $z^{(0)}_i=15$ nm, while adhesion complexes such as ICAM-1/LFA-1 are much larger, $z^{(0)}_i=42$ nm, as it was confirmed experimentally \cite{Choudhuri2005,Choudhuri2007}.

This model makes several predictions that were validated experimentally. First, it explains how the formation of the characteristic bullseye pattern emerges from the passive sorting of molecules based on their size, upon coupling with the mechanical deformation of the surface membrane. Second, the model predicts an optimum for the lifetime of the pMHC-TCR complex for which the accumulation of antigen in the center of the synapse is maximal. For weak antigens, the membrane energy is dominated by the binding of adhesive complexes, and few antigen-TCR bonds can form because of molecular crowding. For very strong antigens,  at intermediate time points ($\approx$ 5min) or with ligands of intermediate affinity, the membranes of the T-cells adopt an inverted bullseye pattern with TCR being in the p-SMAC and integrins being in the c-SMAC. Such patterns can be explained when the binding affinity of the integrin binding is larger than the binding energy of the weak ligands. In this case the integrins populate the center of the c-SMAC and make the deformation of the membrane ''affordable'' energetically (see Figure~\ref{fig:immunological_synapse}).
While this ``bell-shaped'' activation as a function of antigen affinity has been documented experimentally \cite{Valitutti1995a,Kalergis2001}, it has not been  consistently observed and there is a sufficient number of exceptions~\cite{Holler2000,Lever2016} to diminish its significance, especially at the time when clinicians are engineering potent chimeric antigen receptors with unphysiologically--large affinity for their ligands~\cite{Posey2016,Schmitt2017}. 
Third, the state of the  membranes of both the T cell and antigen-presenting cell is expected to affect the synapse and thus T-cell activation, as was observed in experiments where the cytoskeleton has been depolymerized \cite{Valitutti1995b}.

On the other hand, the model also fails to explain more recent observations, such as the ability of T cells to respond to a single agonist ligand while ignoring the presence of many non-agonist ligands. More recently, its interest has been rekindled by new observations emphasizing the importance of mechanics for ligand discrimination.

Using a biophysical setup to measure forces and lifetimes of individual pMHC-TCR pairs under tension, Zhu and colleagues \cite{Zhu2013,Liu2014} discovered two types of bonds depending on the peptide sequence: slip bonds and catch bonds.
The lifetime of slip bonds decreases monotonously under tension: in other words, they break more easily when pulled. By contrast, the lifetime of catch bonds {\em increases} under tension. Molecular dynamics simulations \cite{Sibener2018,Wu2019} suggest that the bonds tension helps expose new residues in pMHC, inducing a better contact with the TCR.
At high forces, catch bonds do ultimately break as well, so there exists an optimal force of around 10 pN for which the catch bond lifetime is longest. 

This distinction between slip and catch bonds has been suggested to have functional significance in terms of T-cell activation. For instance, it was shown \cite{Sibener2018} that TCR-pMHC pairs with similar binding constants elicited different activation modes in T cells, with the stimulatory pair forming a catch-bond and the non-stimulatory a slip bond.

How these catch bonds relate to signalling remains to be explained in detail, but this new effect may be critical for predicting binding pairs, and immunogenic neo-antigens of crucial relevance to immunotherapies (see Sec.~\ref{rec_antigen_spec}).
However, the ability of pMHC to activate the TCR signalling pathway might still be set by the lifetime of the pMHC-TCR complex, but with a ``catch'': this lifetime would have to be assessed under the tension exerted by the membrane dynamics, e.g. 10 pN. This would also explain the dependence of activation upon membrane properties such as stiffness.
The Ginzburg-Laudau model of Eq.~\ref{eq:GLmodel}
will need to be revisited to account for the existence of catch bonds, e.g. by letting the $k_i^{\rm off}$ of pMHC-TCR pairs depend on the tension $\lambda_i(z-z_i^{(0)})$, possibly affecting how ligands get sorted by mechanical forces in the immunological synapse.

Additional measurements in recent years (in particular, using super--resolution microscopy \cite{Cai2017}) emphasize the active role that cytoskeletal rearrangements may play in concentrating TCR in the synapse \cite{Dustin2010} and the complex interplay between membrane ruffling, receptor sorting, and mechanical tensions. A more complete quantitative model would help identify key limiting steps deciding how strongly T cells get activated in different contexts --- different antigen-presenting cells, or different co-stimulatory contexts.

\subsection{B cells}\label{Bcellsignaling}

As mentioned before, the affinity of B-cells with their cognate antigen undergoes a Darwinian selection process that improves their affinities from $K_D=10\,\mu$M down to 100pM. Simple counting of the number of occupied receptors in equilibrium could be sufficient to enforce antigen discrimination in B-cells.
As for TCR binding to pMHC, the association rate of BCR to antigens is constant for all antigens. But unlike TCR-pMHC binding, which is subjected to an activation barrier, this rate is essentially diffusion limited, $k_{\rm on}>10^6 M^{-1} s^{-1}$.
Strong and weak antigens thus only differ by their binding lifetimes $\tau_{\rm off}$, which varies between $0.1 s$  and $10^4$ s, suggesting a possible kinetic proofreading scheme as for TCR~\cite{Tsourkas2012}. However, since the number of potential phosphorylation sites in the BCR complex is small compared to TCR, such a mechanism may not be as important.

The previously mentioned monomeric version of the BCR, IgG, is itself actually a dimer (so that IgM is a pentamer of dimers), with two binding sites.
By analogy with other dimeric receptors on the surface of cells (e.g. Epidermal Growth Factor Receptors or Insulin Receptors~\cite{Blinov2006,Lemmon2010}), one could assume that this dimerization would help concentrate kinases (the enzymes responsible for phosphorylation) around the receptors, causing them to form clusters.
This ``cross-linking model'' of BCR activation is appealing because it is consistent with the formation of microclusters of BCR on the surface of B cells in response to antigens, as observed by super-resolution microscopy. 

An alternative model proposed by Reth and colleagues~\cite{Yang2010} assumes that receptors cluster even in the absence of antigens. When in clusters, BCR inhibit each other and do not signal, but they can assemble and disassemble dynamically. Monomers coming out of these clusters are more prone to signaling, and antigen-binding stabilizes this monomeric state supposedly because of steric hindrances preventing the return into the clustered/inhibited state.
This ``dissocation-activation'' model explains how B cells limit spurious activation of its $\sim 120,000$ surface BCR through their clustering, with signaling made only possible by isolated, antigen-bound monomers. The measurements of Reth and colleagues pose a theoretical challenge in terms of understanding the role dynamic clustering-release-re-clustering of BCR during B cell activation. A full theoretical model of such process and its impact on antigen discrimination remains to be proposed.

One observation that must be taken into account when considering B cell activation is the role of membrane spreading and contraction. Upon initiation of the signaling cascade, B cells rapidly ($<$ 10 min) reorganize their cytoskeleton and membrane, first by spreading to capture as many antigens on the presenting cells as possible, second by contracting to ``concentrate'' the active receptors. The quantitative model proposed in~\cite{Fleire2006} accounts for binding and unbinding events driving cell spreading and contraction. The model predicts that such a dynamic process can increase dramatically the number of agonist ligands that get captured, compared to a static interface, enhancing the difference between weak and strong ligands, while weaker affinity antigens fail to concentrate.
A critical aspect of the model is that, if B cells fail to trigger a sufficient number of BCR by 1 min (a hard cut-off), they terminate the process and shut down their signaling response. Alternatively, the activation response switches to a contraction phase with the surface area decaying according to the phenomenological law $A_{max} t^{-0.35}$  with $A_{max}$ being the largest area the B cell spreads to, $t$ is the time, and 0.35 is an experimentally-determined exponent.
This model is phenomenological as it does not model explicitly the biochemical mechanism driving the spreading and contraction, and makes {\em ad hoc} assumptions about their behaviour. Yet it illustrates quantitatively how such membrane dynamics can help discriminate ligands.

To conclude, although the issue of antigen discrimination in B cells may not be as stringent as for T cells, it poses interesting quantitative issues for a different parameter range of binding constants ($K_D=100$ pM - 10 $\mu$M), over longer timescales ($>$10min), and using different cellular mechanisms: receptor clustering and membrane dynamics. It is worth recalling that this initial recognition process connects to additional processes over longer timescales, such as validation by helper T-cells, and dynamics within germinal centers. We will come back to these questions in Sec.~\ref{affinity_maturation}.

\subsection{Coarse-graining of molecular details and model reduction}

As we go deeper into the molecular details of immune recognition, the number of molecular species, reactions, intermediates, and therefore parameters explodes. This poses a challenge for a number of reasons. First, the effective behaviour of the system as a whole may still be relatively simple, suggesting that simpler phenomenological models may describe these processes equally well. While the variables of such models may be hard to relate directly to molecular entities, they are easier to interpret and allow for better analytical progress and predictions. Second, even assuming that the full complexity of all interactions is needed, large numbers of parameters are likely to lead to overfitting problems, meaning that many parameters or combinations of parameters are underdetermined, undermining the accuracy of predictions. And even when they can be determined, it is not always clear which ones need to be fine-tuned to ensure proper function, or what are the broad design principles presiding over their choices. In Sec.~\ref{modelselection}, we briefly review the principles of model selection -- the classical approach of reducing model complexity from statistics. However, that approach requires to have first defined a hierarchy of models to test, from least to most complex. Besides, model selection relies on goodness of fit as a criterion to evaluate models, while in many cases we may be more interested in capturing the principal features of a biological function, rather than fitting all the data.

In another approach to reducing complexity, Fran\c cois and Hakim \cite{hakim-2004} developed an method for generating simple molecular networks {\em in silico} that realize a desired biological function, simply based on a genetic algorithm that selects the ``best'' solutions. Applied to the problem of absolute ligand discrimination reviewed in Sec.~\ref{Tcelldiscrimination}, this method infers a class of network motifs, called ``adaptive sorting'', that recapitulates known features of T cell activation \cite{Lalanne2013}. In particular, it predicts the emergence of kinetic proofreading and biochemical adaptation. However, the chemical species and reactions of the networks produced by that method may generally not be directly related to the known actors of the phenomenon under study.

To reduce model complexity while keeping close to the details of actual biochemical reactions, one can instead start from a complex biochemical network described by many ordinary differential equations, and ``prune'' its parameters by setting them, individually or by their combinations (ratio or products), to 0 or infinity \cite{Proulx-Giraldeau2017}. Applying this strategy to a complex model of T cell recognition \cite{Lipniacki2008}, which contains close to 100 parameters, shows that its behaviour can be boiled down to just three coupled differential equations highlighting its main features of adaptation and discrimination, and reveals the broad design principles that implement these features.

Beyond this particular example, this kind of approach has great potential for helping to make sense of complex biological systems with many entities and interactions, as is often the case in the immune system.

\section{Cell-to-cell communication through cytokines}\label{cytokines}

Cytokine communication is critical to synchronizing the activation of various immune cells and to bridging multiple spatial-temporal scales in immune responses, from local or individual cell activation to global, systemic responses. Cytokines are small glycoproteins that get produced and secreted by all cells, immune or not, with varied dynamics, amplitude and frequency. These cytokines then diffuse and bind to receptors present on the surface of adjacent cells to elicit a signaling response that trigger a gene regulatory response. In short, cells use cytokines to communicate between themselves. In this section, we discuss quantitative models that have been introduced to model how individual leukocytes respond to cytokines (\eg using the JAK-STAT pathway) over multiple time- and length-scales. 

\subsection{Cytokine signaling and the JAK-STAT pathway}

Cytokine receptors are heterodimers, or more rarely heterotrimers, so that they carry at least two intracellular signaling domains. The JAK-STAT pathway is the dominant pathway engaged by cytokine signaling. Upon binding by its cognate cytokine, the receptor undergoes a conformational change that presents phosphoryl attachment sites on the JAK (Janus Kinase) receptor to face the kinase domain that induces phosphorylation, which in turn leads to activation. Then a protein called STAT (Signal Transducer and Activator of Transcription) interacts with the intracellular domains of the receptors, gets phosphorylated by the activated JAK, and dimerizes upon release from the receptor. Dimers of STAT then translocate into the cell nucleus, bind to the chromatin in specific sites and elicit a transcriptional response. Such signal transduction is one of the simplest biological pathway connecting the extracellular environment and its messenger cytokines to a transcriptional response (Figure~\ref{fig:JAK_STAT}). Its complexity is encoded in the large number of pairs of cytokines ($>40$) and their receptors that cells can express, as well as the 7 forms of STAT and $7^2=49$ homo- or hetero-dimers that they can form upon activation \cite{Altan-Bonnet2019}. Additionally, the dynamics of cytokine signaling enriches the biology of these pathways, with the existence of positive and negative feedback regulations: cytokine signaling inducing the expression of additional cytokine receptors, cytokine degradation, cytokine receptor endocytosis, expression of negative regulators such as Suppressor of Cytokine Signaling (SOCS). Here we present simple derivations that can help understand the quantitative regulation of cytokine communication within the immune system, which underlies the coordination and orchestration of the immune response.

\begin{figure*}
\includegraphics[width=\linewidth]{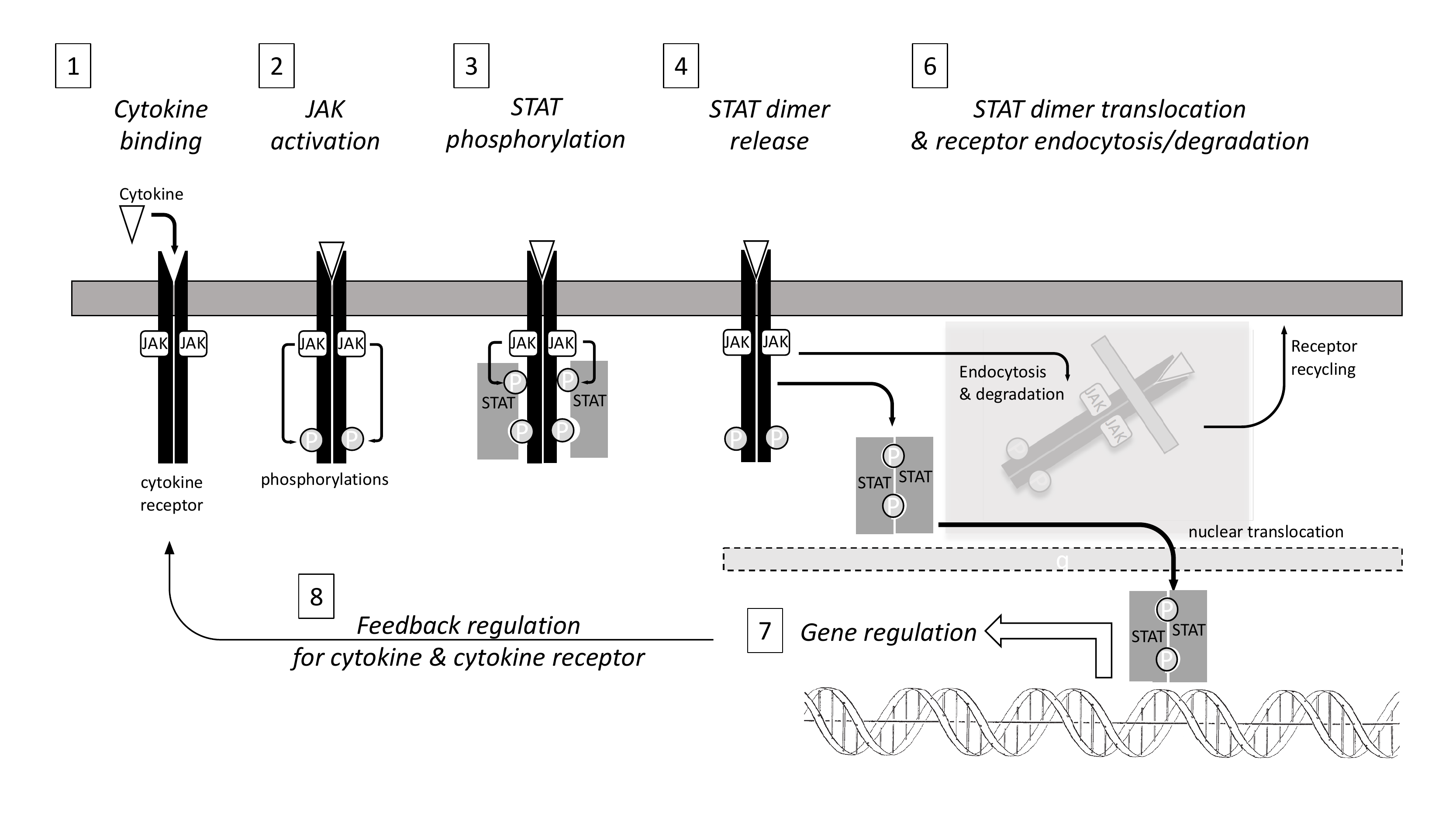}
\caption{\textbf{Model of the regulation of JAK-STAT signaling in response to cytokines. This model in the adiabatic regime can be solved analytically.}}
\label{fig:JAK_STAT}
\end{figure*}

\subsubsection{Cytokine binding and signaling at equilibrium}

In this section, we discuss the contribution of the field of quantitative immunology to understanding the JAK-STAT pathway as a signal transduction cascade in well-mixed conditions. Spatial heterogeneities in cell-to-cell communication via cytokines will be tackled in section~\ref{sec:cytokine_comm}. 

Within an individual cell, the biochemical reactions of the JAK-STAT pathway can be treated as well-mixed:  the typical concentrations of molecules are fairly high, with $\sim 1000$ molecules per cell of diameter 10$\mu$m, which translates into concentrations $\sim 100$nM. We also assume that most reactions are essentially diffusion-limited, as discussed before: $k_{\rm on}>10^7 M^{-1}s^{-1}$. The signal transduction cascade for JAK-STAT cascade can generically be modelled in a step-wise manner, as sketched in Figure~\ref{fig:JAK_STAT}.

The kinetics of cytokine binding to its receptor activating JAK into ${\rm JAK^*}$  can be modeled as a simple bimolecular process:
\beq\label{LR_basic}
\frac{d}{dt}{{\rm \#Receptor^*}}=k_{\rm on}[{\rm Cytokine}]{\rm \#Receptor}-k_{\rm off}{\rm \#Receptor^*},
\eeq
where ${\rm \#Receptor^*}$ is the number of cytokine-engaged receptors,
$[{\rm Cytokine}]$ is the cytokine concentration. Because binding is very strong (typically $K_D=k_{\rm off}/k_{\rm on} \approx 10pM$ with $k_{\rm off}^{-1}\approx 3600$ s and $k_{\rm on}\approx 3 \cdot 10^7 M^{-1} s^{-1}$), most cytokine molecules bind at a diffusion-limited speed with very strong binding.
We assume a large extracellular volume, so that cells do not deplete cytokines as they bind them. 
Receptors are usually pre-loaded with JAK, waiting for the conformational change associated with cytokine binding, to induce phosphorylation. From Eq.~\ref{LR_basic} the number of activated JAK (JAK$^*$) at steady-state is: 
\beq\label{eq:JAK}
{\rm JAK^*}={\rm JAK}_{\rm total}\frac{[{\rm Cytokine}]}{K_D+[{\rm Cytokine}]},
\eeq
where ${\rm JAK}_{\rm total}$ is the total JAK receptor concentration. The kinetics of STAT phosphorylation into pSTAT is described using a classical Michaelis-Menten equation:
\beq\label{eq:pstat}
\frac{d}{dt}{\rm [pSTAT]}=k_{\rm cat}{\rm [JAK^*]}\frac{\rm [STAT]}{{\rm [STAT]}+K_m}-k_{\rm dephos}{\rm [pSTAT]},
\eeq
where ${\rm [JAK^*]}={\rm JAK^*}/(V\mathcal N_A)$ is the concentration of activated JAK within the cytoplasmic volume $V$,
${\rm pSTAT}$ is the phosphorylated form of STAT,
$k_{\rm cat}$ and $k_{\rm dephos}$ are phosphorylation and dephosphorylation rates, and $K_m$ is the dissociation constant of STAT to the receptor, which controls the specificity between different receptors and different variants of STAT.
Note that, for most signaling networks, the identity of phosphatases that take care of dephosphorylating receptors and kinases remain often undetermined (because of overlapping and pleiotropic activities): their activity is modeled phenomenologically as a single rate reaction. 

This system reaches steady-state within minutes~\cite{Vogel2016,Altan-Bonnet2019} such that one can solve $d{\rm [pSTAT]}/dt$ in Eq.~\ref{eq:pstat} for pSTAT (with the conservation of matter condition [pSTAT]+[STAT]=$S_{\rm total}$) as a quadratic equation. Combined with Eq.~\ref{eq:JAK}, the result gives a direct expression of the response STAT as a function of the input cytokine concentration. 
This model serves as the basic building block to tackle the dynamic complexity of cytokine responses.

\subsubsection{Tunability of cytokine responses.}
In the previous section we assumed that cytokines, which are formed of several sub-units, were in a pre-assembled configuration.
When that is not the case, cytokine binding to the cytokine receptor can be modeled as a two-step scheme. 
\beq
{\rm L + R_1 + R_2 
\begin{array}{c}
k_1^{\rm on}\\
\rightleftharpoons \\
k_1^{\rm off}
\end{array}
LR_1 + R_2
\begin{array}{c}
k_2^{\rm on}\\
\rightleftharpoons \\
k_2^{\rm off}
\end{array}
LR_1R_2.}
\eeq
The cytokine ligand L binds to the cytokine receptor composed of two parts R$_1$ and R$_2$. A simple equilibrium model can be solved, with the dissociation constants $K_i=k_i^{\rm off}/k_i^{\rm on}$: 
\begin{eqnarray}
K_1[LR_1] &=& [L][R_1]=[L]\left([R_1]_{\rm total}-[LR_1]\right) \nonumber\\
K_2[LR_1 R_2] &=& [LR_1][R_2]=[LR_1]\left([R_2]_{\rm total}-[LR_1 R_2]\right)\nonumber
\end{eqnarray}
 that yields the total concentration of fully-engaged and signaling receptors:
\beq\label{eq:cytokine_binding}
[LR_1R_2]=\frac{[R_2]_{\rm total}}{1+\frac{K_2}{[R_1]_{\rm total}}\left(1+\frac{K_1}{[L]}\right)}.
\eeq

This two-step model was shown to be valid for one of the most studied cytokines, IL-2 \cite{Feinerman2010}. IL-2 binds weakly to the abundant $\alpha$ chain of the IL-2 receptor (${\rm R_1=IL2R}\alpha$), before being ``locked'' into a stable and signaling complex by association with the two other chains (${\rm R_2=IL2R}\beta$-$\gamma_C$) to form the full IL-2$-$IL2R complex. In that case, $K_1 [L] \ll 1$ such that equation~\ref{eq:cytokine_binding} simplifies to:
\beq\label{eq:cytokine}
[LR_1R_2]=[R_2]_{\rm total}\frac{ [L]}{ [L]+\frac{K_1K_2}{[R_1]_{\rm total}}}.
\eeq
This equation reveals key insights for IL-2 and other cytokines~\cite{Cotari2013}: the amplitude of cytokine signaling is proportional the number of $\beta-\gamma$ part of the IL2 receptor (the limiting part), and reaches half of its maximum at $[L]_{50}=K_1K_2/ [R_1]_{\rm total}$, inversely proportional to the $\alpha$ chain of the IL-2 receptor (the non limiting-part).

Such tunability of both the amplitude and sensitivity of the dose response of cytokines based on the number of cytokine receptor chains can be critical to achieve immune plasticity, as demonstrated in the context of competition for limited amounts of cytokines. For example, the tug-of-war for IL-2 between effector T-cells and regulatory T-cells is mediated by the exact levels of IL2R$\alpha$ on the surface of these cells. {Stimulated {\it effector} T-cells initiate an immune response, whereas stimulated {\it regulatory} T cells (Treg)  downregulate the response to suppress auto-immune responses (effector response to self-antigens). In general, Treg cells are thought to bypass negative thymic selection despite their strong recognition of self-antigens. These cells then act as pre-activated sentinels that respond to inflammatory cues (such as the upregulation of self-antigens) while not contributing to inflammation itself through cytokine secretion: Tregs constitutively express transcription factor FoxP3, a general downregulator of cytokine production.

Whenever an effector T-cell initiates a spurious response to self-antigens, Tregs can extinguish it by consuming cytokines (\eg IL-2) and by downregulating their inflammatory impact. In this tug of war, whichever cell type, effector or regulatory, expresses more receptors boosts its ability to capture the cytokine and deprives the other cells of this key cytokine for anti-apoptosis and proliferation in the T-cell compartment. Quantitative models of such competition for IL-2 based on differential expression of IL-2R$\alpha$ receptors have illuminated how self vs. non-self discrimination can emerge from such IL-2 tug--of--war~\cite{Busse2010,Feinerman2010,Hofer2012}.

Some cytokine receptors have an extracellular domain that gets secreted extracellularly, in a soluble form. These can ``pre-bind'' the cytokine in the extracellular milieu,  and deliver it to the complementary chain bound to the cell membrane: such molecular event can have positive or negative effects on cytokine signaling depending on the context. If the pre--formed soluble cytokine/receptor complex binds to the cell to form an incomplete receptor which lacks the intracellular signaling domain of the soluble cytokine receptor, the JAK misses its trans-phosphorylation partner and the cytokine/cytokine receptor fails to signal--- ``no clap from one hand.''  Such soluble complexes act as decoys that antagonize cytokine response (as with viral analogues of IFN receptors, or IL-1 receptors), limiting inflammation to the benefit of viruses \cite{Levine2004}.  

Alternatively, secretion of soluble cytokine receptors can trans-activate cytokine signaling and result in a boost in cytokine response (as with IL-1, IL-2 or IL-6). For example, in the case of IL-6,  the soluble portion of the IL-6 receptor (sIL-6R) gets secreted to stabilize IL-6 in the extracellular medium (most cytokines are very small proteins with short half-lives) and to accelerate the assembly of a complete IL-6/IL-6R/gp30 signaling complex by binding to gp130 dimers  on the surface of cells: such activation of the IL-6 signaling response thus does not require the presence of the IL-6 receptor in the membrane of the receiving cells, but only the ubiquitous presence of gp130: this is called cytokine trans-activation~\cite{Rose-John2004}.

Hence, depending on the exact molecular details, secretion of cytokine receptors in the extracellular environment can trigger or antagonize the cytokine signaling response. The role of quantitative immunology in that context is then to tease out the physico-chemical parameters of such regulation to better understand in which regime inflammatory cues are regulated.

\subsubsection{Regulation  by cytokine consumption}
Immune responses need tight regulation to avoid spurious auto-immune activation. This is particularly important for cytokine regulation as overabundant cytokines can be extremely deleterious to the organism as a whole. For example, high concentrations of cytokines induce a ``capillary leak syndrome,'' whereby tissues lose their barrier against blood serum: this results in septic shocks or viral hemorrhagic fevers. One simple mode of regulation of cytokine signaling by the JAK-STAT pathway is simply to limit the availability of cytokines in the extracellular medium. Most cytokines are very small proteins (with molecular weights in the 10-20kDa range) such that their half-life in the bodily fluid is short. For example, the half-life of IL-2 injected intravenously for the treatment of renal cell carcinoma is at most 3 h in the serum. More significantly, even in the extracellular environment of lymphoid organs, cytokines have a very short half-life due to rapid binding and consumption.

Given the strong binding of cytokine to their receptors (typically $K_D<100$ pM or $k_{\rm off}<(3000s)^{-1}$), immune cells can rely on binding to switch off the cytokine signaling response through buffering. This resetting process relies on cytokine consumption by endocytosis of the cytokine with its receptor, trafficking them towards lysosomes, release of the cytokine because of low pH within the lysosomal compartment, and degradation. Note that the signal transduction may persist after receptor endocytosis as long as the cytokine-receptor pair remains engaged \cite{Cendrowski2016}. What happens to the cytokine receptor in this process varies based on the cytokine identity. For example, the endocytosed IL-2 receptor dissociates and has its chains sorted towards different compartments: the IL2R$\gamma_C$ chain goes to the lysosome and gets degraded, while the IL2R$\alpha$ chain goes to early endosomes and gets recycled. This process of endocytosis, sorting, recycling and degradation can be very complex at the molecular level. Yet, a coarse-grained model of this process as a single-biochemical step can be sufficient when modeling IL-2 availability~\cite{Tkach2014,Voisinne2015}: the typical rate for this step has been measured to be of the order of (900 s)$^{-1}$.

One functional consequence is that cytokines get rapidly consumed while triggering signaling. At the more global level, one can integrate a dynamic equation for production/consumption that accounts for the rise and decay of inflammatory signals in the immune system:
\beq
\label{eq:cytokine_accumulation}
\frac{d}{dt}[{\rm Cytokine}]=\kappa_{\rm prod}-\kappa_{\rm consum} .
\eeq
 In the most simple case, $\kappa_{\rm prod}=N_{\rm prod}k_{\rm prod}/(V {\mathcal N}_A)$ is fixed by the number $N_{\rm prod}$ of activated cells at a given time, the rate of secretion $k_{\rm prod}$ per individual cells, and the extracellular volume $V$. The rate $\kappa_{\rm consum}$ is determined by the rate of binding of the cytokine to its receptor on the surface of the cells:

$\kappa_{\rm consum}=k_{\rm on}N_{\rm consum} N_{\rm R}[{\rm Cytokine}]/{V {\mathcal N}_A}$,

where $N_{\rm consum}$ is the number of consuming cells and $N_{\rm R}$ the number of receptors per cell.
The time dependency of $N_{\rm prod}$, $N_{\rm consum}$, $\kappa_{\rm prod}$, and $N_{\rm R}$ can be arbitrarily complicated and must be parametrized for each immunological setting under consideration. For example, in the case of IL-2 in the early events of an immune response within a lymph node, $V \approx 50\mu\ell$ (free volume), $k_{\rm prod}=10{\rm s}^{-1}$, $k_{\rm on}=3 \cdot 10^7 M^{-1}s^{-1}$, $N_{\rm consum}=10,000$ Treg cells, each endowed with typically $N_{\rm R}=3,000$ receptors. For IL-2 to accumulate and reach a significant concentration (typically, $10^{-11} M$ to trigger STAT5 phosphorylation), one must have $\kappa_{\rm prod}>\kappa_{\rm consum}$, i.e.  $N_{\rm prod}>1,000$. {This estimate illustrates that there exist thresholds of activation for immune responses, whereby the systems needs a critical mass of activated, cytokine-secreting cells to overcome consumption and drive activation and differentiation \cite{Polonsky2018}.} The estimate above does not take into account some of the intricacies of IL-2 regulation: positive feedback in IL-2 secretion, recycling of IL2R$\alpha$ chains, upregulation of IL-2R in T$_{\rm reg}$ cells have been documented. To account for these, Eq.~\ref{eq:cytokine_accumulation} must be solved numerically in more complex settings~\cite{Tkach2014,Voisinne2015}. 

\subsubsection{Other regulations}
Cytokine consumption is one major mechanism to limit the duration of availability of cytokines in the extracellular medium, but there exist additional, cell-intrinsic mechanisms that also limit or expand the duration of JAK-STAT signaling in cells, as illustrated by the following examples.

Certain cytokine pathways rely on intracellularly stored pools of receptors that get recruited upon receptor engagement and JAK-STAT signaling. This recruitment further fuels JAK-STAT signaling by avoiding saturation of the initial receptors on the membrane surface. This process was analyzed quantitatively in Ref.~\cite{Becker2010} to reveal how a certain type of leukocyte called erythroid progenitors can sense a cytokine called Epo over a large range of concentrations ($>1000$-fold range), where classical ligand-receptor binding would predict fast saturation with increasing cytokine concentration.

Another positive feedback in JAK-STAT signaling exists in the IL-2 signaling pathway: The $\alpha$ chain of the IL-2 receptor gets upregulated upon IL-2 signaling, lowering the cytokine concentration at which activation is half-maximum (by virtue of Eq.~\ref{eq:cytokine}), hence driving
further IL-2 signaling \cite{Busse2010,Feinerman2010,Cotari2013}. Such positive feedback was shown to be critical to commiting cells to long-term JAK activation and proliferation \cite{Voisinne2015}.

Alternatively, JAK-STAT pathways are also endowed with negative regulators, such as SOCS and Cytokine Inducible SH2-containing  (CISH) proteins, which compete with STAT for JAK binding, preventing activation \cite{Raue2011}. This can be quantitatively modeled with SOCS or CISH binding competitively with a stronger $K_m$ than STAT onto the receptor (Eq.~\ref{eq:pstat}). Varying the exact value of $K_m$ allows SOCS and CISH to negatively regulate JAK activity in two separate regimes of concentrations of cytokines~\cite{Bachmann2011}. Such dual feedback regulation of signaling was also shown to extend the range of cytokines that trigger STAT5 phosphorylation while avoiding saturation, and ultimately control cell survival. The modeling of such complex regulation of the JAK-STAT pathway involves adding biochemical steps 
that can be integrated numerically. The challenge in this context is to acquire a large number of system-specific biochemical parameters. This task has become more amenable as the field progresses and quantitative approaches have been delivering more and more estimates: protein abundances, biophysical parameters, enzymatic rates \cite{Cotari2013,Karr2015,Shi2016,Mitchell2018}.
 
To conclude, in this section we have reviewed the basic equations governing the regulation of the cytokine-activated JAK-STAT pathway in the immune system. While JAK-STAT signaling can be solved analytically in the adiabatic limit, the long-term dynamics of cytokine accumulation and consumption are complicated by the multitude of feedback regulations triggered in leukocytes. Such rich dynamical complexity will need to be tackled quantitatively and systematically to deliver a more comprehensive understanding of cytokine communication in the immune system \cite{Altan-Bonnet2019}.

\subsection{Communication across space and time, and cytokine niches}~\label{sec:cytokine_comm}

Cytokines allow immune cells to communicate and to modulate their response collectively. In section~\ref{sec:celldiff}, we will discuss how lymphocyte differentiation can be decided by toggle switches encoded in gene regulatory networks. Most of the positive feedbacks in these toggle switches are in fact associated with a response to cytokines, and are thus intrinsically collective. Tackling the nonlinearities and spatio-temporal aspects of cytokine communication at a more quantitative level is critical to expanding our understanding how antigen recognition leads to such a collective response. We now focus on the dynamics of cytokine secretion and capture in space and time, over the relatively short timescales (minute to hours) over which cytokines convey information between cells.

\subsubsection{Cytokine dynamics as diffusion-degradation}
The propagation of cytokines within tissues is governed by reaction-diffusion equations which generalize Eq.~\ref{eq:cytokine_accumulation}:
\beq
\label{eq:diffusion_consumption}
\frac{\partial c(\vec{r},t)}{\partial t}=D \nabla^2 c(\vec{r},t) +\kappa_{\rm prod}(\vec{r},t)-\kappa_{\rm consum}(\vec{r},t),
\eeq
where $c(\vec{r},t)$ is the spatio-temporal profile of the concentration of cytokine, $D$ is the diffusion coefficient for the cytokine in the extracellular medium, $\kappa_{\rm prod}$ and $\kappa_{\rm cons}$ are volumic rates of cytokine production and consumption respectively.
Cytokines are produced by discrete cells, and the production rate $\kappa_{\rm prod}(\vec{r},t)$ concentrates on the cell surface.
The consumption rate can be estimated by taking into account that cytokines bind tightly to cytokine receptors on the cell surface and get consumed by endocytosis of the engaged receptors. The limiting step for this cytokine consumption is then the binding to receptors, so that
\beq
\kappa_{\rm consum}=k_{\rm on}N_{\rm R} n_{\rm consum} c(\vec{r},t)=k_c c(\vec{r},t),
\eeq
where $n_{\rm consum}$ is the density of consuming cells.
This expression is valid at low concentrations, when cytokine receptors are not saturated and the kinetics of binding remain linear with the concentration of cytokines. It also assumes that consuming cells are uniformly distributed, in a mean-field manner, rather than modelling the precise location of their surfaces.
In more general cases, $\kappa_{\rm prod}$ and $\kappa_{\rm consum}$ may have additional, nonlinear dependencies on $c(\vec{r},t)$, for instance because of spatial heterogeneity or receptor saturation. Little can be done analytically in these cases and one must resort to numerical solutions \cite{Thurley2015}.  In the linear regime, Eq.~\ref{eq:diffusion_consumption} becomes:
\beq\label{eq:diffreac}
\frac{\partial c(\vec{r},t)}{\partial t} 
=D \left[\nabla^2 c(\vec{r},t) - {\xi^{-2}}c(\vec{r},t)\right],
\eeq
with a characteristic length $\xi=\sqrt{{D}/{k_c}}$ over which cytokines diffuse before being consumed.
In the following section, we focus on the simple homogenous and linear case to derive insights about the quantitative aspects of cytokine communication. 

\subsubsection{Screening by cytokine-consuming cells}
Diffusion and consumption of cytokines regulates how the cytokine gradient  spreads i.e. the lengthscale for cell communication. 
It was demonstrated that the short timescale to reach stationary concentration profiles for soluble proteins are explained by the first arrival time of the cytokine ligands, rather than the characteristic diffusion timescale for an individual molecule~\cite{Berezhkovskii2010,Kolomeisky2011}. This result implies that the characteristic timescale to reach steady state at a distance $r$ from the secreted cell scales linearly with $r$ rather than with $r^2$. The analytical expression for this local relaxation time $\tau(r)$ is \cite{Berezhkovskii2010}:
\beq
\tau(r)=\frac{1}{2 k_c}\left(1+\frac{r}{\xi}\right).
\eeq
Two relaxation processes are at play in this system: the diffusion timescale ($\tau_d(r)=r^2/2D$) and the reaction timescale ($k_c^{-1}$). The relaxation timescale is dominated by the reaction time at short distances ($r \ll \xi$) where cytokine consumption dominates. However, at large distances ($r \gg \xi$), both diffusion and consumption are important and the relaxation timescale is the geometric mean of the two,
$\tau(r)\sim ({{\tau_d(r) k_c^{-1}}})^{1/2}\propto r$.
Numerical simulations in the context of immune responses~\cite{Oyler-Yaniv2017} validated this theoretical argument and estimated that it takes only a few minutes for the cytokine gradient to reach its steady state. 

The timescale at which immunological regulation by cytokine exchange operates is typically in the range of hours for the activation of gene regulatory elements, to days for the decision to proliferate or to die.  The typical timescale at which cells move is also very slow. In the first phase of an adaptive immune responses, antigen-responding T-cells are essentially stuck on their antigen presenting cells. One can thus assume steady state for the distribution profile of the cytokines, $c(\vec{r},t)=c(\vec{r})$, and solve for $dc/dt=0$ in Eq.~\ref{eq:diffreac}.
Using spherical symmetry  $c(\vec{r},t)=c(r)$ for the Laplace operator, the steady state profile for the cytokine distribution is
\beq\label{eq:Cytokine_distribution}
c(r)=c(R_{\rm cell})\frac{R_{\rm cell}}{r}e^{{(R_{\rm cell}-r)}/{\xi}},
\eeq
and the boundary value $c(R_{\rm cell})$ can be determined by estimating the flux of molecules secreted and consumed at the cell surface:
\beq\label{eq:balance}
\begin{split}
  \kappa_{\rm prod}-\kappa_{\rm consum}&=\oint{\vec{j}.d\vec{r}}  =-4\pi D R^2\left.\frac{dc(r)}{dr}\right|_{R_{\rm cell}}\\
  &=4\pi D c(R_{\rm cell})R_{\rm cell}\left(1+\frac{R_{\rm cell}}{\xi}\right).
\end{split}
\eeq
Typically, cells secrete cytokines at a constant rate between $k_{\rm prod}=$10 to 1000 molecules per second (e.g. IL-2 and IFN-$\gamma$ cytokines) and consumption by the secreting cell is given by $k_{\rm consum}=k_{\rm on}N_{\rm R} c(R_{\rm cell})$. Solving for $c(R_{\rm cell})$ yields:
\beq\label{eq:cbound}
c(R_{\rm cell})=k_{\rm prod}/(4\pi D R_{\rm cell}(1+R_{\rm cell}/\xi)+k_{\rm on}N_R)
\eeq

The characteristic scale
\beq
\xi=\sqrt{D/k_c}=\sqrt{D/(k_{\rm on}N_Rn_{\rm consum})}
\eeq
is analogous to the screening length of electrostatic interactions and is inversely proportional to the square root of the density of consuming cells. 
By this analogy, consuming cells ``screen'' the diffusion of cytokines from the secreting cells and end up determining the extent of cell-to-cell communication in a dense tissue. In other words, the simple diffusion-consumption of cytokines can account for the heterogeneous accessibility of cytokines in dense tissues and the formation of cytokine ``niches,'' defined as the portion of space in where cytokines can be sensed.

$\xi$ is controlled by the density of cytokine-consuming cells, and the number $N_{\rm R}$ of receptors on the surface of consuming cells. In close-packed tissues the density of consuming cells is typically 1 cell every 10$\mu$m, implying $n_{\rm consum}\sim 10^{-3}\mu{\rm m}^{-3}$; for IL-2 $D$ is approximately 100$\mu m^2/s$ (even in close-packed tissues, there is sufficient free space in the extracellular medium for cytokines to diffuse freely as in solution), and each cell consumes typically 10 cytokine molecule per second \cite{Oyler-Yaniv2017}, implying $k_{\rm on}N_R=10^{11}M^{-1}s^{-1}$, so that $\xi \sim 25\mu{\rm m}=2.5$ cell diameters. In more physiological conditions, where cells are more dilute, occupying only 10\% of the space and cells express lower level of receptors, $n_{\rm consum}\sim 10^{-2}\mu{\rm m}^{-3}$, this length scale rises to $\xi\sim 250\mu{\rm m}\sim 25$ cell diameters: there is negligible screening and all the cells within a lymph node or a spleen have access to the cytokine. Note that this 10-fold increase in the screening length $\xi$ implies an increase by 1000-fold in the volume of the cytokine niche and a 1000-fold increase in the number of cells that potentially respond to the cytokine. Such tuneability can be critical to decide which cells (helper T cells or regulatory T cells) win the tug of war for IL-2  and ultimately whether the immune response expands or gets extinguished~\cite{Busse2010,Feinerman2010,Hofer2012}.

\subsubsection{Probability of autocrine capture}~\label{autocrine}
Cells that secreted cytokines often express the associated receptors, and can in principle communicate with themselves.
To quantify this effect, we must estimate the non-zero probability $P_{\rm auto}$ that cells capture their own cytokines rather than let them diffuse and interact with neighbouring cells. This phenomenon, called autocrine signaling,
is particularly significant when cytokine-secreting cells are surrounded by cytokine-consuming cells and compete for their consumption.

At steady state, there is a balance between cytokine production $k_{\rm prod}$, cytokine consumption by the secreting cell $k_{\rm consum}$, and diffusion away from the cell followed by consumption by other cells, as given by Eq.~\ref{eq:balance}. The probability of absorption by the secreting cell is given by:
\beq
P_{\rm auto}=\frac{k_{\rm consum}}{k_{\rm prod}}=\left[1+\frac{4\pi DR_{\rm cell}(1+R_{\rm cell}/\xi)}{k_{\rm on}N_R}\right]^{-1}.
\eeq
We can define the characteristic number of receptors at which this probability reached 50\%, $N_R^*=4\pi DR_{\rm cell}(1+R_{\rm cell}/\xi)/k_{\rm on}\approx R_{\rm cell}/\sigma_{\rm R}$, where $\sigma_{\rm R}=R_{\rm receptor}e^{-\Delta G_{\rm assoc}}$ is the effective cross-section of the receptor-cytokine interaction (Eq.~\ref{eq:association_rate} and \ref{eq:association_full}), and where we assumed $\xi\gg R_{\rm cell}$.

Leukocytes are typically $R_{\rm cell}=5$ to 10 $\mu$m in radius, and $\sigma\sim 0.5$nm, hence $N^*_R\approx 10^4$. If cells have much fewer than $10^4$ receptors (a common situation), $P_{\rm auto} \ll 1$ and most of the secreted cytokines diffuse away without being captured in an autocrine manner. Alternatively, if cells have more than $10^4$ receptors, $P_{auto} \sim 1$ and secreting cells capture a large fraction of their own cytokines. The functional relevance of autocrine capture remains debated in the field: why would a cell need to respond to the cytokine it secretes. There are also molecular mechanisms limiting its impact. For instance, T cells secreting IL-2 upon antigen recognition express low levels of  IL-2 receptors as long as the antigen response lasts to limit autocrine capture \cite{Tkach2014}). In the tug of war between cell types for cytokine consumption, competition between autocrine and paracrine signaling is crucial, and thus the amount of autocrine capture is significant \cite{Hofer2012}. Discriminating between autocrine and paracrine signaling is often difficult and it has been suggested that dilution experiments in vitro can be used to distinguish the two modes~\cite{Marcou2018c}.

\subsubsection{Size of cytokine niches}

Eqs.~\ref{eq:Cytokine_distribution}-\ref{eq:cbound} can also be used to estimate the functional size of a cytokine niche established by a single secreting cell. In the limit of no autocrine capture and long dispersal $\xi\gg R_{\rm cell}$, we have:
\beq
c(r)=\frac{k_{\rm prod}}{4\pi Dr}e^{-r/\xi}
\label{eq:CytokineProfile}
\eeq
The radius of the niche $r_{\rm niche}$ can be defined as the $r$ at which $c(r)$ falls below the characteristic concentrations $K_D$ needed to trigger cytokine signaling, and is the solution to the implicit equation: $r_{\rm niche}=\xi\ln (k_{\rm prod}/4\pi r K_D)$. The logarithmic dependence on the rate of cytokine secretion $k_{\rm prod}$ is significant because that rate  may vary greatly depending on the cytokine or depending on the state of differentiation of cells. For example, CD8+ T cells can secrete up to 1000 IFN-$\gamma$ per second, CD4+ T cells secrete 10 IL-2 per second upon activation from a naive state, and 1000 IL-2 per second as they mature~\cite{Tkach2014} or when get activated from a memory state~\cite{Huang2013}.

In Figure~\ref{fig:niche}, we present the numerical solution for $r_{\rm niche}$ for different secretion rates for cytokines with different screening lengths. Note how variable the volume of the cytokine niche can be, based on realistic parameters of immune responses.

\begin{figure*}
\includegraphics[width=\linewidth]{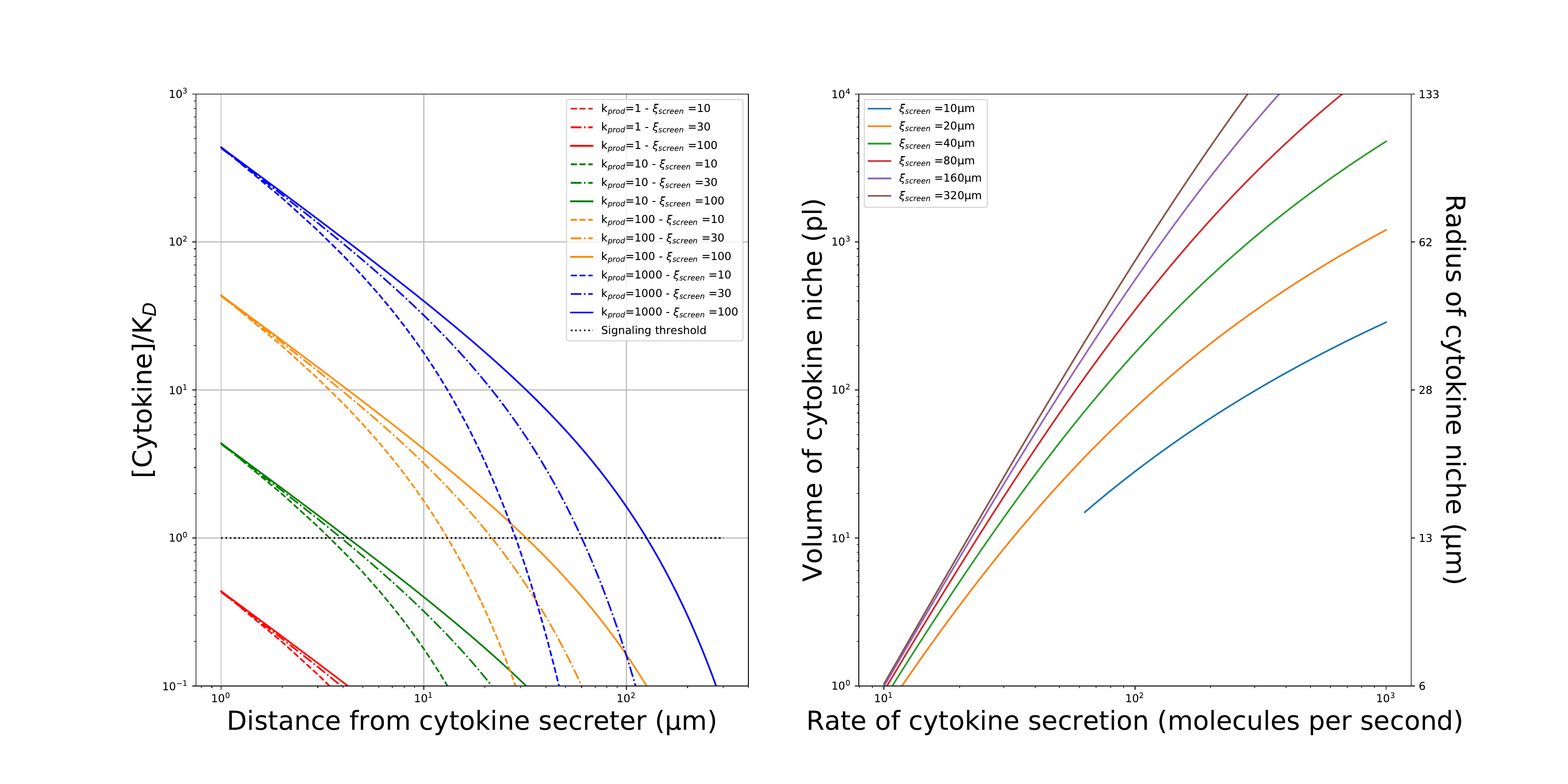}
\label{fig:niche}
\caption{\textbf{Size of cytokine niches for different parameters of secretion/consumption.} (Left) Cytokine profile as a function of the distance from the surface of the secreting cell (normalized by  $K_D$ -- the critical cytokine concentration to induce STAT phosphorylation), for different rates of secretion $k_{prod}$ (in molecules per second), and different screening length $\xi$ (in $\mu$m). (Right) Size of the cytokine niche as a function of the rate of cytokine secretion. For these graphs, $K_D=3$pMol and $R_{cell}=5 \mu$m in equation~(\ref{eq:CytokineProfile}).}
\end{figure*}

To conclude, in this section we discussed how simple equations for the diffusion and consumption of cytokines in dense immunological tissues can quantitatively account for the tunability of cell-to-cell communication in the immune system. It is particularly striking that such active modulation of cell communication  plays a functional role in modulating immune responses. In the first moments of activation, there are very few cytokine-consuming cells, and the field of cytokines extends to the entirety of the lymphoid organ ($\xi > 100 \mu$m), reaching blood vessels and making communications global. Within 24 hours, many cells migrate closer to cytokine-secreting cells. As they do so, they increase their cytokine consuming capabilities by upregulating their receptors,  thereby ``screening'' the diffusion of cytokines  ($\xi \sim 10 \mu$m) and limiting the extent of cell communication to nearest neighbours, within a tight niche around secreting cells.

\section{Cell fate}~\label{cellfate}

There is a large variety of types and subtypes of immune cells, making the field of immunology sometimes difficult for the non-specialist. These fates are acquired either at the very beginning during haematopoiesis, the process by which all types of blood cells are generated by differentiation from stem cells, or following an immune response or challenge, during which cells further specialize to better fight infections. We first describe the physical mechanisms of gene regulation that lead to stable distinct cell fates. We then review modeling strategies for understanding and inferring the hematopoiesis differentiation tree. Finally, we discuss the problem of differentiation during an immune response.

\subsection{Gene regulation and cell differentiation \label{sec:celldiff}}

Many cell decision making processes in the immune system are carried out at the level of gene regulation, whereby expression of key transcription factors decide cell fate and subsequent immune effector function. For example, T-cells sense their inflammatory environment as defined by the combinations of cytokines in the extracellular environment, and drive their signaling response towards expressing key transcription factors that in turn decide which cytokines get produced and secreted. 

To consider a concrete example of gene regulation, we use the classical example of the differentiation of CD4+ helper cells, a particular type of T cells whose function is to regulate the action of other immune cells through the production of cytokines --- the messenger molecules used by immune cells to communicate with each other. In their na\"ive state, T cells are essentially a blank slate which can later differentiate into more specialized cell states to best match the pathogenic threat. Upon antigen stimulation, helper T cells can orchestrate three  types of immune responses. They can elicit a `Th1' response, further unleashing T- and NK cells' cytotoxic response against infected cells. This is particularly relevant when eradicating an intracellular infection such as a viral infection. Alternatively, helper T cells can elicit a `Th2' response to engage with B cells, drive affinity maturation and class switching in antibody production towards secretion of antibodies that can annihilate extracellular (e.g. bacterial) infections.  Finally, T-cells can elicit a `Th17' response that protects mucosal barriers from infection. These three characteristic types of T-helper cell differentiation respond to and amplify distinct cytokine environments. Note that similar decisions are made by other immune cells: for example macrophages  differentiate into `M1' (pro-inflammatory) and `M2' (anti-inflammatory) types, which mirror the decisions and inflammatory outputs associated with the Th1 and Th2 helper T-cell types. 

Immunologists are interested in dissecting the interplay and feedback between inflammatory environments and immune cell differentiation~\cite{Zhu2010}. In that context, quantitative immunology brings the tools of nonlinear dynamics to explain how sharp and cross-inhibitory decisions
can be made by activated T-cells during  differentiation. Many immunological systems involve nonlinearities with self-reinforcing feedback loops, whereby the protein of interest acts as a positive feedback that drives further expression. Additionally, most gene regulation involve mutltimerization of transcription factors onto one gene locus, as in the case with phosphorylated STAT, a transcription factor induced by cytokine signaling. Multimerization implies a nonlinearity because the association rate of an n-mer scales with the n$^{\rm th}$ power of the monomer's concentration.
These nonlinearities and their functional consequences for cell fate in the immune system can be studied using nonlinear stability analysis. Many transcription factors auto-amplify their production in a direct loop or, more generically, in a more convoluted cytokine-mediated manner. For example, the transcription factor Tbet drives the secretion of a cytokine called interferon $\gamma$ (IFN-$\gamma$) that signals through pSTAT1 to produce more Tbet in the context of Th1 T-cell differentiation. In section~\ref{genereg} of the Methods, we recall the basic formalism of nonlinear analysis, which explains the generic occurence of bimodal distributions of expression of transcription factors, cytokines and surface markers that immunologists encounter in their single-cell measurements (cytometry or single-cell transcriptomics). We discuss classical models of gene regulation and feedback regulation that account for the molecular programs enforcing sharp cell differentiation in the immune system.

\subsubsection{Th1/Th2 differentiation}

Early experimental evidence demonstrated that CD4+ helper T-cells commit to two distinct and incompatible states of differentiation, Th1 and Th2,
upon antigen activation and response to inflammatory cues. A model describing this system is based on two inhibitory loops (see Sec.~\ref{genereg}) whereby transcription factor $P_1$ inhibits the production of $P_2$ and vice versa (Figure~\ref{fig:Th1_Th2}A \& B). 
\begin{equation}\label{eq:bimodal}
\begin{split}
\frac{dR_1}{dt} & =  v_1 \frac{K_2^{n_2}}{K_2^{n_2}+P_2^{n_2}}-\gamma_{R_1} R_1 \\ \nonumber
\frac{dP_1}{dt} & = k_{1} R_1 - \gamma_{P_1} P_1\\ 
\frac{dR_2}{dt} & =  v_2 \frac{K_1^{n_1}}{K_1^{n_1}+P_1^{n_1}}-\gamma_{R_2} R_2 \\ 
\frac{dP_2}{dt} & =  k_{2} R_2 - \gamma_{P_2} P_2 ,
\end{split}
\end{equation}
where $R_1$ and $R_2$ are number of mRNA leading to the production of $P_1$ and $P_2$, $\gamma$ decay rates, $k_i$ and $v_i$ are production rates, $K_i$ are binding dissociation constants, and $n_i$ are Hill coefficients.
At steady state, 
\begin{equation}\label{eq:null}
\begin{split}
P_1 & \propto \frac{K_2^{n_2}}{K_2^{n_2}+P_2^{n_2}} \\ 
P_2 & \propto \frac{K_1^{n_1}}{K_1^{n_1}+P_1^{n_1}},
\end{split}
\end{equation}
to which some background expression may be added to account for leaky gene regulation.

\begin{figure*}[!ht]
 	\centering
	 {\includegraphics[clip,width=1\linewidth]{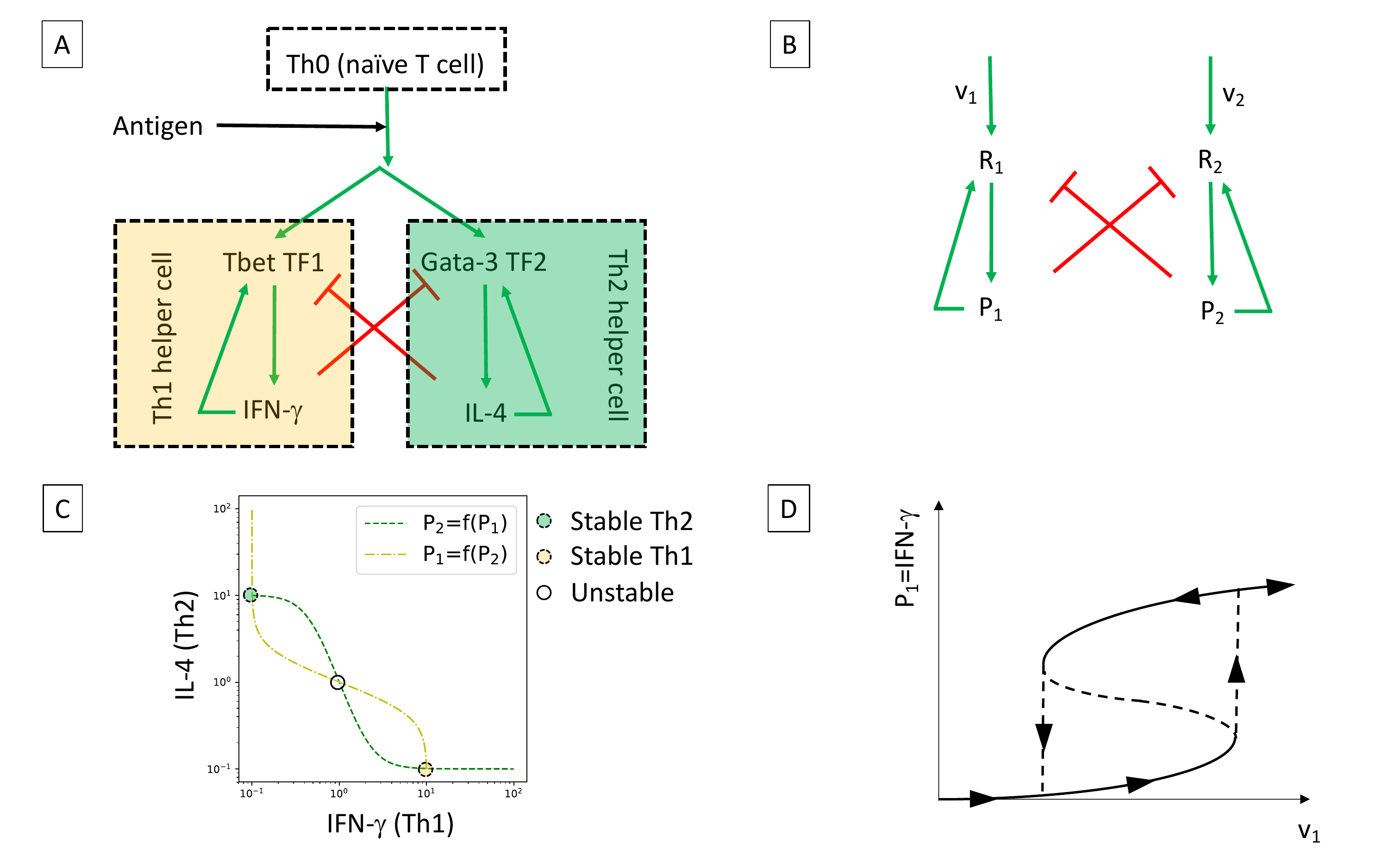}}
	 \caption{Modeling Th1/Th2 differentiation. A. Cartoon model of the cross-regulation of Tbet and GATA-3 transcription factors (TF) whose expression determines T cell fate in terms of Th1/Th2 differentiation. B. Diagram of the interactions of TF deciding cell fates. C. Nullcline for the regulation of TF ($P_1$ and $P_2$): three fixed points can emerge in the dynamics of the system, two stable, one unstable. D. Dynamics of expression of $P_1$ when the tonic rate of expression $v_1$ (driven by antigen response) is increased: note the hysteresis in the levels of $P_1$ whereby a bimodal distribution can be anticipated for intermediate values of $v_1$.}
\label{fig:Th1_Th2}
\end{figure*}

The nonlinearities, which are controlled by the Hill coefficients $n_i>1$, originate from the multimeric control of transcription by the transcription factors. A graphical inspection of the nullclines (Eq.~\ref{eq:null}) for this dynamical model (Figure~\ref{fig:Th1_Th2}.C) reveals that there exist three fixed points for this gene regulatory system, two stable ones, $(P_1^{high},P_2^{low})$ and $(P_2^{high},P_1^{low})$, and one unstable fixed point $(P_1^{high},P_2^{high})$.  This simple dynamical system reveals how cross-inhibition of transcription factors yields a classical toggle-switch that yields two incompatible states. Note that such a model predicts hysteresis in T cell differentiation (Figure~\ref{fig:Th1_Th2}D) as has been observed experimentally~\cite{Chaouat2004,Yates2004,Antebi2013}.

\subsubsection{Other differentiation switches}

A similar formalism has been applied to account for the differentiation of CD4+ T-cells in Th17 and regulatory T-cells (Treg). Th17 cells are pro-inflammatory and critical to eradicate infections in mucosal tissues; regulatory T cells  are anti-inflammatory and limit overzealous immune responses, such as auto-immune disorders caused by a strong response against self antigens, or runaway immune responses such as septic shock. These two states of differentiation for CD4+ T cells can be accessed upon exposure to the cytokine TGF$\beta$. However, that cytokine has been shown to be both pro- or anti-inflammatory depending on the context. Tyson \& coworkers~\cite{Hong2011} proposed a model to account for such bimodality in TGF-$\beta$ action.
In that model, the concentration of each protein species $P_i$ generically follows:
\beq
\frac{dX_i}{dt}=\gamma_i\left(X_i^{\rm eq}\frac{1}{1+e^{-x}}-X_i\right)
\eeq
where $\omega_i=\omega_i^0+\sum_{j}\omega_{j\rightarrow i}P_j$, $P_i^{\rm eq}$ is the characteristic concentration of $P_i$ in differentiated cells at steady state, $\gamma_i$ are degradation rates, $\omega^0_i$ is the constitutive level of expression of regulation of $X_i$, $\omega_{j\rightarrow i}$ is the (possibly negative) influence of $X_j$ on $X_i$, and $\sigma_i$ is a parameter modulating the steepness of the regulation. Note that proteins can regulate their own production, $\omega_{i\to i}\neq 0$.

There are three main species involved in Th17/Treg differentation: $P_1$=FoxP3, $P_2=$ROR$\gamma$t (the transcription factors controlling Treg and Th17 differentiation, respectively), and $P_3=$TGF$\beta$ (a cytokine that regulate T cell differentiation), which acts as an externally fixed stimulus. Their interactions are represented in Figure~\ref{fig:Th17_Treg}A.
For instance ROR$\gamma$t$^+$ is activated by TGF$\beta$  ($\omega_{3\to 2}>0$),
as well as by itself ($\omega_{2\to 2}>0$), and but is repressed by ROR$\gamma$t  ($\omega_{1\to 2}<0$).

Graphically, one can represent the nullclines ($dX_i/dt=0$) for both transcription factors as shown in Figure~\ref{fig:Th17_Treg}. There are three stable fixed points where the nullclines meet: ROR$\gamma $t$^+$FoxP3$^-$ for the Th17 cells, FoxP3$^+$ROR$\gamma$t$^-$ for the Treg cells and ROR$\gamma $t$^+$FoxP3$^+$ for a mixed state whose existence was later confirmed experimentally~\cite{Hong2011}.
This formalism is interesting because it handles coupling and feedback phenomenologically, through the effective regulation parameters $\omega_{i\to j}$, rather than through detailed mechanisms of how these regulations are implemented.

\begin{figure*}[!ht]
 	\centering
	 {\includegraphics[clip,width=1\linewidth]{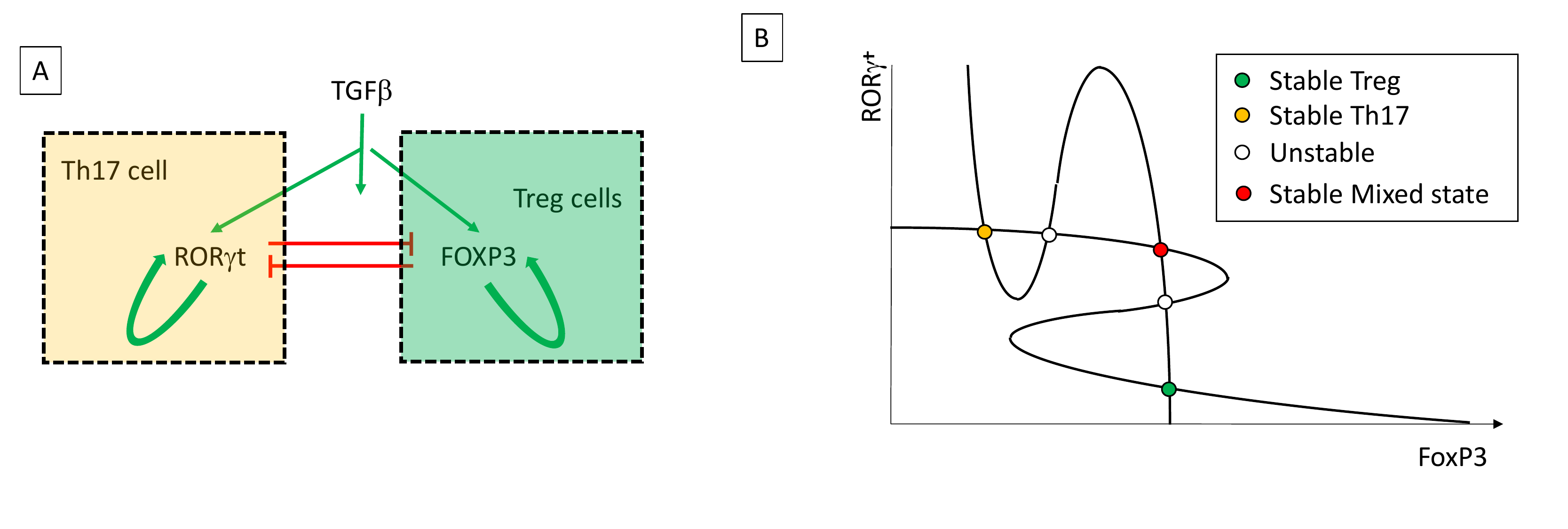}}
	 \caption{Modeling the trimodal distribution of T-cell fate upon exposure to TGF$\beta$. A. Cartoon of the genetic network associated with the regulation of ROR$\gamma$ t and FoxP3 upon TGF$\beta$. B. Nullclines for the dynamics of regulation of ROR$\gamma$t and FoxP3. Each intersubsection of the nullclines corresponds to a fixed point; graphical considerations can be used to deduce the stability of each fixed point.}
\label{fig:Th17_Treg}
\end{figure*}

\subsubsection{Experimental test of bistability in cell differentiation}

The elegant models presented in the two previous subsections (Th1/Th2 and Th17/Treg differentiation) account qualitatively for the observation that many activation conditions lead T-cells to commit to clear, separate and stable cell fates. The crux of these models is the existence of toggle switches based on cross-inhibition of transcription factors and, optionally, self-reinforcing positive feedback loops for transcription factors. Similar models have been proposed in the context of hematopoietic differentiation~\cite{Laslo2006,Kueh2013}, and they provide a quantitative framework to account for the bimodal distribution of transcription factors in immune cells. How do these models compare with experiments?

A theoretical inspection of the Th1/Th2 model of T-cell differentiation by Antebi {\it et al.}~\cite{Antebi2013} highlighted how the bistable solution of the model presented in Figure~\ref{fig:Th1_Th2} does not generically produce a bistable solution. A simple scan of parameters within the physiological range for induction of Tbet and GATA-3, the candidate $P_1$ and $P_2$ in Eq.~\ref{eq:bimodal}, demonstrated that the ratios of rates for their self-reinforcement and inhibition most commonly predicted a single stable fixed point, and only rarely more than two.
In addition, the induction of transcription factors is known to be inherently stochastic, because of intrinsic noise in gene induction~\cite{Walczak2005b, Walczak2005c,Friedman2006, Elowitz2002} or extrinsic noise in the levels of expression of signalling molecules in T cells~\cite{Feinerman2008}. Overall, the ``simplest'' model of a toggle switch in T-cell differentiation, though elegant in its inception, is not consistent with observations. This case is typical in quantitative immunology, where initial insight is often derived from genetic manipulations with severe consequences as in the case of the Th1/Th2 system where knocking out Tbet leads to a reduction/maintenance  of induction of Th1/Th2 cytokine secretion, respectively~\cite{Szabo2000}.
These knockout experiments must be balanced and reinterpreted in nonperturbative settings, e.g. by monitoring the spontaneous induction of Tbet in differentiating T-cells. Similar reassessments are happening in many models of leukocyte differentiation, such as in the M1/M2 model of macrophage differentiation~\cite{Palma2018}, or PU-1-controlled differentiation in the myeloid compartment~\cite{Kueh2016}). Future work will require better parametrization an innovative modeling approach that can embrace the combinatorial and dynamic complexity of cytokine communications and gene regulation in the immune system. 

Such ``murkiness'' in immunology must be embraced as it opens up functional possibilities. For example, Peine {\it et al.} \cite{Peine2013} tracked the formation of Tbet$^+$GATA-3$^+$ mixed phenotypes, i.e. with characteristics of both Th1 and Th2, so in apparent contradiction with bistability. Their presence was demonstrated to limit the deleterious impact of all-out, non-mixed Th1 or Th2 inflammation. Thus, a better understanding of noise and stability in immune cell differentiation will be key to understanding virtuous and pathologic inflammation.

\subsection{Hematopoiesis}\label{sec:hemato}

\subsubsection{Timescales}

Where do immune cells come from? A human being has $\sim 5 \cdot 10^{13}$ cells in the body and more than half of them are made by hematopoietic stem cells (HSC) found in the bone marrow. There are an estimated order of magnitude $\sim10^{16}$ cell divisions per human, which gives $\sim 10^6$ cell divisions per second ~\cite{Moran2010,MiloPhillips}, and most of them are linked to the HSC cells. HSC number $\sim10^4$ in mice (there are no reliable numbers for humans), many of them already made in the fetus. Through a series of differentiation and phenotypic commitment events (HSC$ \rightarrow$ Short Term HSC (ST-HSC) $\rightarrow$ Multipotent progenitor (MPP), these cells give rise to different kinds of cells found in the blood, including red blood cells and all immune cells (see Fig.~\ref{fig:hematopoiesis}). The first branching decision, whose precise timing is currently being questioned, is about becoming a myeloid progenitor, that will give rise to red blood cells, mast cells, thrombocytes (so not immune cells) but also macrophages, granulocytes (neutrophils, basophils and eosinophils) and dendritic cells -- so cells of the immune system that eat up cells and proteins non-specifically. They function as cells of the innate immune system, simply eliminating bacteria and other pathogens, but some of them (e.g. dendritic cells) also play an important role in adaptive immunity as antigen presenting cells (APC). The remaining branch of the differentiation tree leads to lymphoids that include cells of the adaptive immune system: B and T-cells as well as natural killer (NK) cells. The decision about becoming a myeloid or a lymphocyte is widely assumed not be an autonomous decision but is believed to be influenced by external signal (see below for a detailed discussion).

General quantitative questions about differentiation apply to the differentiation of cells in the immune system. For example, given that in mice there are of the order of $10^4$ HSC, what kind of dynamics results in $\sim 10^7$ very short lived granolocytes (of the order of a day) while at the same time producing $\sim 10^{11}$ lymphocytes that can live even for years? How long does it take to produce these cells? What kind of differentiation process can produce this kind of diversity, and how is it regulated to produce the right numbers of cells? A lot of information about these differentiation processes has be gained from two types of experiments. The first involves exposing cells at a given upstream stage with died or radioactive markers (i.e. bromium, deuterium) that get taken up --- and then diluted --- upon cell division. Analysis of the decay curves is informative about cellular lifetimes. The second type of experiment --- an adoptive transfer experiment --- is experimentally harder, since it involves transplanting new marked cells into an animal (typically mouse) and then tracking them. To study cells during the early stages of hematopoiesis the mouse must first be cleared of its natural cells, and recent  results suggest that the dynamics of differentiation after transplantation may be very different from regular dynamics~\cite{Busch2015}.  Lymphocyte divisions can be studied without killing the host's own immune system. The principle of these experiments is simple --- after some time $\delta t$ the marked cell will appear in a given compartment.

\begin{figure}
\includegraphics[width=\linewidth]{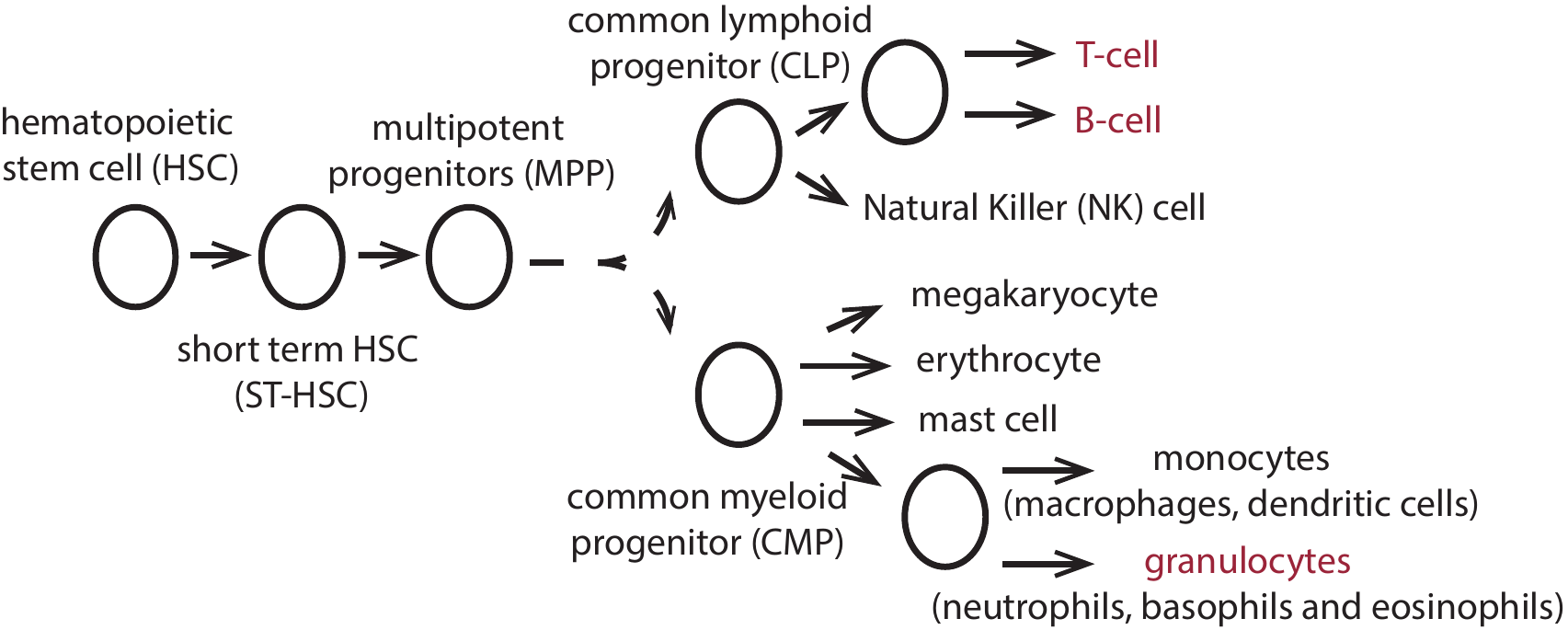}
\caption{\textbf{Hematopoiesis}. {A cartoon representation of the outline of the hematopoiesis process that leads to the formation of immune cells. The cells that are discussed in some detail in this review are marked in red.}}
\label{fig:hematopoiesis}
\end{figure}

At the population level, given that cells at stage $i$ of differentiation can proliferate with rate $\lambda_i$, differentiate with rate $\alpha_i$ or die with rate $\delta_i$, the dynamics of a given cell type $n_{\rm R}$ in the differentiation process follows~\cite{Hofer2016, Busch2015}
\beq
\frac{d n_{\rm R}}{dt} = \alpha_{\rm u} n_{\rm u} - (\alpha_{\rm R}+ \delta_{\rm R} - \lambda_{\rm R})n_{\rm R},
\eeq
where $n_{\rm u}$ describes the upstream (pre-differentiation) cell type. Cells can also migrate, which  introduces a spatial component to the equations. In general, all these rates are functions of the concentrations of the different populations $n_i$, resulting in non-linear equations. $\alpha_{\rm R}+ \delta_{\rm R} - \lambda_{\rm R}=k_{\rm R}$ define a collective timescale of decay of the $n_{\rm R}$ population. This highlights the identifiability problem that without specially planned out experiments guided by theory, it may be hard to tease apart the timescales of biological interest. 

Commonly in this kind of analysis it is assumed that the cellular differentiation of the adult individual is in steady state. We will return to this assumption when we learn the timescales for the process. Within this assumption the steady state ratio of cells in the compartment of interest, compared to the upstream compartment is ${n_{\rm R}}/{n_{\rm u}}=  \alpha_{\rm u}/( \alpha_{\rm R}-  \beta_{\rm R})$, with  $\beta_{\rm R}= \lambda_{\rm R}- \delta_{\rm R}$. This kind of approach has been used to learn proliferation rates for T-cells~\cite{Zilman2010}. Stochastic versions of such models have also been considered in detail~\cite{Perelson2002, Yates2007, Zilman2010}, showing that theories of age-structured cell populations~\cite{Perelson2002} are equivalent to branching processes~\cite{Zilman2010} (see section~\ref{branching}). For hemapoietic stem cells, a HSC in the bone marrow divides once every 50 days ($\lambda_R=1/110 {\rm days}$) and a HSC becomes a granulocytes after $\sim1$ year (and then lives for a day). This is known from so-called fate-mapping experiments, consisting of tracking labeled cells, which is easier for granulocytes since they are known not to proliferate. It is also known that granulocytes have no other cell source, unlike other cell types. In this case the label from the HSC appears in a granulocyte after a year. Since a lifetime of a mouse is $\sim 2$ years, a $1$ year timescale to produce a new cell may put the steady state assumption into question. The fetal dynamics is very different, the steady state assumption cannot be used, and a new granulocyte is produced in a few weeks~\cite{Busch2015}. 

Technically, measuring cell counts directly is not reliable because of sampling bias. Instead, one should measure cell ratios, $f_i=n^{i}/n^{\rm total}$, as was shown in the previous example. In steady state, the fraction of cells of a given type evolves according to $d{f}_i/dt=k_i (f_{i-1}-f_i)$, with $k_i=\alpha_i-\beta_i$.
By looking at subsequent compartments it is possible to disentangle differentiation and proliferation. This gives an estimate of how {\it self-renewing} the compartment is ($\beta_i$), and how much comes from upstream differentiation ($\alpha_i$). For a perfect stem cell we expect $k_i \rightarrow 0$. Real stem cells (e.g. HSC) are essentially self-renewing with $\tau$ (in mice) on the order of the lifetime of the mouse. Other differentiation compartments are clearly transitional, with $k_i^{-1} \sim 1$ day. HSC divide once every $110$ days but they also differentiate once every $110$ days, so they seem to have a lifetime of  (they replace themselves every) $110$ days (this remains to be verified with more direct means). Short Term HSC (ST-HSC, see Fig.~\ref{fig:hematopoiesis}) die at a rate of 1 per month and Multipotent progenitor (MPP) of 1 per day. This produces a strong self-renewal gradient from HSC to MPP. At the same time the proliferation rates of these cells go up from HSC to MPP with a similar gradient. Some have speculated that these inverse trends of proliferation and self-renewal protect the organism from malignancy by making sure the right amounts of cells are produced. Myeloid lymphomas --- a condition where the organisms produces too many myeloid cells --- have been linked to disturbances in this balance and the condition is only visible in aged individuals when the overproduced cells have accumulated. 

Most HSC become myeloid cells, with only one in 100 cells becoming a lymphoid cell. Lymphoid cells are very long lived and the thymus can function for long time without input from new HSC. In general, due to the slow timescales of the whole process, the system has a lot of inertia. If all HSC are killed in a young adult mouse, there are no visible phenotypic effects for $\sim 5$ months, unless the hemopoietic pathway itself is stressed~\cite{Schoedel2016}.

There are {\em a priori} no biological reasons for downstream cells in the hematopoietic pathway not to influence upstream cells, however this has not yet been directly observed. Probably certain feedback mechanisms are at play. 

\subsection{Inferring the hematopoiesis differentiation tree}\label{stoch_hema}

Analyzing experiments from mice with progenitor cells carrying unique barcodes with a maximum likelihood approach (see section~\ref{inference}) coupled to a stochastic model, Peri\'e et al~\cite{Perie2014}, uncovered a more complex picture of myeloid and lymphoid differentiation. They considered the lymphoid primed multipotent progenitors (LMPPs) differentiation process during which each LMPP is called an MDB, since it has the potential to become a myeloid (M), dendritic (D) or B cell. They built a stochastic branching process model of cell differentiation (see section~\ref{branching}), where in each step of the decision tree a cell can loose one or two of its potentialities: an MDB can become an MD, MB, DB or directly an M, D, B. Similarly an MD can become an M or a D etc. They contrasted this with a traditional differentiation model, where the MB phenotype is forbidden, and an MDB cannot loose two potentials at once. Using mice with progenitor cells carrying unique barcodes, they build lineage trees, traced each offspring cell to its parent and counted how many offspring of each type ($N_i$  for $i\in \{M,D,B\}$) were produced by each MDB cell. The generating function $G(z_M, z_D, z_B)=\sum_{m, d, b} z^m_M z^d_D z^b_B P(m, d, b)$ for the probability of a cell to give rise to a given combination of offspring cells $ P(N_M, N_D, N_B)$ can be calculated from a convolution of simple branching process generating functions~\cite{Perie2014}):
\beq\label{gen_perie}
G(z_M, z_D, z_B)=	H(p_{\rm MDB}, G_{\rm MDB}(z_M, z_D, z_B)),
\eeq
where $H(p,z)$ is given by the generating function,
 in  Eq.~\ref{branchproc_genfun} in section~\ref{branching}, and
 \begin{eqnarray}\label{recursivegen}
\begin{split}
 &G_{\rm MDB}(z_M, z_D, z_B) =\sum_{i \in \{M, D, B\}}p_{MDB \rightarrow i} z_i \\
  &\qquad+ \sum_{i,j \in \{M, D, B\}, i\neq j} p_{MDB \rightarrow ij} H(p_{ij},G_{ij}(z_j, z_j)),
\end{split}
\end{eqnarray}
 $p_{\rm MDB}, p_{\rm MB}, p_{\rm MD}, p_{\rm BD}$  are the probabilities of not differentiating from (staying in) the $MDB, MB, MD, BD$ states and $p_{MDB \rightarrow ij} $ describes the transition probabilities from the $MDB$ state to the $ij \in \{MD, MB, DB \}$ states. The remaining functions are defined recursively as in Eq.~\ref{recursivegen}, e.g.
 for $ij=MD$: $G_{\rm MD}(z_M, z_D) = p_{MD \rightarrow M }z_M+   p_{MD \rightarrow D }z_D$ etc. 

The marginals of this distribution give the probability that a given barcode appears in each of the 7 possible cell types. Following the equal loss of potential model (ELP)~\cite{Ogawa1983, Tsuji1989}, the rate of loss of the cells' ability to make different cell types was assumed to be constant and independent of current cell type, which gives three rates $\alpha_M, \alpha_B, \alpha_D$, that parametrize the transition probabilities between the cell types $p_{i \rightarrow j}$ (for all allowed combinations of $i, j \in MDB, MD, MB, DB, M, D, B$) and reduces the number of parameters. The parameters are determinied by carrying out a maximum likelihood fit $\vec{p}_{\rm max} = {\rm argmax}_{\vec{p}} \left(\sum_i n_i \pi_i (\vec{p} )\right)$ 
, where $n_i$ is the number of observed barcodes in state $i$ and $ \pi_i (\vec{p}) $ is the model probability (from Eq.~\ref{gen_perie}) that an ancestral barcoded cell gives rise to offsprings in states $i$, where $i \in MDB, MD, MB, DB, M, D, B$ calculated from the generating function:  $\pi_{MDB} (\vec{p}) =\sum_{m,d,b\geq 1} P(m, d, b)$, $\pi_{MD} (\vec{p}) =\sum_{m,d\geq 1} P(m, d, b=0)$, $\pi_{M} (\vec{p}) =\sum_{m\geq 1} P(m, d=0, b=0)$ etc.

This approach showed that the classical model of sequential loss forbidding the MB state is not consistent with data. In fact, the MB state is very unlikely but is essential to explain the observed barcode distribution. Simulation of the dynamics of these models showed that the differentiation process between LMPPs and the committed cell types requires $\sim 20$ rounds of differentiation, which translates into about $2$ weeks. Recent barcoding experiments combined with similar inference methods that looked at earlier stages of differentiation from HSC cells~\cite{Pei2017} showed that most HSC cells give rise to many cell fates.

Similar techniques of branching processes coupled to likelihood inference methods were initially used by Yates et al~\cite{Yates2007}, to analyse fluorescent dye experiments of differentiating T-cells and estimate cell division and death rates. 

\subsection{Cell fate during the immune response}

\subsubsection{Choice and timing of cell fate under stimulation}~\label{cyton}

Lymphocytes, both T and B-cells, start growing and dividing upon stimulation during an infection. When they divide they also acquire new specialized functions, make many decisions at the individual cell level, e.g. whether to divide further, whether to switch class for immunoglobulins (Ig) molecules, whether to become memory cells for both T-cells and B-cells, or to become plasmablasts for B-cells. As their number increases, they eventually make another decision to stop dividing and start dying. At the population level we observe a decrease of the population size, after the large expansion due to proliferation, and a return to pre-infection cell counts. Since each cell can in principle adopt a different decision path, at the population level we see a large combinatoric diversity of cell states: from cells that have divided a small number of types and become memory cells, to cells that have divided many types to leave no offspring. 

This diversity is even more staggering for B-cells, since they undergo a phenomena called class (or isotype) switching recombination (CSR), where the variable region (discussed in Sec.~\ref{repertoires}) of the antibody does not change but the constant region of the receptor is modified. This process does not change the affinity of the antibody for the antigen, but it means the antibody cannot interact with signaling molecules (see sec.~\ref{Bcellsignaling}). B-cells do this by expressing a specific gene located in the heavy chain locus, organized in a specific order. During CSR, segments of genes in this locus are removed, and the remaining DNA is recombined to encode a different isotype. Since the double stranded breaks occur at conserved nucleotide motifs, the identity of the expressed isotype-encoding genes are conserved between cells and individuals. Since the non-expressed isotype-encoding genes are deleted from the locus, in general the order of cycling through the classes is conserved, although inter-chromosomal translocation from the other allele can add deleted loci~\cite{Laffleur2014}. 
 
How are these decisions made by cells? One view is that cells integrate signals from the environment, which act as cues to trigger decisions. While this is certainly true \cite{Murphy2007, Murphy2000}, and experimentalists can induce certain cell fates in vitro using cytokine cocktails (see sec.~\ref{cytokines}), an alternative (yet not contradictory) idea was proposed in the ``cyton'' model \cite{Hawkins2007}. The basic idea behind this model is that each cell in the population makes a stochastic decision, choosing from one of the accessible cell fates. To do this it uses an intrinsic pre-programmed timer, telling it when it should divide and die. However since the timer in each cell picks a time from a distribution of times, each cell will go through a different decision scenario. Putting together the different decisions each cell can make this leads to a heterogeneity of cell fates in one population. Yet, for each trait (division times, becoming a memory cell), the population level distribution is reproducible. 

The model can be and has been adapted to different cells \cite{Hawkins2007, Marchingo2014, Marchingo2016, Duffy2012} but let us present it on the simple example of three possible options: to divide,  to die or do nothing even if stimulated. Given an experimentally characterized distribution of division time $\phi_i(t)$ and death times $\psi_i(t)$ for each division round $i$, and assuming that a fraction $F_i$ of cells will divide when stimulated in division round $i$, the number of cells dividing or dying per unit time is a given round $i=1,...,m$ of divisions is:
\begin{eqnarray}
n_i^{\rm div} (t) &=& 2 pF_i \int_0^t dt' n_{i-1}^{\rm div} (t') \left(1\!-\!\! \int_{0}^{t-t'}\!\!\!\!\! dt'' \psi_i(t'') \right) \phi_i(t\!-\!t'),  \nonumber\\
n_i^{\rm die} (t) &=& 2  \int_0^t dt' n_{i-1}^{\rm div} (t') \left(1\!-\!\!F_i \int_{0}^{t-t'} \!\!\!\!\! dt'' \phi_i(t'')) \right) \psi_i(t\!-\!t'),\nonumber
\end{eqnarray}
where the terms in parantheses account for cells that would have divided (died) at time t but had previously died (divided). The total number of cells in each division round is then calculated by summing gains and losses in each round:
\begin{eqnarray}
N_0(t)& =& N- \int_0^t dt' \left(n_0^{\rm div} (t)  + n_0^{\rm die} (t) \right), \\ \nonumber
N_i(t) &= & \int_0^t dt' \left(2 n_{i-1}^{\rm div} (t)  - n_{i}^{\rm div} (t)- n_i^{\rm die} (t) \right),
\end{eqnarray}
for $i=1, ...m$. These integrals can be performed analytically for exponential distributions but since the experimental data are better described by log-normal distributions \cite{Deenick2003,Tangye2003}, the integrals are performed numerically.

The cyton model and its generalizations have been shown to be able to fit the number of cells in experiments with and without stimulation \cite{Hawkins2007}. Of course, one could note that the cyton model takes the experimentally observed distribution of division and death times and simply calculates the number of surviving cells. Yet the power of the model lies in the paradigm shift that assumes there are rules and, while the immune system functions stochastically and relies on heterogeneity, we can nevertheless predict and understand its behaviour. The idea of a programmed stochastic system, as opposed to a black box integration of cues, is extremely powerful. 

Using data from transgenic mice infected with the influenza virus, Marchingo et al~\cite{Marchingo2014} showed that T-cell receptors and costimulatory signals impose an intrinsic division fate on each cell, and cells then count through generations, as defined by the cyton model, before returning to the pre-stimulation state. This initial heritable priming can be later modified by dose dependent cytokine signling, which is also integrated by the cells and influences their the response.  As a result different combinations of costimulatory signal and cytokines can generate signals of similar magnitude. These experiments also show that the cyton model is not in contradiction with signaling-based decision making.

\subsubsection{An aside on cell types}
Cell types in immunology are defined using surface markers. Cells communicate by binding and unbinding signaling molecules (cytokines), for which they need specialized receptors. A cell that commits to participate in a given communication channel, expresses a surface receptor and the combination of surface receptors gives the spying experimenter an idea of how this cell is bound to behave when triggered -- this is what we call a cell type. This phenomenological approach has been very successful in the history of immunology and provides a way to classify different cells and give us some idea about their function and properties. In practice, FACS (Fluorescence-activated cell sorting) sorting experiments segregate cells according into ones that have a high concentration and a low concentration of a given surface marker on their surface. By doing this in many dimensions, one can zero in on a very specific cell type. Technically, this leaves the problem of deciding where the boundary between high and low is -- in practice cells are heterogenous and the experiment produces a distribution of marker concentrations, which, if we are lucky, is bimodal. This problem of finding the separatrices between cell types is called {\it gating}, and is an art implemented in analysis softwares. Advances in machine learning will alleviate the need for expert-guided gated and make the definition of cell types more automatic and easy-to-validate. Being aware and using the heterogeneity of the population data is often useful. Looking at the population distribution has made people realize that often there is continuum of phenotypes and a binary approach is not valid. This has been amplified by recent high throughput cell sorting experiments by mass cytometry --so-called CYTOF~\cite{Bendall2012}, which make it clear that cell types are more continuous and dynamic than traditionally assumed. With that in mind, we can now try to learn something about how cells acquire and switch identity.

\subsubsection{Inferring cell fate timelines during the immune response}

The problem of infering the timeline of cell differentiation during T-cell response can also be treated using the cyton theory~\cite{Hawkins2007, Marchingo2014, Marchingo2016, Duffy2012}. 
Here we present another more data-driven approach to the problem described in Buchholz, Flossdorf et al~\cite{Buchholz2013}. We will concentrate on the example of CD8 T-cells. A clone, here defined as all the TCR that respond to the same antigen, starts with $\sim 10-100$ long-lived naive cells (lifetime of about 80 days in mice). During an infection this clone grows to $\sim 10^{7}$ short-lived effector cells. After infection clearance, there remains $10^3$ memory cells with an intermediate lifetime  (about 15 days in mice). These memory cells can then respond more robustly in a subsequent infection. Memory and effector cells also proliferate within their classes to maintain their pool. This is done through small antigenic signals they constantly receive for memory, and strong antigenic signals for effector cells.

The above textbook description provides a simplified picture, but is unable to put measurable numbers behind each of the described populations. More importantly, this picture was painted by looking by eye at bulk population data summing over many cells. Can inference and  data discrimination approaches help us infer the differentiation pathways? Population-level adoptive transfer experiments can be used to infer the parameters of known models, but it is much harder to infer the topology of the network, especially from single snapshots. Flossdorf et al~\cite{Buchholz2013, Flossdorf2015} analyzed single cell adoptive transfer (single cell fate mapping) experiments to discriminate between different differentiation tree topologies and find the underlying model parameters. They considered a model with four cell types: naive (N), effector (TEF), effector memory precursor (TEMp) and central effector memory precursor (TCMp) and wrote down the most general network diagram, in which every non-naive cell type could differentiate into any other cell type, all of which also could proliferate and further differentiate. Naive cells could become any of the other cells, but no other cell could become a naive cell. This led to 302 possible models, defined by their network topologies, as well as their parameters. The precise definitions of the particular cell types in the experimental data was done using surface markers (see discussion above) can be found in the original paper. The immunological meaning of these cell types is under debate -- but we are mainly interested here in the power of the theoretical method.

The considered models where simple stochastic dynamics for each cell type that considered differentiation ($\vec{\alpha}$) and proliferation ($\vec{\beta}$) reactions. As mentioned above, a lot of heterogeneity was observed within naive cells and in vitro experiments show heterogeneity in the reaction rates, which had implications on the model. Differentiation rates where chosen from underlying distributions: an exponential and gamma distribution of differentiation rates were considered. The data resolution did not allow to discriminate between the two, with an exponential giving a sufficiently good fit. The proliferation rates were taken to be fixed, since longitudinal correlations of proliferation events were previously shown to decay rapidly~\cite{Dowling2014}. The considered models ignored the heterogeneity in the phenotype of naive cells, but since the heterogeneity in the number of naive cells would render model selection impossible, the experiment was tuned to start with one naive cell ($N_{\rm naive}=1$). 

The model consisted of a Master equation for the evolution of the joint probability of the different cell types:
\beq
\partial_t P(\vec{N}) = F\left(P(\vec{N}), \vec{\alpha}, \vec{\beta}, n_{\rm naive}, \vec{r} \right),
\eeq
where is F a linear operator describing the proliferation and differentiation dynamics of the different cells types,  $\vec{N}= \left(N_{\rm TEMp}, N_{\rm TCMp},  N_{\rm TEF} \right)$ are the numbers of cells of each type, and $\vec{r}$ is the death rate vector of all cell types.  The analytically calculated first and second moments of these equations are fit to the variances ($CV_{i}(t)$) and  covariances ($\Sigma_{ij}(t)$ and mean concentrations ($\av{N}_i(t)$) for each cell type $i$ from cell fate mapping data using a mean squared error minimisation.  
  The model selection is then finalized by comparing the AIC (see section~\ref{inference}) of the best fit for each topology. The necessity to use correlations in the inference procedure, and not just population averages, was demonstrated on synthetic data generated with known parameters and topologies. For instance, both a linear ($A\rightarrow B\rightarrow C$) and tree ($C\leftarrow A\rightarrow B$) topology can give rise to the exact same evolution of the mean number of cells as a function of time, for well chosen parameters. However covariances between these numbers give clearly different signatures, allowing us to distinguish the different topologies. These data-based, model-selection approaches are similar to those  developed for stem cell hemapoiesis~\cite{Perie2014}, described in section~\ref{stoch_hema}.

As a result, from the 302 possible models, two strongly stand out as potential candidates to describe the cell differentiation of CD8 T-cells responding to a listeria expressing chicken ovalbumin (to which these T-cells were specific) in mice. These two models are:
\begin{eqnarray}
&&{\rm naive} \rightarrow {\rm TCMp}  \rightarrow  {\rm TEMp} \rightarrow {\rm TEF}, \\ \nonumber
&&{\rm naive} \rightarrow {\rm TCMp}  \rightarrow  {\rm TEMp} \rightarrow {\rm TEF}\ \&  \ {\rm naive} \xrightarrow[]{\text{10 \%}} {\rm TEMp},\\ \nonumber
\end{eqnarray}
where each cell type except for the naive one proliferates. The models have well defined parameters that can easily be identified. This bacterial response is an acute infection and generates exponential cell growth, and the inferred proliferation rates form a gradient from ${\rm TEF} \rightarrow {\rm naive} $ cells with naive cells proliferating the least. Both of these models correctly predict the time dependent dynamics of the mean concentrations of the different cell types. Even if learned on single-cell progeny data at only one time point, the first linear diversification model is able to predict the phenotypic composition of the expanding population at earlier time points. 
 These kinds of approaches shed light on the possible differentiation dynamics and allow us to rule out possible scenarios. To give just one example, for this system the cell differentiation is largely ($\sim 90 \%$) symmetric: the two offspring of a cell are usually the same, although not the same as the mother, as opposed to asymmetric divisions where the two daughter cells have different cell types which happen about $\sim 10 \%$ of times. Also, it turns out that the differentiation and proliferation rates must   depend on time to explain the data. These time-dependent rates are the same in a response to both bacterial (listeria) and viral challenges, suggesting universal response dynamics.

Using similar methods to those proposed by Flossdorf et al~\cite{Buchholz2013, Flossdorf2015}, Miles et al~\cite{Miles2018} showed that the model is consistent with the data of Buchholz~\cite{Buchholz2013} et al, but applied to the data of Kinjyo et al~\cite{KinjoNatComm2015} suggests that memory precursors are produced before effector cells. This shows that this type of inference is robust but the biological interpretation of the results depends both on the experimental conditions and model assumptions.
 
 \subsubsection{Quorum sensing}
 Recent experiments have revisited the question of naive T-cells differentiation into different memory cell types upon antigen stimulation, by considering collective effects through cell-cell communication. Polonsky et al~\cite{Polonsky2018} tracked live cells tagged with antibody markers over time in microwells with different initial cell numbers. By continuously tracking individual cell differentiation states and proliferation, they are able to overcome the limitations of bulk experiments where local cell densities are hard to control.
Live-cell imaging showed that the decision to differentiate into progenitor central memory (pTCM) is collective: in equal medium conditions, cells surrounded by more cells had a larger differentiation rate. That rate depended solely on the instantaneous number of cells in the well, as shown by data collapse of the differentiation rate onto a single curve as a function of the number of cells, for various conditions (different media, different initial number of cells in the well).
The curve shows that cells are more likely to differentiate into pTCM if there are more than $N_c\sim 30$ cells in the cluster. The process can be described by a stochastic differentiation model, in which cells divide and die, and differentiate with a rate $R$. 
However, to explain the data $R(N)$ must be made to depend on the number of cells $N$.
Such a stochastic model also captures the observed well-to-well variability. The collective decision could be linked to the IL-2 and IL-6 cytokine communication: blocking IL-2 reduced the maximum of the universal differentiation rate curve, while blocking IL-6 increased $N_c$ without significantly altering the plateau value.   Interestingly, IL-2 and IL-6 receptors where found to cluster on cell surfaces, with receptor patches directed towards neighbouring cells. This experiment shows that the percentage of cells that differentiate into memory cells and effector cells depends on the instantaneous local T-cell density, independently of the additional post-stimulation influence of T-cell receptor signalling strength, or the effect of antigen presenting cells. 
 
Polonsky et al~\cite{Polonsky2018} give a threshold in terms of absolute number of cells in a micro-well needed for collective decision making, $N_c\sim 30$. However, that number must depend on the size of the well, if cells do communicate through cytokine diffusion. As we have discussed in section~\ref{sec:cytokine_comm} in the context of CD4+ T-cell differentiation into Tregs and T-helper cells, signal propagate to short distances as they are taken up by neighboring cells~\cite{Oyler-Yaniv2017}. For pTCM differentiation, this is facilitated by the IL-2 and IL-6 orientation of the receptors towards secreting cells. But at large distances, diffusion starts playing a role. Active clustering seems to be a mechanism that puts cells in the non-diffusive regime.

The scale of cytokine communication has also been shown to depend on the type of signal: an anti-inflammatory IL-10 signal produced by dendritic cells is long-ranged, while the pro-inflammatory TNF$\alpha$ signal produced by the same cells is short ranged. These cells are also known to cluster in cultures and in-vivo. A two step model for the integration of these two opposing signals was proposed to explain the nonlinear response of dendritic cells when integrating pro- and anti-inflammatory information~\cite{Marcou2018c}: an initial intermediate molecule integrates both signals in an indiscriminate way until a threshold value, providing a bottleneck. This way either the pro- or anti-inflammatory signal can stimulate the system. But the output of the this bottleneck is later down-regulated by only the anti-inflammatory signal. This combination of collective decision making with a modulated bottleneck safeguard was proposed as a way to control possibly excessive immune responses.

Collective decision making, also called ``quorum sensing'', was previously proposed theoretically as a way  for immune cells to help solve the self non-self discrimination problem~\cite{Butler2013a}. According to that theory, which we discuss in more detail in Sec.~\ref{sec:thymic}, communication allows cells to make a census-based decision by integrating the signal read out by many clones in order to correct mistake made by individual cells. This is a cellular implementation of an error-correcting code, similarly to kinetic proof reading discussed in section~\ref{proofreading}. The context of that proposal is different from that of 
\cite{Polonsky2018}, where the T-cell population is monoclonal, while the idea of error correction relies on taking a census of different T cell clones that make independent decisions. 
This difference in clonality does not change the general similarity in the nature of the collective decision. The experimental results are likely to hold for different T-cell clones with with similar affinities for the stimulating antigen. The T-cell response is multiclonal (see section~\ref{sec:rep_response}) and any quorum-sensing mechanism is likely to involve multiple distinct clones. Understanding the details of how the memory and effector cells pools are controlled in a multiclonal setting remains an open question.

\section{Repertoires}\label{repertoires}

\subsection{Size of immune repertoires}\label{repertoiresize}

For the adaptive immune system to protect us against all the different pathogens we may encounter, including ones that may not exist when we are born, we need a large set of different immune receptors. 

There are of the order of $4\cdot 10^{11}$ T cells circulating in the human body \cite{Jenkins2010}, and of the same order of B cells \cite{Glanville2009}, each expressing a single type of receptor to a first approximation~\cite{Han2014,Dupic2018}.
However, T and B cells divide, meaning that many cells can carry the exact same receptor, defining a ``clonotype''. The repertoire size, or number of clonotypes, is therefore smaller than the number of cells.
Early work \cite{Arstila1999} based on a subset of the repertoire gave an estimate of $\sim 10^6$ unique $\beta$ chains in one human individual.

However, these estimates were indirect, and have been updated recently thanks to
the advent of repertoire high-throughput sequencing (RepSeq) techniques in 2009 \cite{Weinstein2009,Robins2009,Boyd2009a}, discussed in a number of recent reviews \cite{Benichou2012,Six2013,Robins2013a,Georgiou2014,Heather2017,Minervina2019}. This method, which we briefly discuss in section~\ref{repseq}, allows one to obtain the sequences of a non-exhaustive but fairly substantial portion of all receptors in a biological sample.
As different technologies have been emerging, the need for standarisation of RepSeq data reporting has appeared, and the Adaptive Immune Receptor Repertoire (AIRR) community~\cite{NatureAIRR} has been organizing meetings and producing web-based and journal publications to promote data reporting unification. 

RepSeq experiments on human blood samples, combined with statistical extrapolation estimators, report unique TCR$\beta$ numbers ranging from $4\times 10^6$ \cite{Robins2009} (using Fisher's Poisson abundance model \cite{Fisher1943}) to the order of $10^8$ \cite{Qi2014} (using the Chao2 estimator \cite{Chao2002}). For BCR heavy chains, estimates obtained with the Poisson abundance model yields $1$-$2\cdot 10^9$ \cite{DeWitt2016}. One should take all these empirical estimate with great caution, as they make unverified, and probably wrong, assumptions about the shape of the clonotype abundance distribution when doing the extrapolation from small blood samples to the entire organism (more on that in Sec.~\ref{diversity}). Theoretical estimates based on population dynamics models assuming that naive cells divide little or not at all logically give much higher estimates, closer to the number of cells itself, $10^{10}-10^{11}$ \cite{Lythe2015,Mora2016e}.

How large is large enough?
De Boer and Perelson \cite{Boer1993} proposed that the minimal repertoire size is ultimately determined by self-tolerance. We revisit their argument in the light of modern estimates, with simplified notations, starting with the easier case of B-cells.
B-cells interact directly with antigens, unlike T-cells that interact with short peptide fragments presented on the multi-histocompatibility complex (MHC). Call $p\ll 1$ the probability that a randomly chosen receptor recognizes a randomly chosen epitope.
During the self-tolerance selection process, the initially generated repertoire of size $R_0$ is reduced to $R$, $R=f R_0$, where $f$ is fraction of receptors that survive negative selection, i.e. that do not recognize any self-epitope, so that 
\beq\label{eq:f}
f=(1-p)^{n}\approx e^{-pn},
\eeq
where $n\sim 10^5$-$10^6$ is the number of self-epitopes (in humans).
The probability that an epitope escapes the immune system is then given by
\beq
p_e=(1-p)^{R}\approx e^{-pR}.
\eeq
This gives the repertoire size $R=-(1/p)\ln p_e$ as well as the pre-selection repertoire size
\beq
R_0=-(1/p)e^{pn}\ln p_e.
\eeq
Assuming that evolution has optimized the recognition probability $p$ so as to minimize $R_0$, solving for $\partial R_0/\partial p$ an optimal $p^*=1/n$, and an optimal repertoire size 
\beq
R^*=R_0 f*=-n \ln p_e,
\eeq
with an optimal selected fraction $f^*=\exp(-1)\approx 0.37$.

\begin{figure}
\includegraphics[width=\linewidth]{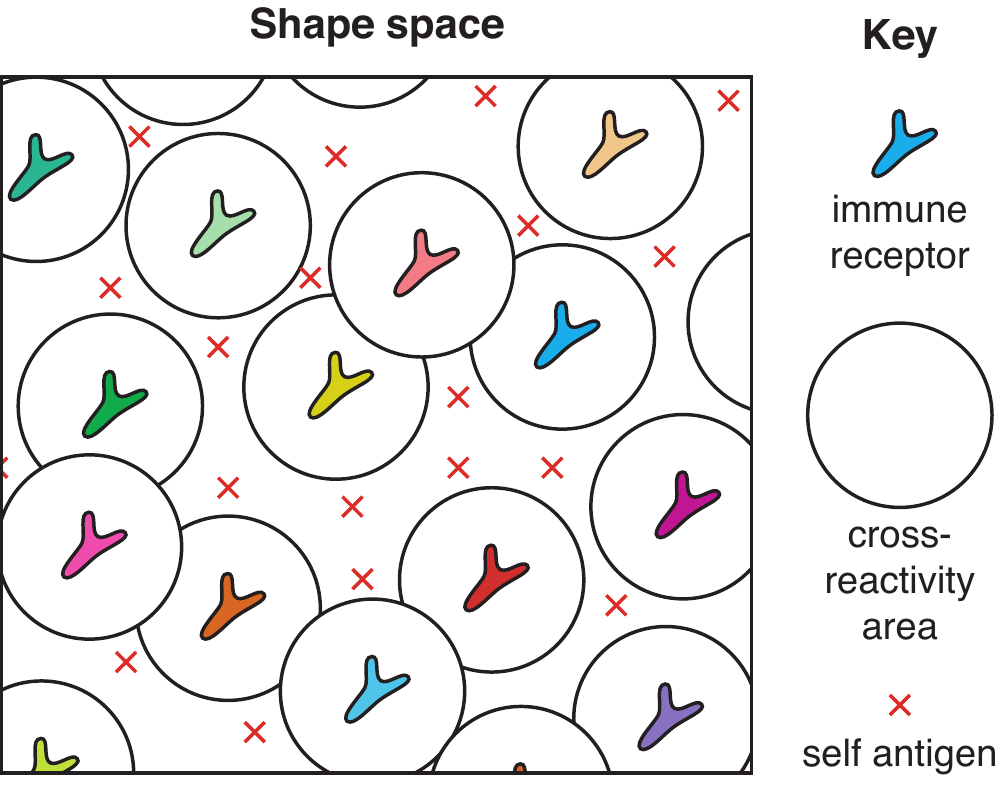}
\caption{\textbf{The minimal repertoire size is set by the number of self-antigens through negative selection}. {A two-dimentional cartoon of the recognition ``shape space'', an abstract space which summarizes the main bio-chemical properties of antigen-receptor binding into a unique space \cite{Perelson1979}. Antigen-receptor pairs that are close are likely to bind and trigger an immune response, while distant pairs do not interact.} Each expressed receptor of the repertoire covers a ball of cross-reactivity where antigens are recognized by that receptor. Negative selection ensures that self-antigens (red crosses) are not included in any of the cross-reactivity balls of the repertoire. To cover shape space efficiently with that constraint, the repertoire size (number of balls) must scale with the number of self-antigens to avoid.
}
\label{fig:covering}
\end{figure}

This estimate suggests that the number of self-epitopes is the main determinant of the minimal repertoire size, with which it scales linearly. Another prediction is that the recognition probability should be inversely proportional to the number of self-epitopes. We can intuitively understand this result by thinking about the repertoire as a covering problem: viewing the set of epitopes recognized by a single receptor as a ball
in an abstract phenotypic space, how many such balls does one need to cover the entire space of foreign epitopes, while avoiding a finite number of self-epitopes?
The simple argument given above, as well as the schematic of Fig.~\ref{fig:covering}, tells us that the volume of each ball, $p$, should be inversely proportional to the number of self-epitopes, $n$, implying in turn that their number ($R$) should scale with $n$.

Empirical estimates suggest $p\sim 10^{-5}$ and $n\sim 10^5-10^6$ \cite{Boer1993}, consistent with the theoretical prediction $p\sim 1/n$. The prediction of $f\approx 37\%$ of cells passing negative selection is consistent with recent experimental estimates $25\%-45\%$ \cite{Wardemann2003}.
The predicted B cell receptor diversity crucially depends on $p_e$, which is hard to estimate. De Boer and Perelson estimated that 99\% of antigens, presenting each $10$ epitopes, were recognized by at least one antibody of the immune system. This implies $p_e^{10}\sim 10^{-2}$, and $R^* \sim 5\cdot 10^4 -- 5 \cdot 10^5$, which is much smaller than current experimental estimates of BCR diversity ($\sim 10^9$).

These estimates become a bit more involved for TCR since epitopes are presented by the Multi-Histocompatibility Complexes (MHCs), which translates into an additional level of sampling. As already mentioned in previous sections, there are 2 types of MHCs relevant to adaptive immunity: MHC type I complexes present peptides from inside the cell to CD8+ (also called killer) T-cells, whose goal is to kill infected or cancerous cells; MHC type II complexes present peptides from outside the cell to CD4+ (helper) T-cells whose role is to stimulate B-cells during affinity maturation process (see section below).

Each epitope, or peptide derived from the antigen of interest (self or foreign) must both be presented by one of the $m=6$ MHC genes for each MHC type, and be recognized by one of the $R$ TCR expressed in the body. This implies that the probability of escape reads:
\beq
p_e=(1-q+q(1-p)^{R})^m\approx \exp\left[-mq\left(1-e^{-pR}\right)\right],
\eeq
where $q\ll 1$ is the fraction of peptides that can be presented by a given MHC molecule. 
Eq.~\ref{eq:f} is still valid, with $n$ being the total number of possible MHC-self-peptide complexes. The repertoire size is then given by
$R_0=-({e^{pn}}/{p})\ln[1+({\ln p_e}/{mq})]$,
and the optimal size is obtained through $\partial R_0/\partial p$:
\beq\label{eq:RTCR}
R^*=-n\ln\left(1+\frac{\ln p_e}{mq}\right)
\eeq
with $p^*=1/n$ and $f^*=\exp(-1)$. Thus, the scaling and selection probability are predicted to be the same as for B cells.

The probability of presenting a given peptide on MHC class I complexes is relatively well known: deep learning algorithms have been very succesful in building algorithms that predict which peptide is likely to be presented. 
These methods are implemented in software packages such as netMHC \cite{Buus2003,Andreatta2015}, which account for known biological features to guide the learning. The same prediction task has been implemented for MHC class II \cite{Jensen2018}, but the prediction is much harder, partly because unlike MHC class I presented peptides, which have a relatiely fixed length of 9-11 amino acids, MHC class II presented peptides have varying length, from 13 to 25. More data than is currently available is needed to successfully train the neural networks, and to improve prediction. The models predict that about 1\% of peptides are presented by any given particular MHC allele, i.e. $q=0.01$.

The number of MHC-self-peptide complexes is given by $n=qm n_{\rm p}$, where $n_{\rm p}$ is the number of peptides in the human proteome. Each individual has 6 MHC alleles for each class, i.e. $m=6$.
$n_{\rm p}$ is roughly the number of proteins, $\sim 3\cdot 10^4,$ times their average length, $\sim 400$, i.e. $n_{\rm p}\sim 10^7$, and $n\sim 6 \cdot 10^5$. The recognition probability, $p$, has been estimated to be in the range $10^{-4}$-$10^{-6}$ \cite{Moon2007203,moon-2012}. These numbers are consistent with the theory $p\sim 1/n$. The fraction of negatively selected cells, predicted to be $f^*\approx 37\%$ by the theory, is estimated to fall in the range $20$-$50\%$ \cite{Yates2014}.
As for B cells, predicting the size of the repertoire is hard without a good estimate of $p_e$, but current estimates of $R=10^8$-$10^{10}$ are probably much larger than predicted by \eqref{eq:RTCR} regardless of the estimate of $p_e$.

In summary, these theoretical estimates still seem to be relevant today for the relationship between recognition probability, $p$, and number of self-epitopes, $n$, both for T and B cells. Recent estimates of TCR and BCR diversity, however, are much larger than predicted by these simple theories.
However appealing, a major limitation of these estimates and predictions is that Eq.~\ref{eq:f} assumes that each receptor is tested against all possible self-epitopes, while in practice this is impossible due to the limited duration of lymphocyte maturation. We will come back to this point in Section \ref{sec:thymic} devoted to modeling thymic selection.

\subsection{Inference of the stochastic repertoire generation process}\label{pgen}

Antigen receptors are proteins, which must be encoded as genes in the DNA. Humans have of the order of $10^4$ protein-coding genes. Directly encoding the whole diversity of immune receptor genes ($\sim 10^8-10^{10}$) in each genome would make it impossible for the DNA to fit in the nucleus. The immune system has solved that problem by stochastically creating receptors in each cell, in a process called V(D)J recombination that combines combinatorics and randomness to generate diversity.
B- and T- cell receptors are made of two chains, light and heavy for BCR, and $\alpha$ and $\beta$ for TCR. The  genome encodes a certain number of gene templates, called V, D and J for the heavy and $\beta$ chains, and V and J for the light and $\alpha$ chains.
Upon creation of a chain, DNA is edited and one of each of these gene templates is chosen per receptor.
The combinatorics of templates typically results in $\sim 10^3$ different receptors. To obtain the (much larger) observed diversity, nucleotides are randomly inserted in a non-templated way and deleted at the junctions between the V and D and D and J genes (or V and J genes for $\alpha$ and light chains). This process of generating ``junctional diversity'' in fact accounts for most of the diversity of the repertoire \cite{Murugan2012, Elhanati2015}. 

Characterizing the above described process of receptor generation in quantitative detail has been made possible thanks to the development of RepSeq methods (see Sec.~\ref{repseq}). This can be done using out-of-frame sequences (with a frameshift due to the random number of additions and deletions at the junctions), which are nonproductive and hence a raw product of recombination, as they are free of selection effects. These sequences survive selection because they are in the same cell as an in-frame sequence. Since each cell has two chromosomes, out-of-frame sequences come from cells where the first rearranged chromosome resulted in an out-of-frame sequence and the second rearranged chromosome resulted in an in-frame sequence. 
Due to the random insertions and deletions of nucleotides, it is impossible to reliably determine how a given receptor was formed (V, D, J assignments, as well as distinguishing inserted from templated nucletoides) from its mere sequence.
Instead one can consider a list of scenarios (which includes V, D and J gene choice plus a number of insertions and deletions at each of the junctions) and sum over them weighted by their probabilities as determined self-consistently using a probabilistic model. Concretely, in the case of the simpler $\alpha$ or light chains, the probability of a given recombination scenario $r$ is:
\beq\label{eq:pr}
P_\text{\rm rearr}(r) = P(V,J) P(\text{del}V|V) P(\text{del}J|J) P({\rm ins}VJ),
\eeq
where $\text{del}V$ and $\text{del}J$ denote the number of deletions at the V and J ends, and ${\rm ins}VJ$ is the list of inserted nucleotides. A similar expression can be written for the $\beta$ or heavy chains by adding the D gene and the related deletions and insertions.  The model is the most general factorizable distribution that is consistent with the data and the known biological constraints (such as the relative positioning of the genes in the genome). 
The overall generation probability of a given sequence $s$ is obtained by summing of the probabilities of all scenarios that could have given rise to this sequence:
\beq\label{eq:pgen}
P_{\rm gen}(s)=\sum_{r\to s} P_{\rm rearr}(r).
\eeq
The learning of the model parameters, encoded in the probability distributions in the factorized form \eqref{eq:pr}, is performed by maximizing the likelihood of the sequences, using the Expectation-Maximization algorithm to deal with the sum over the hidden variable $r$ (see Sec.~\ref{inference} for details about Maximum Likelihood and Expectation-Maximization).

This inference procedure has been applied to variety of immune receptors and species, TCR $\beta$ \cite{Murugan2012} and $\alpha$ \cite{Pogorelyy2017} chains, BCR heavy \cite{Elhanati2015} and light chains \cite{Toledano2018} in humans, as well as TCR$\beta$ in mouse \cite{Sethna2017} and BCR heavy chains in trout \cite{Magadan2018}. It is implemented in the IGoR software \cite{Marcou2018} which can be used to learn additional models from other species and locus combinations. IGoR can also be used to generate synthetic sequences from the $P_{\rm gen}$ distribution, and to estimate $P_{\rm gen}(s)$ of an arbitrary nucleotide sequence $s$. Another method, OLGA \cite{Sethna2019}, was designed to estimate the probability of amino-acid sequences, using dynamic programming to deal with the enormous sums involving the enumeration of all possible nucleotide variants. Other methods relying on hidden Markov models (see Sec.~\ref{inference}), were also proposed to handle the high dimensionality of hidden variables \cite{Munshaw2010,Elhanati2016,Ralph2016,Ralph2016a}.

Repertoires generated {\it in silico} using the learned model confirm that the inference is able to call the correct recombination scenario in at best $\sim 25\%$ of cases, making the deterministic annotation of sequences impossible and necessitating  a probabilistic inference approach. A perhaps counterintuitive consequence is that the model parameters can be inferred with arbitrary accuracy despite our inability to annotate any sequence reliably.

The results of the inference show that the probability of generation of receptors is incredibly reproducible between individuals of the same species. The distribution is dominated by insertions and deletions, whereas gene choice contributes relatively little to the overall probability. Among the features of V(D)J recombation, the gene choice distribution varies the most between unrelated individuals \cite{Murugan2012}, even when corrected for single nucleotide polymorphisms (SNP) in V, D and J alleles~\cite{Wang2011}. As far as can be determined, the generations of the $\alpha$ and $\beta$ chains are largely independent, and the overall generation probability of a receptor is well approximated by the product of the generation of its two chains~\cite{Dupic2018}. Although no similar analysis has been performed for BCR, it is likely that the formation of its two chains are also independent.

Although these recombination models ignore selection effects, to be discussed in the section below, it is a good predictor of the abundance of particular TCR$\beta$ in a human population \cite{Sethna2019}, and it has been used to detect signatures of immune responses as deviations from this baseline distribution \cite{Pogorelyy2018a,Pogorelyy2018b}.

\subsection{Thymic selection and central tolerance}\label{sec:thymic}
After the receptors are generated, they undergo an initial selection step, known as thymic selection in T-cells. A similar process called central tolerance occurs in B-cells maturation \cite{Wardemann2003}, but let us focus on T-cell for concreteness. For a more comprehensive survey of models of thymic selection, we refer readers to the review by A. Yates \cite{Yates2014}, and focus here on recent development involving quorum sensing and data-driven models of sequence specific selection.

T-cells mature in the thymus, an organ that contains only proteins that are native to the host organism, termed self-proteins. The newly generated receptors are expressed on cell surfaces and their binding properties are tested against the peptides from the self-proteins, presented by the MHC complexes. If the receptor fails to bind any self-peptide, even weakly, it will probably fail to bind any protein and the cell carrying these receptors is discarded -- a process called positive selection which removes $\sim 80\%$ of immature cells. Conversely, if a receptor binds any one self-protein too strongly, it is also discarded as a result of negative selection, since it is likely to bind self-proteins later on and trigger auto-immune diseases. Negative selection was at the core of the argument for the optimal recognition probability and repertoire size presented in Sec.~\ref{repertoiresize}, and it is estimated to remove $50\%-80\%$ of cells. The detailed process of T cell selection, complete with the timing of the $\alpha$ and $\beta$ chain recombination events, is summarized in Fig.~\ref{fig:thymic_sel}. During positive selection, cells with TCR that recognize peptides presented by the MHC class I commit to becoming CD8+ cells, whose main function is to kill infected cells, while cells with MHC class II TCR specificity become CD4+ cells, whose main function is to help B cells in their affinity maturation, and to regulate the immune response. There are about 4 times as many CD4+ cells as CD8+ cells. A small subclass of CD4+ T cells, called regulatory T cells (Tregs), actually suppress the immune response, and play an essential role to prevent auto-immunity \cite{Wing2010}. Tregs are selected for higher affinities to self-epitopes than regular T cells, which allows them to selectively suppress immune responses to self-antigens, although the picture seem to be quite complex (see \cite{Bains2013b}).

\begin{figure}
\includegraphics[width=\linewidth]{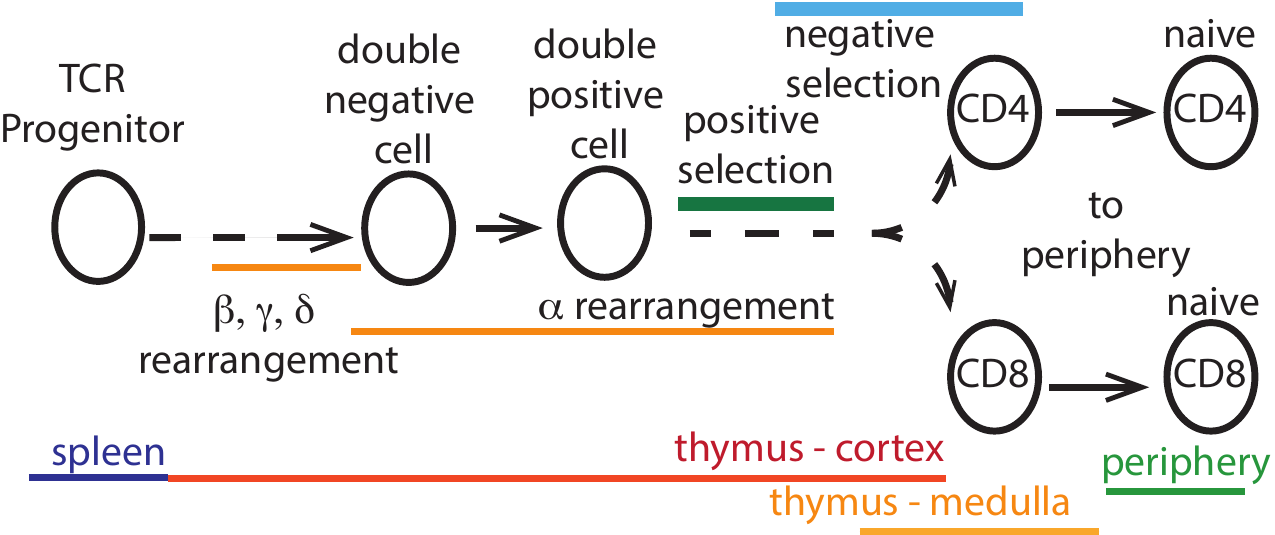}
\caption{\textbf{Thymic selection}. {A cartoon representation of the basic processes that take the cells from the bone marrow to the thymic cortex and medulla and finally lead to the generation of functional naive CD4+ and CD8+ T-cells that are exported to the periphery. The duration of the processes is not drawn to scale -- it is meant solely as an indication of temporal order and overlap.}}
\label{fig:thymic_sel}
\end{figure}

To be able to ensure self tolerance, each receptor should in principle be tested against all possible self-epitope, which would take an impractically long time. In practice, experiments suggest that each T cell may encounter around 500 antigen presenting cells \cite{LeBorgne2009}, while theoretical arguments estimate the number of presented self peptides around a few thousands \cite{Butler2013a} (these numbers are not inconsistent because each antigen-presenting cell may present many different peptides in single encounter with a T cell). These numbers are much lower than the estimated diversity of presentable self peptide-MHC complexes, $5\cdot 10^5-5\cdot 10^6$ (see Sec.~\ref{repertoiresize}). Calling $x$ the fraction of presented self peptides during negative selection, the probability that none of the selected TCR are self-reactive is $(1-p)^{RNx}$, which quickly goes to 0 as $R$ becomes even moderately large, even if $p$ scales as $1/N$.

A recently proposed solution to this problem is to assume that a minimal number $t$ of TCR must recognize a peptide to trigger an immune response \cite{Butler2013a}. By virtue of the law of large numbers, the number of T cells responding to a self-peptide is distributed as a Gaussian with mean and variance $=Rp(1-x)$, while the number of T cells responding to a foreign peptide is of mean and variance $=Rp$. Good discrimination is achieved when the two distributions are well separated, giving the condition $Rp x^2\gg 1$, with an optimal discrimination threshold $t^*=Rp(1-x/2)$ in the limit $x\ll 1$. This condition is barely satisfied by $x=10^{-2}$, $R=10^9$ and $p=10^{-5}$, for which the optimal threshold for the number of TCR involved is $t^*\sim 10^4$. This argument suggests another, stricter lower bound on the diversity $R$ of the T cell repertoire that is necessary to make such a collective decision. It also implies a ``quorum sensing'' mechanism by which responding T cells have a way of estimating how many other cells are involved in order to commit to a response. Recent work suggests that such quorum sensing does occur locally, probably using cytokine signaling \cite{Polonsky2018}.

To model thymic selection in more quantitative detail, a common strategy has been to use additive models of binding free energy similar to \eqref{eq:specificity} \cite{Detours2000,Kosmrlj2008,George2017}. In these models, the only receptors that  survive are those whose maximal affinity $E^*$ to any self peptide falls within a range $(E_p,E_n)$ corresponding to the positive and negative selection thresholds. This condition can be mapped onto an extreme value statistics problem, allowing for a statistical mechanics treatment \cite{Kosmrlj2009}. Under this framework, it was shown that the sequence composition of negatively selected TCR was biased towards weakly binding residues \cite{Kosmrlj2008}, and the theory was subsequently applied to explain clinical data on ``elite controllers'' of HIV expressing a particular type of MHC class I molecule, HLA-B27 \cite{Kosmrlj2010}. A similar theory was used to study the sensitivity of TCR that target tumor cells \cite{George2017}.
The parameters of these models are not inferred empirically, but are instead picked from popular but unrealistic interaction potentials between amino acids such as the Miyazawa-Jernigan matrix \cite{Miyazawa1996}. Yet many conclusions of these studies are relatively insensitive to the details of the interaction matrices in the extreme value statistics regime. 

Statistical inference methods based on immune repertoire sequencing can also be used to estimate the probability that a particular receptor passes selection based on its sequence. We can define a selection factor corresponding to the ratio of probabilities to find a sequence in a selected repertoire, $P_{\rm sel}(s)$, relative to its probability in the unselected repertoire, as given by the recombination model $P_{\rm gen}(s)$: $Q(s)=P_{\rm sel}(s)/P_{\rm gen}(s)$. In practice, it is impossible to evaluate $P_{\rm sel}(s)$ directly, as the number of possible sequences to be considered is too large, spanning more than 20 orders of magnitude in generation probabilities. However, simplifying assumptions can be made on the form of $Q(s)$.
Specifically, Elhanati et al~\cite{Elhanati2014} considered:
\beq\label{eq:Q}
Q(s)=q(L)q(V,J)\prod_{i=1}^{L} q_{i|L}(a_i),
\eeq
where $(a_1,a_2,\ldots,a_L)$ is the amino-acid sequence of the Complementarity Determining Region 3 (CDR3, which forms a loop important for antigen recognition, and covers the most variable part of the receptor ranging from the end of the V to the beginning of the J segments). Selection is assumed to act independently on the length of the CDR3 region $q(L)$, the V gene,  $q(V)$, and the J gene, $q(J)$. The $D$ gene is not taken into account seperately, but selection on amino acids in the CDR3 is considered  explicitly. The parameters of the model are inferred by maximizing the likelihood using Expectation Maximization (Sec.~\ref{inference}). This approach has been applied to the initial selection process of $\beta$-chain sequences \cite{Elhanati2014}, $\alpha$-chain sequences \cite{Pogorelyy2017}, and BCR heavy chain \cite{Elhanati2015} and light chain \cite{Toledano2018} sequences. For many of these cases, the $Q(s)$ selection factors did not differ substantially from individual to individual. Instead, they reflected an global selection for general biophysical and biochemical features of amino acids (suggesting positive selection for proper protein function), rather than individual-specific removals in the repertoire caused by negative selection. Similar observations were made for BCR by direct comparison of pre-mature and mature repertoires \cite{Kaplinsky2014}. Modeling negative selection is much harder, as it requires to learn the ``holes'' that selection pokes into the repertoire, whose post-selection landscape looks like a dense golf course. A simple way to model negative selection is to remove a random fraction $1-q$ of the sequences from the selected repertoire, boosting the likelihood of surviving ones by a factor $1/q$ \cite{Elhanati2018a}. In practice, in simulations one can use a hashing function, which associates a quasi-random number to each sequence, and select sequences whose hashing number falls below a chosen threshold, so that selection is reproducible, yet agnostic to the features of the sequences.

Note that models of the form \eqref{eq:Q} are not restricted to thymic selection, and can be used to describe the sequence-wide selection pressure during any process, by comparing the initial distribution before selection to the final distribution after selection, as was for instance done to characterize responsive receptors to yellow fever vaccination \cite{Pogorelyy2018b}. For instance, different T cell subsets (e.g.  CD4+ and CD8+ cells), different BCR isotypes, in different organs or environments, are probably characterized by distinct selective pressure, each of which could be modeled using similar $Q(s)$ factors.

\subsection{Diversity and the clone size distribution}\label{diversity}
The distributions of TCR and BCR span a very large space, and are highly skewed.
It is often useful to summarize these distributions by a single summary statistics that quantifies their diversity. Diversity measures in the context of immune repertoires have been thoroughly discussed in a recent review \cite{Mora2016e}. Diversity is important for understanding how well our immune repertoire prepares us against a wide range of pathogenic threats.
As we will see, diversity measures are deeply linked with the distribution of sizes of immune clones (number of cells with the same receptor), which contains important information about the dynamics of immune cells in response to environmental challenges (see Sec.~\ref{dynamics}). In addition, it determines the amount of overlap one expects between repertoires of distinct individual, and underlies the concept of ``public repertoire'' shared by all individuals (see Sec.~\ref{sharing}). All these aspects can be formalized within the common langage of statistical mechanics through the definition of a density of state, which we introduce below.

The most general family of diversity measures for a distribution $p(s)$ is given by Renyi entropies:
\beq
H_\beta=\frac{1}{1-\beta}\ln\left[\sum_s p(s)^{\beta}\right].
\eeq
This definition reduces to the Shannon entropy for $\beta=1$, $H_1=-\sum_s p(s)\ln p(s)$ (see Sec.~\ref{infotheory}), to the total number of receptors for $\beta=0$, $R=\exp(H_0)$, and to the probability of drawing the same receptor twice with replacement for $\beta=2$, $\exp(-H_2)=\sum_s p(s)^2$, also equal to the inverse of the the Simpson index. More generally, the family $H_\beta$ recapitulates the entire clone size distribution, defined in its cumulative form as the number $G(E)$ of receptor sequences with $-\ln p(s)<E$, through a Laplace transform:
\beq\label{eq:renyi2}
H_\beta=\frac{1}{1-\beta}\ln\left[\int dG(E)\,e^{-\beta E}\right].
\eeq
Note that $G(E)$ is formally equivalent to a cumulative density of states in statistical physics. $E=-\ln p$ is sometimes called the ``surprise'' and is also formally similar to an ``energy'' by analogy with Boltzmann's law, $p\sim \exp(-E)$, making $\beta$ analogous to an inverse temperature. The mapping between $G(E)$ and $H_{\beta}$ is formally analogous to the equivalence of the canonical and micro-canonical ensembles, and is amenable to a statistical physics analysis \cite{Mora2016d}. The lower the temperature (the higher the $\beta$), the more the R\'enyi entropy concentrates on low-energy, high probability sequences.

$G(E)$ can also be interpreted as the rank (ordered from most probable to least probable) of sequences with probability $p=e^{-E}$. 
The distribution of clone sizes is often presented in terms of a clone-size frequency rank distribution, where clones are ranked ordered by their sizes, and their normalized frequency  is plotted as a decreasing function of its rank.
$G(E)$ precisely encodes this rank-frequency relation.

It is important to distinguish between the potential diversity, corresponding to a theoretical distribution $P_{\rm gen}(s)$ or $P_{\rm sel}(s)$, and the realized diversity, corresponding to the actual distribution of a finite number of receptors in a particular sample, $p(s)=n(s)/N_{\rm tot}$, where $n(s)$ is the number of molecules or cells with receptor sequence $s$, and $N_{\rm tot}=\sum_s p(s)$.

Let us start with the potential diversity derived from the probability distribution $P_{\rm gen}(s)$ (see Sec.~\ref{pgen}). For human TCR $\beta$, the total number of possible sequences, $e^{H_0}$, is infinite for all practical purposes ($>10^{39}$ \cite{Mora2016e}), as it is larger than the total number of TCR$\beta$ receptors having ever been produced by a human being. For this reason, it makes more sense to compare Shannon entropies. These can be computed using \eqref{eq:renyi2}, where $G(E)$ is evaluated from the probability density distribution of $E$ among randomly drawn sequences according to $p(s)$, $\rho(E)=(dG/dE)e^{-E}$, so that $G(E)=\int_0^{E}dE'\, \rho(E')\exp(-E')$.

Entropy estimates for both nucleotide and amino acid sequences are reported in \cite{Sethna2019}. Human nucletoide TCR $\beta$ diversity is $H_1\sim 44$ bits (bits refer to $\ln(2)$ units) for nucletoides and $30$ bits for amino-acids, TCR $\alpha$ divesity is $\sim 30$ bits for nucleotides and $25$ bits for amino acids. 
The BCR heavy chain entropy is $67$ bits for nucleotides and $53$ for amino acids. These numbers are very large. If the distribution were uniform, that would correspond to $2^{55}\approx 3.6\cdot 10^{16}$ distinct amino acid TCR$\alpha\beta$ sequences. Because the distribution is non-uniform, the pooled repertoire of all humans having ever lived do not exhaust the potential diversity. As noted before, most this diversity is due to random insertions and deletions.

Thymic selection and central tolerance effectively reduce these potential diversities in $P_{\rm sel}(s)$, by around $9$ bits for TCR$\beta$ \cite{Elhanati2014}, $4$ bits for TCR$\alpha$ \cite{Pogorelyy2017}, and $12$ bits for BCR heavy chain \cite{Elhanati2015}, but the corresponding diversity number remain very high. This diversity loss does not mean that e.g. $2^{-13}\approx 10^{-4}$ of TCR must be discarded; the probability of thymic selection $Q(s)$ is in fact correlated with generation probability $P_{\rm gen}(s)$, meaning that diversity is reduced by selecting already likely sequences.

\begin{figure}
\includegraphics[width=\linewidth]{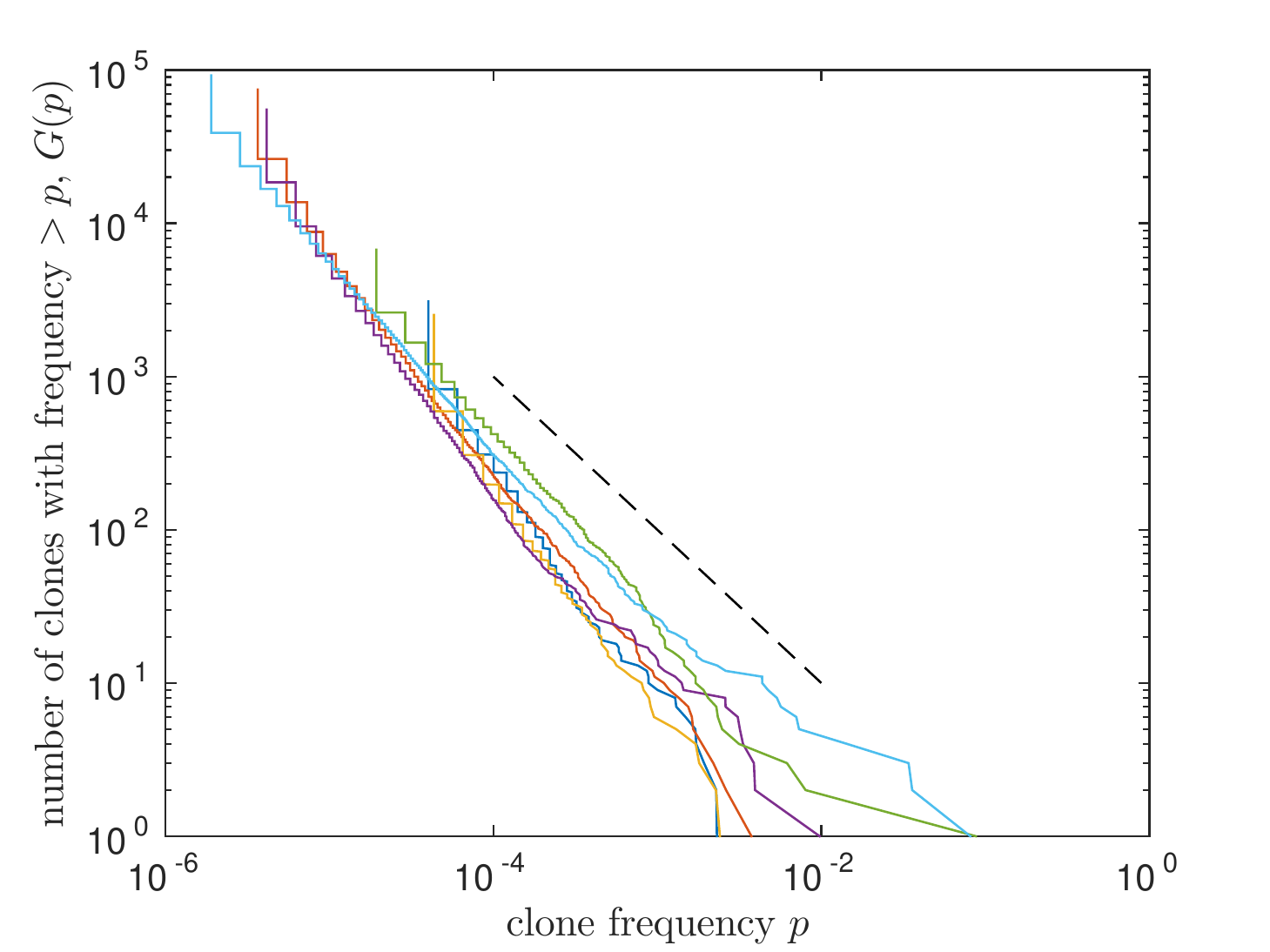}
\caption{\textbf{Distribution of clone sizes follow power laws}. Cumulative distribution $G(p)$ of clone frequencies $p$ of unfractionned TCR $\beta$ sequences sampled from the blood of 6 human donors \cite{Pogorelyy2017}. The dashed line represents slope $-1$.
}
\label{fig:clonesize}
\end{figure}

Let us now turn to the realized diversity in a sample or an individual.
This diversity is best described by the whole clone-size distribution encoded by $G(E)$. Based on recent high-throughput sequencing experiments, the distributions are often long-tailed, spanning several orders of magnitude (see Fig.~\ref{fig:clonesize}) \cite{Mora2010,Desponds2016,Desponds2017,Mora2016e}. These tails seem to be mostly due to the memory fraction of the repertoire \cite{Greef2019}.
Initial worries that these long tails may arise from noise is the Polymerase Chain Reaction, which amplifies differences exponentially \cite{Best2015b}, were put to rest thanks to the introduction of unique molecular barcodes associated to each initial mRNA molecule \cite{Vollmers2013,Shugay2014a,Heather2017}.
Specifically, the clone size distribution follows a power-law:
\beq
G(E)\sim e^{\hat \beta E} \sim \frac{1}{p^{\hat\beta}},
\eeq
with $\beta\leq 1$ but close to 1. The probability distribution of clone frequencies, 
\beq\label{eq:powerlawfrequencydistribution}
\rho(p)=-\frac{dG}{dp} \sim \frac{1}{p^{1+\hat\beta}},
\eeq
is also a power law of exponent slightly below 2. Another way to look at the distribution is to consider the inverse relationship,
\beq\label{eq:zipf}
p\sim \frac{1}{G^{1/\hat\beta}},
\eeq
which also follows a power law, called Zipf's law in this particular context when its exponent $\hat\beta^{-1}$ is close to 1. Since $G$ is the rank of the clone (from most frequent to least frequent), this relation is called the rank-frequency relationship. 

Because of these long tails and limited sampling, very few of the diversity measures can be estimated reliably.
One way to overcome these issues is to build a statistical model that is then parametrised by the experimental data. In Sec.~\ref{repertoiresize} we already mentioned Fisher's Poisson abundance model and the Chao estimators, which use this idea. More sophisticated methods have been proposed for immune repertoire, e.g. using Expectation Maximization (see Sec.~\ref{EM}) \cite{Kaplinsky2015} or using rarefaction curves \cite{Laydon2015}, but all these methods are susceptible to huge errors when the clone size distribution follows a power law \cite{Haegeman2013}. We expect the reported numbers to grossly underestimate the true diversity of both BCR and TCR. To make progress, models of lymphocyte population dynamics should be leveraged to reliably extrapolate the tails of clone-size distributions. We will discuss some of these models in Sec.~\ref{dynamics}.

\begin{figure}
\includegraphics[width=\linewidth]{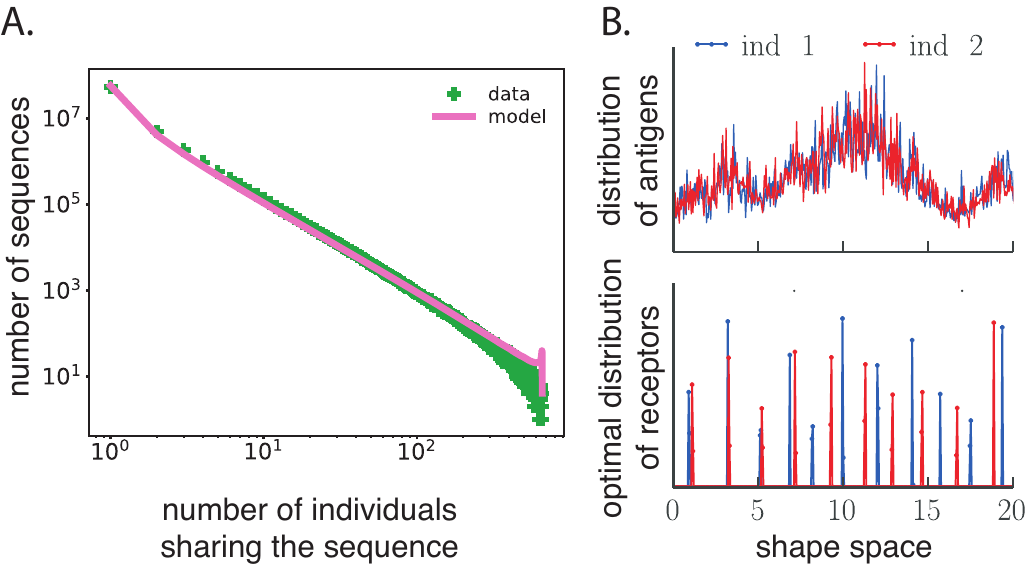}
\caption{\textbf{Overlap of immune repertoires between individuals}. {A. The amount of shared lymphocyte receptors (TCR $\beta$ chain) between individuals  is consistent with pure chance \cite{Elhanati2018a} (data from \cite{Emerson2017a}).} B. To explain that observation, a simple model of optimal immune coverage of pathogens \cite{Mayer2015} predicts that even very similar immune environments experienced by two individuals (top) can lead to very different (and peaked) distributions of protections (bottom).}
\label{fig:sharing}
\end{figure}

\subsection{Repertoire sharing}\label{sharing}

Receptor sharing is at the heart of the public vs private debate in immunology. Public receptors are those shared between individuals in a given cohort, while private are ones that are seen only in individuals. If one studies a group of individuals with a specific condition (e.g. cytomegalovirus or CMV \cite{Emerson2017} or Ankylosing Spondylitis \cite{Faham2017}), receptors that are shared between these people can be expected to be linked to a disease. This justifies the search and investigation of ``public'' sequences. Both humans and mice do share a non-negligible numbers of receptors in functional repertoires, regardless of their lifestyles, environmental factors and family ties. However, receptors are also expected to be shared by chance, because some sequences are likely to recombine independently in different individuals \cite{Venturi:2006hk, Venturi:2011co, Elhanati2014, Madi2014a}. One can try to correct for this convergent recombination by using an outgroup --- a group of individuals that are as similar to the condition-specific group but do not have this condition --- and identify sequences shared more in the studied group than in the control group, as was done for CMV \cite{Emerson2017}.
Alternatively, one can use the knowledge of immune repertoire diversity encoded in $P_{\rm gen}(s)$ to estimate our null expectation.

All sharing properties are recapitulated by the cumulative density of states $G(E)$. Given $n$ independent repertoires of sizes $N_1,\ldots,N_n$, the expected number of sequences shared by exactly $m$ repertoires, $M_m$, is given by the following generating function \cite{Elhanati2018a}:
\bea
g(x) & = &\sum_{m = 0}^{n} M_m x^m \label{eq:sharing}\\
&=&\int_0^{+\infty} dG(E) \prod_{i=1}^n \left[ e^{-N_i e^{-E}} + \left(1 - e^{-N_i e^{-E}} \right) x \right]\nonumber
\eea
For example, for $n=2$, in the limit of small samples, $N_{1,2}e^{-E}\ll 1$, the expected number of shared sequences is
\beq
\begin{split}
M_2\approx &  N_1 N_2\int dG(E)\,e^{-2E}\\
\approx & N_1 N_2\sum_s p(s)^2 = N_1 N_2 e^{-H_2},
\end{split}
\eeq
reducing to simple biased birthday problem related to Simpson's diversity and thus R\'enyi entropy of order 2. More generally, sharing between $m$ small samples is governed by $\int dG(E)\,e^{-mE}$ and to R\'enyi entropies of order $m$. The ensemble of sequences shared between $m$ samples thus corresponds to sampling the generation probability at temperature $1/m$, focusing on more and more likely sequences as $m$ increases.

This calculation shows that in such large ensembles of cells as the B or T cell repertoire, which contain upward of $10^8$ unique receptors, we expect the most common receptors with a high probability to be independently generated multiple times in different people. The number of shared TCR$\beta$ sequences among sampled repertoires from a cohort of 658 human donors \cite{Emerson2017} was well predicted by the model in Eq.~\ref{eq:sharing}. By itself, the generation probability $P_{\rm gen}$ was enough to build a classifier that can determine whether a given sequence will be shared by a minimal number of people, with a $>95\%$ accuracy for $m>10$.
This analysis also brings our attention to the definition of public repertoires. A receptor that is shared between $m=2$ and $m=1000$ people has a very different notion of publicness. For this reason it is better to talk about degrees of publicness. In addition, sharing and publicness depends very sensitively on the collected sample size $N_i$. A reasonable choice for a single definition of receptor publicness could that the sequence must be present on average once in each individual, implying $P_{\rm gen}(s)>1/R$.

Interestingly, the theory can also identify receptors that are shared for other reasons than sheer chance. These receptors have a high propensity to be shared, but a relatively low generation probability. The discrepancy between chance prediction and observed sharing was used to discover shared T cell clones in identical twins \cite{Pogorelyy2017}, to find candidate TCR sequences associated with conditions such as CMV or diabetes without a control group \cite{Pogorelyy2018b}, or to assess the public BCR response to vaccination in trout \cite{Magadan2018}.
 
Identifying common, even unusual receptors does not necessarily mean they are responding to the same antigen or  pathogen, and even when it does it does not tell us what the target antigen is.
As discussed in Sec.~\ref{rec_antigen_spec}, immune response is a complex phenomenon that involves the binding different epitopes to immune receptors, followed by signal propagation, cell commitment, and cell-cell communication through messenger molecules. Yet tools based on the statistical expectations of an unbiased repertoire allow us to identify deviations from this baseline, and propose candidate responding receptor sequences to be tested in other experiments. 

\subsection{Optimal receptor distribution}\label{optimal}
In Sec.~\ref{repertoiresize} we discussed theoretical arguments for an optimal repertoire size, but recent data suggests that actual repertoire sizes are much bigger than these theoretical predictions. However, as stressed in Sec.~\ref{diversity}, the distribution of receptors is itself highly skewed, suggesting that the relative abundance of receptor types, rather than their absolute number, is an important factor of repertoire design. Can we find an optimization principle over the composition of the repertoire? Take a distribution $Q_a$ of antigens to be recognized, where $a$ lives in a theoretical high-dimensional ``shape space'' encompassing both antigens and receptors, and in which proximity reflects receptor-ligand affinity \cite{Perelson1979,Perelson1997}, through a recognition or ``cross-reactivity'' matrix $f_{ar}$ between antigen $a$ and receptor $r$ which encodes the probability that a random encounter between the two results in an immune response. For a given distribution of receptors, $P_r$.
The expected cost of an infection linked to $a$ is a decreasing function $c(\cdot)$ of the probability that a chance encounter with a receptor is successful, $\sum_r f_{ar}P_r$, so that the expected cost for a random infection reads \cite{Mayer2015a}:
\beq\label{eq:cost}
C(\{P_r\},\{Q_a\})=\sum_a Q_a c\left(\sum_r f_{ar}P_r\right).
\eeq
Minimizing that cost with respect to the receptor distribution $\{P_r\}$ gives an interesting multi-peak structure (Fig.~\ref{fig:sharing}B). Because of the degeneracy of recognition allowed by cross-reactivity, similarly well protecting repertoires in different individuals may exist, even if they sample essentially the same pathogenic environment (Figure \ref{fig:sharing}B). As a result, we should not be surprised by great differences at the phenotypic level, let alone at the sequence level, when looking at repertoires responding to specific threats. In other words, even if convergent selection is at play, each immune repertoire may find a different molecular solution to it, which may explain the observation that most repertoire sharing seem to occur by chance. 

\subsection{Repertoire response to an immune challenge}~\label{sec:rep_response}
RepSeq technologies allow one to track changes in the repertoire in time and in response to environmental changes, infections, or vaccinations. In many animal models such as mice or fish, it is generally not possible to sample the repertoire in the same individuals. Instead, the repertoires of distinct but isogenic individuals are sampled at the different times before and after an immune challenge. By construction, this strategy restricts the analysis to public features of the repertoire response. To identify common sequences features of repertoires that responded to a particular challenge, machine learning techniques based on low-level features \cite{Thomas2014b} or support vector machines \cite{Cinelli2017} have been used in mice to classify responding repertoires. Enrichment of particular V classes and sequences features have also been reported in the BCR repertoires of trout following immunization \cite{Castro2013,Magadan2018}.

In humans, the BCR response to influenza vaccination \cite{Vollmers2013,Laserson2014}, BCR and TCR response to varicella-zoster vaccination \cite{Wang2015,Qi2016}, and the TCR response to yellow fever vaccination (a model for an acute infection) \cite{DeWitt2015,Pogorelyy2018} have been studied at the repertoire level using RepSeq. These studies are mostly descriptive, but they pave the way towards a better characterization of the specificity, reproducibility, and publicness of the immune response at the repertoire level. A common observation is that the response is mostly private, although responding clones are more shared than would expected by chance. Responding sequences also tend to be clustered in sequence space \cite{Pogorelyy2018}. This clustering gives a criterion for identifying responding sequences in a single repertoire snapshot, based on the density of similar sequences (differing by at most one amino acid) in the repertoire, relative to this density expected by the recombination model, $\sum_{s',|s-s'|\leq 1}P_{\rm gen}(s')$ \cite{Pogorelyy2018b}.

\section{Lymphocyte population dynamics}\label{dynamics}

T and B cells are organized in clones of cells that express the same immune receptor. These clones grow and decay as individual cells divide or die either spontaneously or in response to external signals --- e.g. an antigen recognition event triggering cell proliferation, or cell-cell communication through messenger molecules called cytokines. The correct way to approach and to model these complex lymphocyte dynamics, as well as their implication for experimental observables such as clone size distributions or clone expansion, is still a largely open question. There is a vast body of literature on modeling lymphocyte dynamics for broad subpopulations with no clonal information, using models ordinary differential equations. Ref.~\cite{Perelson2002} provides a useful entry point into that literature.
 Here we will focus on stochastic models population dynamics, where the identities and sizes of individual lymphocyte clones are tracked. This approach is in line with recent developments in high-throughput repertoire sequencing (Sec.~\ref{repertoires}), which allows for such a fine-grain description of population dynamics, and for which much work remains to be done, both experimentally and theoretically. While progress on this topic is still in its infancy, it promises to give us better insight into the collective decision-making and dynamics of populations of immune cells, and to extract important {\em in vivo} parameters from data, at a moment when most studies are based on cultured cells or mouse models.

We also briefly review the application of differential equation models to HIV dynamics, in what constitutes perhaps the most successful application of modeling to translational immunology.

\subsection{Neutral dynamics}\label{neutralclonesize}

Let us fist consider a simple stochastic model of repertoire evolution in the absence of any antigenic proliferation. Such a model is relevant for the naive pool, which is often believed to be unaffected by external stimuli (although this is also debated, see below), but can also serve as a ``neutral'' baseline against which to compare other dynamics or longitudinal data. While some of the results discussed in this section were presented in \cite{Desponds2017}, here we expand on them and add a few original results, notably on the establishment of the steady state, and on estimates of the total number of clones.

In this simple model, new clones of size $C$
come out of the thymus (for T cells) or bone marrow (for B cells) with rate $\theta_C$, so that the total cell output $\theta=\sum_C C\theta_C$.
Then each cell can divide with rate $\nu$, and die with rate $\mu>\nu$.

The total number of cells $C_{\rm tot}(t)$ follows the simple differential equation:
\beq\label{eq:Ctotdyn}
\frac{dC_{\rm tot}}{dt}=\theta - (\mu-\nu)C.
\eeq
This simple equation is at the basis of many studies aiming at quantifying lymphocyte population dynamics, using experiments with isotope labeling \cite{Borghans2007,DeBoer2013} or other markers of cell divisions \cite{Bains2009a}.
When thymic output, division and death rates are constant, Eq.~\ref{eq:Ctotdyn} is solved by (assuming no cell at $t=0$):
\beq\label{eq:totalcells}
C_{\rm tot}(t)=\frac{\theta}{\mu-\nu}\left(1-e^{-t(\mu-\nu)}\right).
\eeq
At steady state, the total number of lymphocytes $C_{\rm tot}=\theta/(\mu-\nu)$ reflects the balance between thymic output, cell division and death. It is believed that the division and death rates ajust themselves so as to keep $C_{\rm tot}$ constant, a process called homeostasis.
For T cells, $\theta\sim 10^{6}$/day for mice, and $\theta\sim 10^{8}$/day for humans, although that number varies with age. Taking $C_{\rm tot}\sim 10^{11}$, this gives an effective decay time of $(\mu-\nu)^{-1}\sim 1,000$ days, which is consistent with lifetime estimates of T cells in human using deuterium water \cite{DeBoer2013}.

However, these numbers are not informative about the repertoire structure and its clone size distribution. These calculations are also unable to disentangle division and death, lumped in the single parameter $\mu-\nu$.
The expected number of clones of size $C$, $N_C(t)$, is governed by the following equations \cite{Desponds2016}:
\beq\label{eq:neutralthymic}
\begin{split}
\frac{dN_C}{dt}=&\nu((C-1)N_{C-1}-CN_C)\\
&+\mu((C+1)N_{C+1}-CN_C)+\theta_C
\end{split}
\eeq
At steady state, $dN_C/dt=0$, the solution for $C>\max\{C:\theta_C>0\}$ is a power-law of exponent 1 with an exponential cutoff:
\beq\label{eq:powerlawneutral}
N_C \sim \frac{1}{C}{\left(\frac{\nu}{\mu}\right)}^C.
\eeq
When birth and death are balanced, $\nu\sim\mu$, the exponential cutoff disappears. Comparing to the power law Eq.~\ref{eq:powerlawfrequencydistribution} in the distribution of frequencies $\rho(p)$ with the correspondance $N_{C}=N_{\rm tot}\rho(p=C/C_{\rm tot})$, this model would predict a exponent $\hat\beta=0$, which is not supported by repertoire data on unsorted or memory cells. This suggests that the dynamics of the memory repertoire cannot be simply neutral as modeled by Eq.~\ref{eq:neutralthymic}, but must be affected by selection, such as expansions events following antigen recognition, as we will see in the next section.

The dynamics of Eq.~\ref{eq:neutralthymic} can in fact be solved analytically using generating functions and the method of characteristics \cite{Anandprivate}. Assuming for simplicity that clones have initial size 1, $\theta_C=\theta \delta_{C,1}$, we obtain (see Sec.~\ref{neutralthymicderivation}):
\beq
N_C(t)=\frac{\theta}{\nu}\frac{1}{C}{\left(\frac{\nu}{\mu}\right)}^C{\left(\frac{1-e^{-t(\mu-\nu)}}{1-(\nu/\mu)e^{-t(\mu-\nu)}}\right)}^C.
\eeq
This relation is still a power law with an exponential cut-off. The power law exponent remains 1 at all times, but the cut-off gets larger and larger as the system reaches steady state.
The total number of clones, $N_{\rm tot}(t)$, then evolves according to:
\beq
N_{\rm tot}(t)=\frac{\theta}{\nu}\ln\left(\frac{\mu-\nu e^{-t(\mu-\nu)}}{\mu-\nu}\right),
\eeq
and the total number of cells, $C_{\rm tot}(t)=\sum_C CN_C(t)$ follows Eq.~\ref{eq:totalcells}. At steady-state, the total number of clones, $N_{\rm tot}=(\theta/\nu)\ln[\mu/(\mu-\nu)]$ depends on both the division and death rates, and not just their difference. In the limit of small division rate, $\nu\ll \mu$, which is often assumed for naive cells, we get $N_{\rm tot}=\theta/\mu\approx C_{\rm tot}$. Conversely, in the limit $\mu-\nu\ll \mu$, one has $N_{\rm tot}\sim C_{\rm tot} \epsilon\ln(1/\epsilon)$ with $\epsilon=(\mu-\nu)/\mu$, meaning that the number of clones may be arbitrarily smaller than the number of cells. Starting with initial clone sizes of $k_0$ does not substantially affect this picture, with $N_{\rm tot}\sim (H_{k_0}/k_0)C_{\rm tot}$ (where $H_n=\sum_{i=1}^n1/i$) in the limit $\nu\ll \mu$, and with unchanged scaling $N_{\rm tot}\sim \epsilon\ln(1/\epsilon)C_{\rm tot}$ in the limit $\mu-\nu\ll \mu$ (see Sec.~\ref{neutralthymicderivation}).

These models can be refined by accounting for convergent recombination, by which clones sizes can also be increased by thymic exports that have the exact same sequence, which happens with probability $P_{\rm gen}(s)$ for a particular sequence $s$. With this correction, Zheng and collaborators \cite{Zheng2017} found fair agreement between the results of a neutral model with a source Eq.~\ref{eq:neutralthymic} and the bulk distribution of clone frequencies observed in the naive repertoire of mice. However, they also found many outliers, {\em i.e.} large naive clones that cannot be explained by the neutral assumption.
A major difficulty of such a comparison is that, as we already mentioned in Sec.~\ref{diversity}, the clone frequency distribution of the full repertoire is heavy-tailed. Any small contamination of non-naive cells into the studied repertoire is likely to introduce spurious outliers in the distribution, confounding the analysis.

\subsection{A note about ``neutral processes''}\label{neutral_discussion}
In this review we discuss four different types of processes that are referred to as neutral in different domains: (i) the neutral model of population genetics (see section~\ref{SFS}), (ii) the neutral clone size distribution model (see section~\ref{neutralclonesize}), (iii) the Yule-Simon process of speciation (see section~\ref{Yule}) and (iv) the neutral ecological model (see section~\ref{GenLV}). Since these models are often confused with one another, sometimes simply for semantic reasons, otherwise because many of them give power law distributions for some observable, we thought it is pedagogical to discuss how they are different and summarize the exponents they give and what kind of distributions they predict. 

Many of these models predict a power law in the distribution of clone or species frequencies, but details differ. Table \ref{tableneutral} gives a summary of these distributions and differences between the model assumptions.

\begin{table}\label{tableneutral}
\begin{tabular}{c|c|c|c}
Neutral model & $N_C$ & Comment & Eq. \\
\hline
\hline
Lymphocyte dynamics & $\sim (\mu/\nu)^C/C$ & Source &Eq.~\ref{eq:powerlawneutral}\\
\hline
Population genetics & $\sim 1/C$ & Fixed pop. size & Eq.~\ref{eq:neutralSFS} \\
\hline
Yule process & $\sim 1/C^{1+\rho}$ & Expanding pop. & Eq.~\ref{eq:yule}
\end{tabular}
\caption{Summary of different neutral models discussed in the review, with their clone size distributions. $\nu<\mu$ refer to the division and death rates respectively, and $\rho=1/(1-\alpha)$, where $\alpha$ the probability of mutation upon division in the Yule process.}
\end{table}

Neutral models of ecology come in different flavours \cite{Hubbell2001}, but the simplest one is essentially equivalent to Eq.~\ref{eq:neutralthymic}, and the corresponding clone size distribution, called the species abundance distribution in that context, has long been known as Fisher's logseries \cite{Fisher1943}, althouth Fisher derived it through purely statistical means using unjustified assumptions.

The neutral model of population genetics is similar to the neutral model of Eq.~\ref{eq:powerlawneutral}, with a major difference: the population size $C_{\rm tot}$ is fixed. Most often, novel types arise from mutations arising in existing types. In population genetics the ``site-frequency spectrum'' corresponds to the distribution of allele frequencies, and is thus the equivalent of the clone size distribution $N_C$. In the neutral model, the site-frequency spectrum is a straight power law of exponent $-1$. The difference with the result of Eq.~\ref{eq:powerlawneutral}, beyond the absence of an exponential cutoff, is that mutants that reach the (fixed) population size $C_{\rm tot}$ become the wildtype, and are thus removed from the dynamics of mutants, while in lymphocyte dynamics no single clone ever takes over the whole population.

The Yule process is a model of an ever expanding population with mutations.
It assumes that individuals of all types divide with rate $\mu$, with a mutation probability $\alpha$ that generates new types. The crucial difference with Eq.~\ref{eq:powerlawneutral}, beyond the absence of death, is that the rate of novely increases linearly with the population size. Also, the population expands exponentially, and the clone size distribution only reaches a quasi steady state, after renormalizing the total population size, $C_{\rm tot}$. The renormalized size distribution asymptotically follows a power law of exponent $-2$ in the limit $\alpha\ll 1$. Although this exponent is tantalizing close to those observed in data, the assumption that new types arise with a rate proportional to the population size is incompatible with the biology of T cells, as new clones originate from the thymus, and not from already circulating cells. On the other hand, the Yule process may be appropriate for modeling population-wide affinity maturation of B cells, as their receptors hypermutate upon expansion.

\subsection{Population dynamics model with external signals}

To explain the power laws observed in the clone size distribution from sequenced repertoires, one needs to go beyond neutral models and introduce external signals affecting the division and death rates --- or fitness --- of cells as a function of their phenotypic state or of their immune receptor.
Several models of clonal dynamics have been proposed to describe the growth and decline of clones in various populations (see section~\ref{GenLV} for a general discussion of competition models). We will not describe all of these models, but try and give the reader an idea of their general features. The key ingredient is that cells divide in response to antigenic stimulation, which depends on the concentration or frequency of available antigens that are susceptible to be recognized. Cells in a clone expressing receptor $r$ die with constant rate $\mu$ as in Eq.~\ref{eq:neutralthymic}, but they divide with clone-specific rate
\beq
\nu_r\left(\{Q_a\},\{C_{r'}\}\right)=\nu \sum_a Q_a f_{r,a}A\left(\sum_{r'} C_{r'} f_{r',a}\right),
\eeq
where $f_{r,a}$ is the cross-reactivity function between an antigen and a receptor, $Q_a$ the frequency of antigen $a$, and $A(x)$ is a decreasing function describing the availability of a given antigen as it is being bound by other receptors than $r$; for example, $A(x)=(1+\epsilon)/(1+\epsilon x)^{2}$ with $\epsilon$ setting the strength of competition. In this model, receptors interact only indirectly through this competition factor.

As a historical note, models with {\em direct} interactions between receptors, namely antibodies binding to each other, were once proposed as a way to generate interesting dynamics on so-called idiotypic networks \cite{Jerne1974}. While idiotypic networks were often used to explain phenomena in the 1980's, and became very popular with physicists because of their link to network theory and the physics of disordered systems \cite{Perelson1997}, it is not currently considered a dominant paradigm in immunology (and already was not by the time Ref.~\cite{Perelson1997} was written, as discussed therein), as its predictions can be explained by clonal selection theory alone without invoking additional elements.

In the linear noise approximation, the size of the clone of cells expressing receptor $r$ is described by a stochastic differential equation:
\beq\label{eq:competition}
\frac{d C_r}{dt} =  \left[ \nu_r\left(\{Q_a\},\{C_{r'}\}\right) -\mu \right]C_r +\xi_r(t),
\eeq
with $\av{\xi_i(t) \xi_i(t')}=\left(\nu_r\left(\{Q_a\},\{C_{r'}\}\right) +\mu \right)C_r \delta(t-t')$ with It\^o's convention. This general form of the model can describe a wide variety of situations. It can be used as a model for the naive T cell repertoire, where $Q_a$ is the distribution of self-antigens for which T cells compete for survival cues. Such models of competitive exclusion \cite{DeBoer1994} have been studied and used to explain biological facts for both T- \cite{DeBoer1997} and B cells \cite{DeBoer2001} repertoires, such as the fact that a larger diversity is beneficial in terms of overall repertoire fitness. Interestingly, the same dynamical equation emerges as a way to approach an optimal repertoire \cite{Mayer2015a}, as defined in Sec.~\ref{optimal}.

Eq.~\ref{eq:competition} can also be used to model lymphocyte population dynamics in response to transient antigenic stimulation. Antigens appear with some rate $s$, and decay as they are being cleared by the immune system as $Q_a(t)=Q^0_ae^{-\lambda (t-t_a)}$, where $t_a$ is the time at which antigen $a$ appeared in the population. In this context, the antigen is not limiting, and is fully available, $A(x)=1$. When $\lambda$ is very large compared to $\mu$ and $\nu$, each clone experiences a series of short spikes of expansion of magnitude $\exp(\nu \<Q^0\> /\lambda)$ with rate $sp$, where $p$ is the probability that $f_{ar}=1$ for a random choice of $a$ and $r$ (and $f_{ar}=0$ with probability $1-p$, in an all-or-nothing approximation).
With the change of variable $x=\ln C$, each expansion event causes an average jump $\Delta x=\nu \<Q^0\> /\lambda$.

In Ref.~\cite{Desponds2016}, expansion events were assumed to be both small and frequent compared to other time scales, which led to an effective diffusion equation for $x$.
Here, as an original result of this review, we present the solution to the general process. A more detailed solution using Laplace transforms is presented in Sec.~\ref{jump}).
The density of clones with logarithmic size $x$, $\rho(x,t)=(dC/dx)N_{C=e^x}(t)=e^xN_{e^x}(t)$, follows the simple jump process with constant negative drift:
\beq\label{eq:jump}
\frac{\partial \rho}{\partial t}=\mu \frac{\partial \rho}{\partial x}+sp \left[\rho(x-\Delta x)-\rho(x)\right]+ \tilde\theta(x).
\eeq
where $\tilde\theta(x)=e^x\theta(C=e^x)$ is the density rate of new clones entering the population with initial logarithmic size $x$. Assuming a power law Ansatz for the clone size distribution,
\beq\label{eq:powerlawfluctuation}
N_C\propto \frac{1}{C^{1+\alpha}}, 
\eeq
translates into exponential decay in $x$, $\rho(x)=\rho_0e^{-\alpha x}$, giving the consistency equation for $\alpha$:
\beq\label{eq:alpha}
\mu\alpha+sp\left[1-\exp(\alpha\Delta x)\right]=0.
\eeq
{\rev The total numbers of cells and clones can be calculated at steady state (see Eqs.~\ref{Ntotsel},\ref{Ctotsel} in Sec.~\ref{jump}), and read in the case of fixed introduction size $\theta_C=\theta \delta_{C,C_0}$: 
\beq\label{CtotNtotsel}
C_{\rm tot}=\frac{\theta \ln(C_0)}{sp\Delta x-\mu},\quad
N_{\rm tot}=\frac{\theta (C_0-1)}{sp(e^{\Delta x}-1)-\mu}.
\eeq
}

If we assume many small expansion events, $\Delta x\ll 1$, expanding the exponential at second order allows us to recover te power-law exponent of \cite{Desponds2016}:
\beq\label{eq:alphajonathan}
\alpha=\frac{\mu- sp \Delta x}{sp\Delta x^2/2},
\eeq
where the numerator is the net decay rate of the clone size obtained from the difference between death and expansion, while the denominator corresponds to an effective diffusion coefficient stemming from the random arrival of expansion events.

The result of Eq.~\ref{eq:powerlawfluctuation} suggests that a wide range of models with clone-dependent, random expansion or selection events produce power laws with arbitray exponents, in agrement with the data on memory or unfractioned lymphocyte repertoires. To distinguish between different types of dynamics, more details of the dynamics of individual infection events, as well as of the kinetics of unstimulated clones should be studied in terms of their impact on the clone size distribution.

The hypothesis that the expansion factor $\Delta x$ is constant, or at least does not depend on the clone size, is questionable. For instance, in secondary immune responses, the expanded cells already have a memory phenotype, and these cells tend to expand less than naive cells upon a primary infection. In addition, the amount of inflammatory signals that promote expansion should in general depend on the number of cells involved in the response. From a design perspective, it would seem beneficial not to expand clones that are already big, as the organism is already well protected against the pathogens that these clones are specific to. The optimal expansion strategy upon each pathogen encounter can be calculated within the simplified framework described in Sec.~\ref{optimal}. Using Eq.~\ref{eq:cost} with a logarithmic cost $c(x)=-\ln(x)$ and a uniquely specific Kernel $f_{ar}=\delta_{a,r}$, combined with Bayesian prediction theory, the optimal dynamics can be shown to approximately follow \cite{Mayer2018}:
\beq
\frac{dC_r}{dt}=\chi \sum_{i\in E_r}\delta(t-t_{i})-\tau_m^{-1}(C_r-\chi \theta_a),
\eeq
where $\tau_m=2\tau/(\chi^{-1}C_{\rm tot}-1)$ (with $\tau$ the effective timescale of the pathogen dynamics), $E_r$ is the set of infection events $i$ occuring at time $t_i$ in response to which lymphocytes carrying receptor $r$ expand, and $\chi$ is a scaling parameter setting the total number of lympthocytes. The main difference between these and previously considered dynamics is that the growth of $C_r$ is not exponential, meaning that regulation or homeostasis mechanisms must exist to tune the magnitude of the expansion as a function of the size of the antigen-specific repertoire subset.

So far we have assumed that selection manifests itself at the level of clones: all cells in the same clone has the same average growth rate.
However, selection could be cell dependent, which is the case if cells respond differently to non-antigenic stimulation signals, such as cytokines. Growing experimental evidence shows the heterogenous response to cytokines, due to differential expression of cytokine receptors, signalling molecules and their diffusion. 
Models with a cell-dependent fitness do not give power law behaviour, but do produce long tails \cite{Desponds2016}. With the current experimental cut-offs, it may not be possible to discriminate between clone-specific and cell-specific fitness models using repertoire data only. Further studies with longitudinal tracking of clone sizes may help settle these questions.

\subsection{In host HIV dynamics.}\label{HIV_dynamics}

Perelson and collaborators~\cite{Perelson2002, PerelsonRibeiro} considered a modified SIS model of an HIV infection. This is one of the most influential examples of how computational models influenced medicine. It was used to predict the effects of anti-viral drugs on HIV: reverse transcriptase (RT) which blocks the ability of HIV to infect a cell, and protease inhibitors (PI) that result in the production of non-infectious viruses. Target cells $T$, that correspond mainly to CD4+ T-cells, become infected ($I$) at rate $ \beta$, are born with rate $\alpha$ and die with rate $\mu$. Infected cells die at rate $\nu$, but they also help the virus ($V$) reproduce and produce $p$ new virions that are cleared with rate $c$.
\beq\label{PerelsonSIS}
\frac{d T}{d t} =\alpha-\mu T - \beta I S \\ \nonumber,
\frac{d I}{d t} = \beta V T - \nu I \\ \nonumber,
\frac{d V}{d t} =p I- cV.
\eeq
 To study the effect of an antiviral drug, the virus equations are modified to account for two viral species: non-infectious viruses ($V_{\rm NI}$) and infectious viruses ($V_{\rm I}$):
\beq
\frac{d V_{\rm I}}{d t} =(1-\epsilon_{\rm PI})p I- cV_{\rm I}, \\ \nonumber
\frac{d V_{\rm NI}}{d t} =\epsilon_{\rm PI}p I- cV_{\rm NI},
\eeq
where $\beta = 1-\epsilon_{\rm RT}$ and $0<\epsilon_{\rm PI}<1$ and $0<\epsilon_{\rm RT}<1$ are efficacies of RT and PI drugs. The total viral populations is held constant $V_{\rm NI}+V_{\rm I}=V=\rm{const}$. For $\epsilon_{\rm PI}=1$ and $\epsilon_{\rm RT}=1$, assuming $T={\rm const}$, the total viral load will decay according to~\cite{PerelsonRibeiro, Perelson2002}:
\begin{eqnarray}
V(t)&=&V(t=0) e^{-ct} +\frac{cV(t=0)}{c-\nu} \times \\ \nonumber
&&\left[ \frac{cV(t=0)}{c-\nu} \{ e^{-\nu t}-e^{-c t} \} -\nu t e^{-c t} \right].
\end{eqnarray}
This decay was fit to data from an HIV-infected patient on anti-viral therapy by adjusting the  $c$ and $\nu$ parameters using least-squares regression, showing very good agreement~\cite{Perelson1996}. Explaining viral dynamics in the case of combination therapy (tri-cocktails) requires introducing long-lived infected cells that act as a secondary viral source~\cite{Perelson1997a}. Fitting these more complex, yet still extremely simple models to data, allowed Perelson and collaborators to estimate the half-lives of the different species in Eq.~\ref{PerelsonSIS}. This in turn made it possible to effectively administer a tri-therapy cocktail treatment in intervals that made it very hard for the virus to escape. This extremely simple calculation saved lives.

\section{Affinity maturation}\label{affinity_maturation}

Upon antigenic stimulation, B-cell receptors acquire somatic hypermutations (SHM) that can help them explore the binding landscape. Based on high-throughput repertoire sequencing data, combinations of clustering and tree building methods~\cite{Mccoy2015, Yaari2015} have been proposed to characterize lineages. Since affinity maturation is a fascinating example of a Darwinian evolutionary process, a lot of effort is going into extracting the details of the evolutionary processes from data, along with more theoretical efforts. We very briefly review tree building, clustering and lineage reconstruction approaches that are available in different software in section~\ref{lineage_reconstruction}. A recent review more thoroughly sumarizes these attempts~\cite{Hoehn2016}, arguing for methods departing from the traditional  assumptions of population genetics, and tailor-made for the hypersomatic mutation scenario. BCR repertoire sequencing provides large amounts of data that allow software packages like Immcantation~\cite{Gupta2015, VanderHeiden2014, Yaari2015, Yaari2015a, Cui2016, Yaari2013}, Partis~\cite{Ralph2016} (see section~\ref{lineage_reconstruction}) SPURF \cite{Dhar2018} or IGoR~\cite{Marcou2018, Mccoy2015} to learn hypermutaiton models (as described below),  and understand the evolutionary process.  As mentioned in section~\ref{lineage_reconstruction} the problem of lineage reconstruction is quite general to many areas in immunology. Incorporating the observation that  more abundant clones are likely to have more offspring allows for reconstructing B-cell lineage ancestries from germinal center imaging experiments~\cite{DeWitt2018d}. 

\begin{figure}
\includegraphics[width=\linewidth]{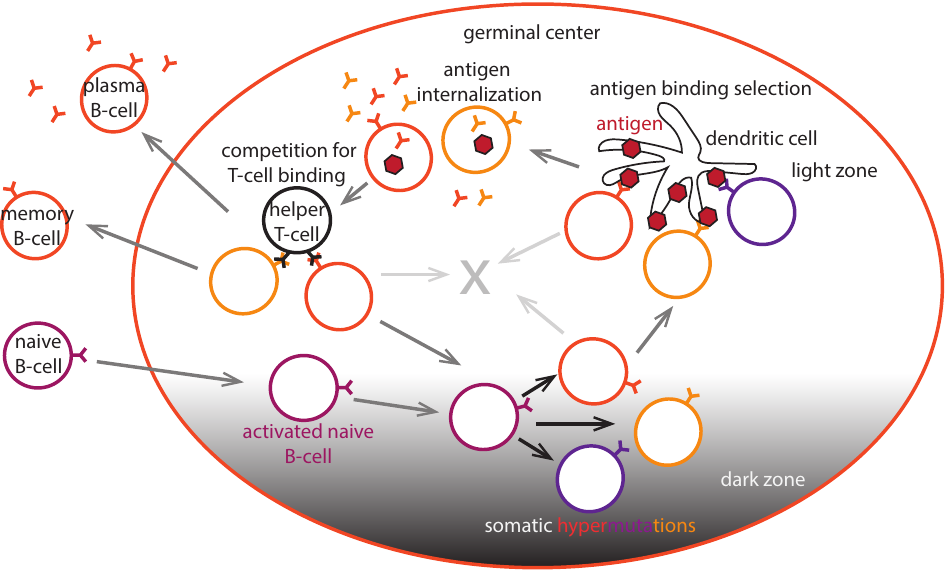}
\caption{\textbf{Affinity maturation in germinal centers}.  Naive B-cells get recruited into germinal centers where they become activated and acquire somatic hypermutations upon proliferation in the dark zone. This produces cells with different BCRs. If the BCR manages to get expressed on the cell surface, it moves to the light zone to undergo selection for binding to antigen presented on follicular dendritic cells (FDC). Cells that bind strongly internalize the antigen and present them to helper T-cells. They undergo a second selection step where they compete to helper T-cell signals. The sucessful cells can undergo another round of affinity maturation by returning into the dark zone to somatically hypermutate, or they can leave the germinal center as a memory B-cell or an antibody secreting plasma cell.}
\label{fig:affinity_maturation}
\end{figure}

Affinity maturation is a sped-up Darwinian evolution process. Upon recognizing an antigen even with low affinity~\cite{Jacob1991}, naive B-cells migrate to germinal centers (GC) in lymph nodes and about 3 days after antigen injection the germinal center reaction starts~\cite{Nieuwenhuis1984}, initially dividing without hypermutations (see section~\ref{cyton}) for about 7 days and reaching a population size of $\sim 1500$ cells~\cite{Jacob1991}. About a week after the infection starts, they first enter the dark zone region of the germinal center, where the Activation-induced cytidine deaminase (AID) enzyme introduces hypermutations as the cells continue to divide. 

\subsection{Modeling the hypermutation process}

Repertoire sequencing analysis of unselected mutations~\cite{Yaari2015, Elhanati2015,Marcou2018} has shown that AID acts highly non-uniformly, with hypermutation hotspots and strong context dependent motifs. The difficulty in learning these models stems from eliminating selection that acted on memory sequences. A 5-mer model, called S5F, was learned from synonymous mutations (that are in principle free of selection) from long heavy chain 
 sequences~\cite{Yaari2015}. Only half of possible 5-mers were observed from data with synonymous mutations, the  remaining ones were inferred by averaging over related observed pentamers. Alternatively, out-of-frame memory sequences were used to learn heavy V  and J segment di- and tri-nucleotide context dependent hypermutation models from low throughput sequencing experiments~\cite{shapiro1999predicting}.

The S5F model learns the probability $P_{\rm mut}(\vec\pi)$ of mutating the central position of a 5-mer motif $\vec\pi=(\pi_{-2},\ldots,\pi_2)$ by counting occurences of synonymous mutations in this motif centered at position $i$ along the sequences $\vec\sigma^{j}$ of a large dataset. Because only synonymous mutations must be taken into account, {\em ad hoc} heuristic normalization rules were used. In practice, mutation rates are given by:
 \beq
P_{\rm mut}(\vec\pi) = \mu \frac{m_{\vec\pi}}{\sum_{\vec\pi'}m_{\vec\pi'}}
\eeq
with $\mu$ an adjustable baseline mutation rate (not learned within S5F), and
\beq
m_{\vec\pi} = \frac{1}{\sum_{j,\pi'}C^j_{\vec\pi'}}\sum_j \frac{C^j_{\vec\pi}/B^j_{\vec\pi}}{\sum_{\vec\pi'}C^j_{\vec\pi'}/B^j_{\vec\pi'}}
\sum_{\vec\pi'}C^j_{\vec\pi'}
\eeq
is the expected number of mutations of each 5-mer in a sequence. This number is obtained by normalizing for being synonymous in each sequence, and then averaging over sequences weighted by their number of synonymous mutations.
In that expression, $C^j_{\vec\pi}$ is the observed number of synonymous mutations in sequence $\vec\sigma^j$ in motif $\vec\pi$, and $B^j_{\vec\pi}$ is a background normalization:
\bea
C^j_{\vec\pi}=\sum_i I_{\vec\tau^j}(i,\vec\pi,\sigma^j_i), \quad B^j_{\vec\pi}=\sum_{i,b} s^{\vec\pi}_b I_{\vec\tau^j}(i,\vec\pi,b)
\eea
where: $\vec\tau^j$ is the germline (unmutated) ancestor of $\vec\sigma^j$;
$I_{\vec\tau^j}(i,\vec\pi,b)$ is an indicator function that is equal to 1
if $\vec\tau^j$ matches $\pi$ over the position range $(i-2,\ldots,i+2)$, and $\tau^j_i\to b$ is a synonymous mutation; finally $s^{\vec\pi}_b=N^{\vec\pi}_b/\sum_{b'\neq \pi_0}N^{\vec\pi}_b$ and $s^{\vec\pi}_{\pi_0}=0$, where $N^{\vec\pi}_b$ is the total number of synonymous mutations to $b$ observed in 5-mer motifs $\vec\pi$ across all sequences in the training dataset.

The S5F model provide a profile of hypermutation hot and cold spots in the absence of selection, which is widely used for analyzing BCR data. In particular, it can used as a baseline for quantifying selection from synonymous mutations \cite{Uduman01072011,Yaari2013a}. 
Note that a 5-mer model was also learned from VH and JH genes sequences in rearrangements engineered not to be productive, in both the heavy and light chains of mouse, without having to use the trick of using synonymous mutations~\cite{Cui2016}.

To avoid for over-fitting, one can further assume that the impact of the motif on mutability is additive. This strategy was applied to non-productive heavy-chain sequences in humans~\cite{Marcou2018}. Mutability of $(2m+1)$-mer motifs $\vec\pi$ is given by:
 \beq
 P_{\rm mut} = \frac{\mu \exp(\sum_{i=-m}^m e(\pi_i))}{1+\mu \exp(\sum_{i=-m}^m e(\pi_i)) },
 \eeq
where $(\pi_{-m},...,\pi_{m} )$ is the $2m+1$-mer sequence context around the mutation $i$, and $e(\pi_i)$ are the inferred elements of the Position-Weight Matrix (PWM), learned within the IGoR framework \cite{Marcou2018} (see Sec.~\ref{pgen}) using the Expectation-Maximization algorithm (see Sec.~\ref{inference}).

The actual hypermutation process is yet more complicated than captured by all these models. It has been pointed out that because of the context dependence of mutations, the order in which they appear matters. Since considering all possible histories is computationally not feasible, Feng et al proposed to Gibbs sample the orderings to learn a hypermutation model that was validated on simulated data~\cite{Feng2017}. More importantly, hypermutations are not independent of each other, and tend to cluster along the sequence \cite{Marcou2018}. Hypermutations result from a lesion of DNA followed by error-prone DNA repair over extended regions ($\sim$ 20 nucleotide) of the sequence \cite{Unniraman2007}. These events may induce several simultaneous mutations at close-by positions along the sequence. It also suggests that context can influence the mutation rate over fairly large distances, since the erroneous repair can occur far from the original damage causing the lesion. Finally, context models cannot explain all of the variance of hypermutation rates, suggesting that other factors, such as length dependence, may be at play.

\subsection{Cycles of selection}

After acquiring hypermutations B-cells pass to the light zone of the germinal center where they undergo the selection step of Darwinian evolution. In this step a B-cell receptor must bind the antigen presented on a follicular dendritic cell (FDC) strongly enough for the B-cell to internalize the antigen and present them to helper T-cells. B-cells compete for binding of their pMHC to these helper T-cells that give them an essential survival signal. Therefore each B-cell passes a two-fold selection step: selection for recognizing an antigen, and selection for competitive binding to a helper T-cell. Binding to T-cells is specific and depends on the peptide presented on the MHC. B-cells that have a higher affinity for the antigen, have a larger internalization rate and a larger probability to succesfuly solicit T-cell help. Helper T-cell stimulation is essential for survival. One unverified hypothesis is that helper T-cells that have previously been trained in the thymus against self-proteins, now discriminate against hypermuations that lead to self-reactive B-cells. B-cells that do not receive a survival signal die.

Theoretical work~\cite{oprea1997, Oprea2000} that was later verified by experiments has shown~\cite{Victora:2012gx} show that a single cycle of passing through the light and dark zones is not enough to generate the numbers of observed high affinity B-cells. This leads to a hypothesis called recycling where $\sim 90 \%$ of B-cells go back to the dark zone. At the end of the affinity maturation process, B-cells increase their affinity to the antigen by $\sim 1000$ fold~\cite{Berek:1987tf,Oprea2000} (or even $\sim 10000$ in rabbits~\cite{Eisen1964}) and the B-cell population is always oligoclonal~\cite{Tas2016}, with a large diversity of different B-cell receptors ~\cite{Kuraoka2016}. Interestingly, a germinal center is binary: either it produces many high-affinity cells, or it fails to produce high-affinity B-cells~\cite{Victora2012}. 

The details of the heterogeneity of GC have been discovered through so-called {\it brainbow} experiments~\cite{Victora:2010kd,Tas2016}, in which individual B-cells were permanently tagged during BCR acquisition with one of 10 possible combinations of four different colored fluorescent proteins. This technique combined with imaging allowed for multicolor fate mapping of a cell's progeny since all cells that come from the same ancestral cell have the same color.  By injecting cells in germinal centers with different dyes and tracking their movement in mice, Victora et al~\cite{Tas2016} have shown that in one lymphnode there are many germinal centers active at the same time, producing different diversities of cells. Most GCs probed at both 6 and 15 days showed a lot of different colored clusters, with lower estimates of $\sim 50$ clones per GC, going up to hundreds. By delaying tag formation until the cells are in the GC,  selection in GCs was shown to keep multicolored clusters, showing that diversity of lineages is always maintained even if a certain dominance of one specific lineage (typically less than $40\%$ of all cells belong to the dominant lineage) is observed at later times. Notably the heterogeneity was huge: some GC had one lineage making up to $80 \%$ of cells, others as little as $\sim 20\%$) two weeks after the infection. The rate of diversity loss was also very heterogenous: some GCs converged to a single lineage within a few days, and others took over two weeks, or never converged. Even in adoptive transfer experiments (see section~\ref{cellfate}), where high affinity clones were introduced and the mouse was challenged with either bacteria or virus antigens, the GCs showed great heterogeneity regardless of the type of antigen - from nearly monoclonal to ones with an effectively neutral distribution of lineage diversity. The GCs expressing one lineage can be traced back to one single ancestral BCR (with possible SHM) that expands in a clonal {\it burst} over a short period of time, leading to a loss of diversity. Interestingly, the affinity of BCRs coming from GCs with varying levels of lineage diversity is similar.
The oligoclonality of antibody repertoires is also confirmed by RepSeq experiments that show that antibodies responding to a specific antigen found at physiologically relevant concentrations can have between $3-147$ distinct CDR3s~\cite{Lee2016}.

\subsection{Evolution of broadly neutralizing antibodies}
Many pathogens such as viruses (influenza, VIH) come in a variety of strains. Typically, antibodies mature to recognize just one type of strain, by targeting easily accessible epitopes on the surface of its proteins. However, viruses can often easily escape immunity afforded by these antibodies by mutating these epitopes, with little fitness cost. Epitopes that are more conserved (i.e. in which mutations carry a significant fitness cost to the virus) would be better targets, but they are usually harder to access by antibodies, as viruses have evolved to hide those conserved regions.
To mature antibodies that bind strongly to more than one strain (called broadly neutralizing antibodies --- BnAbs), the immune system needs to be trained with different antigens from different strains. In the context of a vaccination strategy, a crucial question is in what antigens to present, and in what order. 

To address this question in the context of HIV, Wang et al~\cite{Wang:2014vb} simulated different temporal immunization schemes, by considering three antigens (wildtype HIV antigen and 2 mutants) and modelling their interaction with antibodies using a BCR-antigen binding model such as described in section~\ref{rec_antigen_spec}. The virus was modeled as consisting of a conserved part, in which mutations are deleterious, and a variable part.
The three considered immunization schedules were: all three variants together (scheme 1), WT with mutant 1 at the same time, followed by mutant 2 (scheme 2), WT followed by mutant 1 followed by mutant 2 (scheme 3).
Running these simulations many times, they looked at the antibodies that come out of these in silico germinal centers and scaned them against a standard panel of antigens that are used in affinity maturation experiments.
The breadth of each antibody is defined by how many different test antigens it recognizes. Scheme 1 does not produce any broadly neutralizing antibodies (bnAbs). In fact, it hardly produces any germinal centers that have any output as B cells die quickly during the process.
The frustration of not being able to discriminate between conserved and non-conserved residues results to a situation where the antibodies are not really selected since a BCR is likely seeing very different antigen in each round of selection. 
 Scheme 2 does produce bnAbs, but with very low probability. This is likely what happens during a normal infection because you get infected with a single strain that later diversifies.
In this case, the affinity maturation process in the first round does produce binders to the conserved residues, breaking the frustration.
Finally, scheme 3 produces bnAbs with high probability ($\sim 69\%$). The sequential application allows the evolving BCR to focus on the only non-moving part of the target --- the conserved residues. These predictions have been tested experimentally, showing that mice that were sequentially immunized focus their immune response.

There are effectively two parameters that govern the B-cell germinal center distribution: the probability of surviving selection $P_s$ and the probability of mutation $\mu$.  With probability $P_s(1-\mu)$ the existing B-cells will expand, with probability $P_s\mu$ it will mutate, and with probability $1-P_s$ is will die. Wang et al~\cite{Wang:2014vb} suggest this system is optimally frustrated and the probability of survival puts the system in a special regime: too stringent selection (small $P_s$  will kill all cells), while too lenient selection (large $P_s$) will fill the germinal centers with existing cells, since $\mu$ is small. So only intermediate levels of $P_s$ lead to mutant cells surviving, allowing for many rounds of selection that results in acquiring many mutations that give large breadth. Support for this intermediate level selection comes from simulations that show that intermediate antigen numbers result in the largest number of mutations in a BCR. However experimental validation is still lacking.

Repeated antigen exposure helps increase the affinity of cells through selection, as in Darwinian evolution. In addition, new naive cells are recruited, adding novelty in the selection process other than through mutation. Murugan et al~\cite{Murugan2018} compared the impact of selecting naive clones to that of producing new mutants by SHM. Looking at the response of BCR to the malaria parasite in humans coupled with string models of affinity maturaion, they showed that antigens with low complexity are much more efficient in generating good binders by means of SHM affinity maturation compared to high complexity antigens, which rely on recruitment of new naive cells.
Both routes (SHM and selecting naive cells) where observed experimentally. Influenza vaccine studies show that, upon secondary immunization (booster vaccines), $\sim 7 \%$ of BCR contained no SHM~\cite{Neu2019} (although these experiments showed strong clonal dominance with $\sim 10\%$ of clones representing $\sim 90\%$ of sequences~\cite{Lee2016}).
The model of Murugan et al~\cite{Murugan2018} predicts that long exposure to small amounts of antigen leads to selecting for antibodies with a lot of SHM, whereas short exposure to high antigen concentrations results in selection on existing naive cell variation, which is confirmed in experiments.
These results, which hold for relatively short timescales ($\sim$ few months), are consistent with the argument of Wang et al~\cite{Wang:2014vb} in the case of chronic infections.
Interestingly, both Murugan et al~\cite{Murugan2018} and Neu et al~\cite{Neu2019} experimentally found that the majority of SHM do not necessarily improve antigen binding, and that both routes can lead to equally high affinity mutants, which supports the heterogeneity observed by Tas et al~\cite{Tas2016} in the brainbow experiments. Yet, it is still not clear why some individuals fail to produce any good binding antibodies to the very strong antigenic challenge of malaria.

Additionally, the BCR repertoire changes with age, with both positive and negative selection acting differently in older people~\cite{Ohare2018}, producing longer CDR3s that are more promiscious and likely to bind self-proteins. The rate of observed BCRs with SHM also changes from infancy to adulthood~\cite{Wendel2017}. In influenza vaccine studies~\cite{Neu2019}, people born before 1977 show less  inter clonal diversity but achieve the same affinity as people born after 1977. Adaptation via the SHM route is less likely in older people, and they are more likely to use their larger existing memory pools than younger people whose antibodies acquire a lot of SHM. As a result, the antibodies of the older group target a conserved part (the stalk) of the influenza protein as do BnAbs, whereas younger people's antibodies target the head (aiming for a strain-specific antibody). Younger people can produce BnAbs when hit with a strong challenge (vaccine combined with a pandemic), as predicted by the analysis of Wang et al~\cite{Wang:2014vb}, but the response to subsequent infections is still specific. In summary, these results highlight the complexity of BCR maturation, which involves both naive cells (with no SHM) and highly mutated cells. The details of the response depend on the dose concentration and timing, past infections, age and antigen complexity.

\subsection{Population genetics approaches to affinity maturation}

\subsubsection{Evolutionary analysis of repertoire dynamics}

Recent RepSeq data was used to quantify the evolutionary regime of in-host HIV evolution~\cite{Nourmohammad2018} and in response to influenza vaccines~\cite{Horns2019}. The flu vaccine is a model of an acute infection, which occurs on short timescales. BCRs from 9 timepoint blood samples, before and after the infection over the course of $\sim 2$ weeks, resulted in very skewed lineage trees and a U-shaped site-frequency spectrum (SFS). SFS corresponds to the distribution of frequencies of mutations in a population, and its shape is often informative about the underlying evolutionary process (see section~\ref{SFS}). U-shaped SFS, as observed in BCR lineages after vaccination, are characteristic of strong selection. However, 3 out 5 of the strongly selected BCRs recognized neither the flu vaccine epitope nor the full virus, suggesting bystander evolution, by which non-responsive clonotypes expand in response to a multitude of signals. This could be a mechanism for upkeeping memory clones between infections with the same or similar pathogens. 

Skewed trees and U-shaped SFS were also observed in the BCR lineages of HIV-infected indivuals~\cite{Nourmohammad2018}.
However, the same observations were made in out-of-frame sequences of healthy individuals. This is not surprising because the repertoires of healthy people are under constant selection, and out-of-frame sequences evolve in cells with functional receptors, and their evolution reflects the hitchhiking and selection on the whole cell. However, a finer analysis  exploiting the dynamic trends of synonymous versus non-synonymous mutations is able to distinguish selection due to the chronic HIV infection from overall selection patterns. More interestingly, it finds that the CDR3 region of the BCR of untreated HIV carriers evolves according to a regime known as ``clonal interference'' (see section~\ref{clonal_interference}). In that regime,
several new beneficial mutations arise at similar times and compete with each other. This competition slows down adaptation, as only one the beneficial mutations can survive.
This analysis is based on estimating the probability that a (beneficial) mutation first rises to a threshold frequency $x$, but then is driven to extinction by a competing mutation, $H(x,x_i)=G(0|x)G(x|x_i)$, where $G(x|y)$, called a propagator is the probability of ever reaching $x$ starting at $y$, and $x_i$ is the initial frequency of the mutant. Both $H(x,x_i)$ and $G(x|x_i)$ can be estimated from the data and for various models of evolution. Neutral models cannot explain empirical observation, and neither does a simple model of selection with a fitness advantage. Instead, data can be fit by a model of varying selection  (see section ~\ref{varrying_sel_Moran}), where the fitness advantage fluctuates as competing mutants come and go.

Analysis of phylogentic trees build from RepSeq data from HIV patients combined with S5F hypermutation models~\cite{Yaari2013} also led to identifying the constraints on BCR adaptation~\cite{Vieira2018}. Ancestral sequences in lineages are more likely to mutate their CDR3s than the framework (FWR). The propensity of a given residue to mutate was decreased more in framework (FWR) regions than in CDR3s, but in both cases a decrease in mutability was much more likely than an increase. Although most of the constraints on the residue to mutate in CDR3 that are under strong positive selection to increase their binding affinity, come from nonsynonymous mutations, in all regions up to $21\%$ of loss in residue mutability was caused by synonymous mutations, which are the result of neutral evolution. These results also point to a slow down in adaptation from clonal interference, as expected in the clonal interference regime.

\subsubsection{Models of co-evolution of phenotypic traits}

The fitness function for the co-evolution between HIV and BCR depends on the ability of the BCRs to recognize the virus, which is  a (possibly nonlinear) function of the binding affinity. By focusing on a single antibody lineage and a viral population, Nourmohammad et al~\cite{Nourmohammad2016} derived an effective stochastic equation for the rescaled mean binding energy $\epsilon$ between antibodies and viral epitopes:
\begin{eqnarray}
\frac{d \epsilon}{dt} &=  &- 2 \left[\theta_A +\theta_V (N_A/N_V) \right] \epsilon \\ \nonumber
& &+ s_A \sigma_{A}-s_V \sigma_{V} + \sqrt{\sigma_{A} +N_A/N_V \sigma_{V}} \xi_{\epsilon},
\end{eqnarray}
where $\sigma_i$ are the variances of the binding energy across the viral and antibody populations, $\theta_i$, $s_i$ and $N_i$ are the mutation, selection coefficients and population sizes of  antibodies ($i=A$) and viruses ($i=V$); $\xi_{\epsilon}$ is Gaussian noise and times is measured in units of antibody coalescence time.
This equation was derived by assuming a linear relationship between binding energy and fitness, and using a additive string model of binding energy as a function of genotype (see section~\ref{receptor_antigen_models}). The approximate method to get from the evolutionary dynamics, which are defined on the genotype, to an effective equation for the phenotype (binding energy), is described in a simpler context in ~\ref{quantitative_traits}.

Intuitively, in general mutations on both antibodies and viruses reduce recognition. On the other hand, antibody diversity ($\sigma_A$) increases binding energy by selecting the best binders, while virus diversity ($\sigma_V$) decreases it by selecting the best escapers on the viral side of co-evolution.
The same formalism can be applied to the evolution of BnAbs, which target epitopes on the conserved regions of the virus. In that context, the equations simplify to $\theta_V=0$ and $\sigma_V=0$, because the viral epitope is constant.

\begin{figure}
\includegraphics[width=\linewidth]{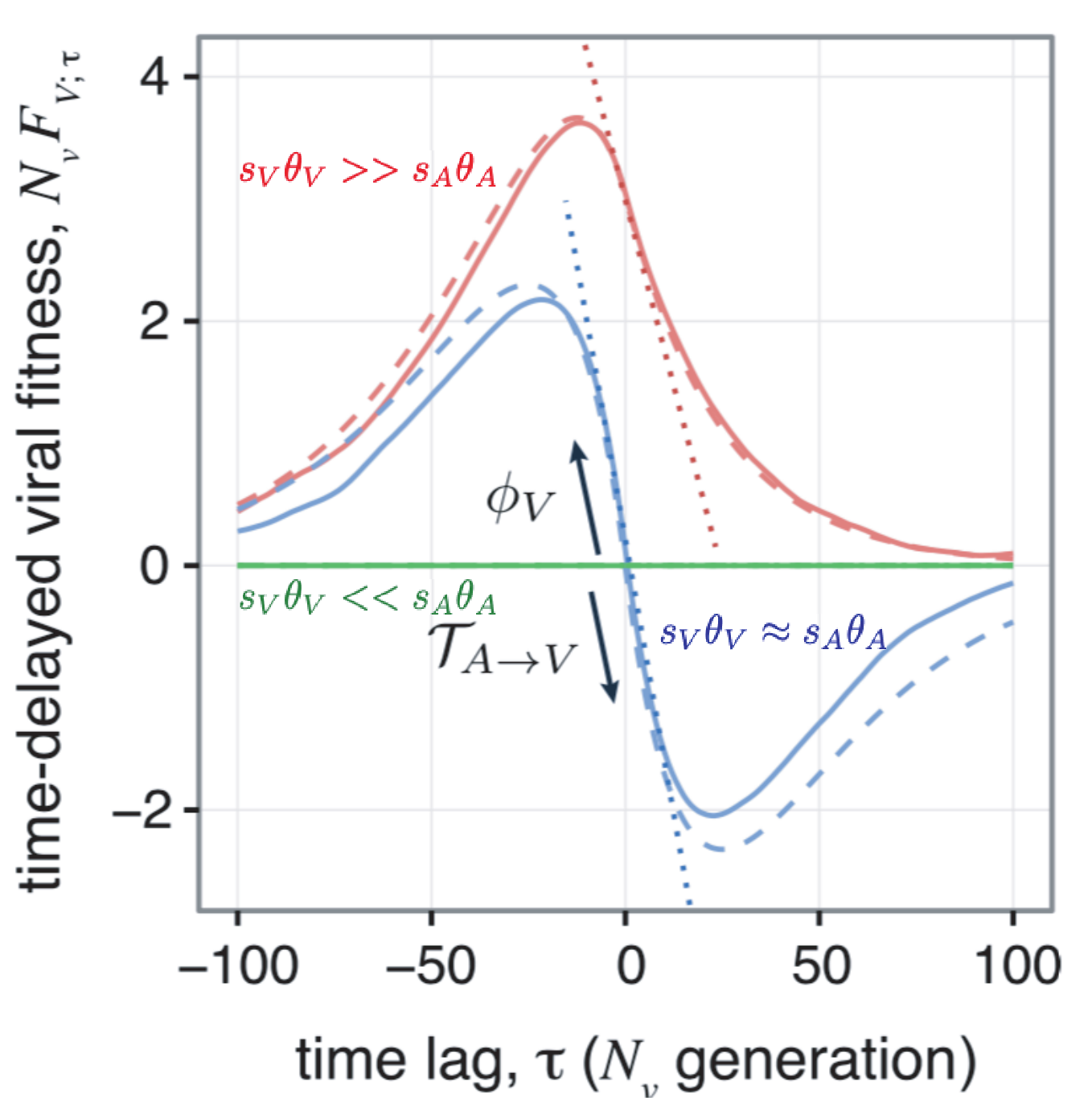}
\caption{\textbf{Time delayed viral fitness}. Adapted from Nourmohammad et al~\cite{Nourmohammad2016}. Viruses have a positve fitness against antibodies from the past, and a negative fitness again antibodies from the future. The slope of the current viral fitness coming from past  antibodies is equal to the fitness flux $\phi_V(t)$ -- a measure of adaptation to existing antibody populations, and the slope  of the current viral fitness coming from future  antibodies is equal to the transfer flux ${\cal T}_{A\rightarrow V}(t)$ that describes the pressure on the viral population from the adaptating antibody population. The blue line describes a co-adapting population, $s_A \theta_A \approx s_V \theta_V$, the red line describes a virus evolving effectively without antibody selection pressure, $s_V \theta_V>>s_A \theta_A$ and the green line is a regime of much stronger antibody adaptation,  $s_V \theta_<<s_A \theta_A$. }
\label{Armita_PloSGenFig4}
\end{figure}

However, tracking the binding energy is not alone a good signature of co-evolution.
Time-shifted statistics, such as the time delayed viral fitness (see Fig.~\ref{Armita_PloSGenFig4}), can be used to identify strongly co-evolving populations. The time delayed viral fitness is defined by the interaction between the viral population at time $t$, with distribution $y^{\gamma}(t)$ of genotypes $\gamma$, and a population of antibodies $\alpha$ taken at a later time $t+\tau$ distributed according to $x^{\alpha}(t+\tau)$: $F_{V,\tau} (t) \sim - s_V(t) \sum_{\alpha\gamma} E_{\alpha\gamma} y^{\gamma} (t) x^{\alpha(t+\tau)}$, where $E_{\alpha\, \gamma}$ is the binding energy from the string model between antibody  $\alpha$ and viral epitope $\gamma$. Note that same-time fitness $F_{V,0}$ is equivalent to $-\epsilon$ up to a selectivity constant.

If the selective effects on the phenotype are comparable in the two populations, $s_A \theta_A \approx s_V \theta_V$, the current virus has highest fitness against antibodies from the near past since it has acquired mutations to escape them, and smaller fitness against antibodies from the future, which have caught up with these escape mutations (Fig.~\ref{Armita_PloSGenFig4}). For long times, mutations randomize  the genome and $F_{\tau} (t \rightarrow \infty)=0$. If the selection pressure of viruses is larger than that of the antibody $s_V \theta_V \gg s_A \theta_A$, the virus is effectively evolving against a neutral antibody population. It still has a high fitness against previously seen antibodies, and genome randomization decreases its fitness advantage against future antibodies, but without a penalty.

The  left slope of the time-shifted viral fitness quantifies the adaptation of the viral population to the existing antibody population and defines the a ``fitness flux'' $\partial_{\tau} F_{V,\tau}(t-\tau)|_{\tau=0-}= \phi_V(t)=\sum_{\gamma} \partial_{y^{\gamma}} F_V (t) (dy^{\gamma}(t)/dt)$. The right slope  measures the pressure on the viral population from the adapting antibody population, which defines the transfer flux of fitness from antibodies to viruses $\partial_{\tau} F_{V,\tau}|_{\tau=0+}= {\cal T}_{A\rightarrow V}(t)=\sum_{\alpha} \partial_{x^{\alpha}} F_V (t) (dx^{\alpha}(t))/dt)$. At stationarity both fluxes sum to zero, and the derivative is continuous, as in Fig.~\ref{Armita_PloSGenFig4}.

The characteristic S-curve shown  for $s_A \theta_A \approx s_V \theta_V$ is indicative of two competing populations. Neutralization measurements~\cite{Blanquart2013, Richman2003, Frost2005,Moore2009} show that viruses are more resistant to past antibodies and more susceptible to future antibodies, which results in the S-curved  time delayed viral fitness~\cite{Nourmohammad2016}. Interestingly, Blanquart and Gandon~\cite{Blanquart2013} also showed that antibodies that evolved in one HIV positive patient are better at targeting the virus found in another person than this person's current antibodies. This result remains unexplained.

While this analysis only involves a single antibody lineage, in germinal centers multiple lineages can compete with each other.
Extending their analysis based on the dynamics of quantitative traits, Nourmohammad et al~\cite{Nourmohammad2016} considered the competition of different lineages during affinity maturation. The change in the frequency of a given antibody lineage (each composed of many genotypes $\alpha$) of size $N_A^C$ in the population $\rho^c=\sum_{\alpha\in C} x^\alpha$ is driven by its mean fitness $F_A^C$ compared to the mean fitness of antibody lineages $F_A=\sum_C F_A^C  \rho^C$:
\beq
\frac{d \rho^C}{dt} = \left( F_A^C- F_A\right) \rho^C +\sqrt{\frac{\rho^C(1-\rho^C)}{N_A}} \xi_C,
\eeq
where $\xi_C$ is a Gaussian white noise. The lineage fitness is an average of the genotype  $A^{\alpha}$ fitnesses $f^{\alpha}_C(t)$ that make up the lineage C, weighted by their frequencies in that lineage $x^{\alpha}_C (t)$, $ F_A^C=\sum_{\alpha} f^{\alpha}_C(t)x^{\alpha}_C (t)$. The mean fitness of a lineage depends on the genotypes within this lineage and also on the frequency of virus $y^{\gamma}$ through $f^{\alpha}_C(t)$.

The probability of fixation $P^C_{\rm fix}$ of lineage $C$ is not only a function of the mean fitness advantage and population size, as in Eq.~\ref{probfix} (Sec.~\ref{extinctionprob}), but also of the ability of the population within the linage to adapt. Its value is given by:
\begin{eqnarray}\label{LINEAGE_FLUX}
P^C_{\rm fix}/P^0_{\rm fix}&\approx&1+\av{N_A \left( F_A^C(t=0)- F_A(t=0)\right)}+\\ \nonumber
&&\frac{N_A^2}{3} \av{\phi_A^C(t=0)-\phi_A(t=0)}\\ \nonumber
&&-N_A N_V \av{\abs{{\cal T}^C_{V\rightarrow A}(t=0)}- \abs{{\cal T}_{V\rightarrow A}(t=0)}},
\end{eqnarray}
where $P^0_{\rm fix}=\rho^C(t=0)$ is the neutral fixation probability, and where $\phi^C_A$ and ${\cal T}^C_{V\to A}$ are the antibody counterpart of $\phi_V$ and ${\cal T}_{A\to V}$ for a specific lineage, and $\phi_A$ and ${\cal T}_{V\to A}$ their average over lineages. The first corrective term is standard and comes from the mean fitness advantage. The second corrective term accounts for the adaptation of each lineage in response to selection of antibodies for viral recognition within lineages. The third corrective term corresponds to the adaptation or escape capability of the viral population and its effect of the lineage.

For typical antibodies targeting a variable region of the virus, a broad viral diversity is detrimental as it allows the viral population to escape immunity (large ${\cal T}^C_{V\to A}$. For BnAbs, which bind to constant regions of the virus, this effect disappears.
In fact the effect is the opposite: higher diversity  increases the probability of BnAb fixation compared to a non-BnAbs antibody, whereas at low viral diversity the probability of fixation of both a non-BnAb and BnAbs is the same. This argument was proposed to explain why BnAbs show up later in the infection when the viral diversity is large.

\section{Population dynamics of pathogens and hosts}~\label{popleveldyn}

\subsection{Viral fitness models}

The immune system of a population of hosts defines a fitness landscape $F$ in which viruses evolve.
Models have been developed to describe how this immune pressure shapes the evolution of both flu~\cite{Luksza2014} and HIV~\cite{Ferguson2013a, Shekhar2013, Barton2016} at the population level. Viruses are usually taken to evolve according to traditional population genetics models.

This formalism can be used to draw short term predictions about the future fate of existing strains, e.g. of circulating influenza over the course of a year.
Viral populations are subject to stochastic Wright-Fisher dynamics, but once they reach large enough numbers they grow exponentially following the fitness function (see section~\ref{popgen:WF}). In that deterministic stage, the size of a given viral strain $\alpha$, $X_{\alpha}(t)$ grows (or decreases) according to the fitness it experiences:
\beq\label{fitnessgrowthflu}
X_{\alpha} (t+\Delta t) = X_{\alpha} (t) \exp(F_{\alpha} \Delta t).
\eeq

Alternatively, one can consider the longer-term prevalance of particular strains of very diverse viruses such as HIV across many hosts, and use that to infer the fitness landscape of the virus. To do this, we can assume that the population reaches an ``equilibrium'' as the result of many mutations and fixation events across time in the entire population of viruses and hosts. The
probability of finding a given viral strain is then assumed to follow Boltzmann's law:
\beq
p_{\alpha}\approx \exp(\beta F_\alpha),
\eeq
where $\beta$ plays the role of an inverse temperature setting the tension between entropy from mutations, or genetic drift, and the fitness advantage.
One possibility is to make use of the equilibrium assumption and learn the fitness Hamiltonian directly from existing viral data. This has been done for the Gag envelope protein of HIV~\cite{Ferguson2013a} using maximum entropy approaches (see section~\ref{inference}), such that the probability of seeing a given HIV strain $\alpha$ defined by its amino acid sequence $\vec{\sigma}$, is $p_\alpha=p(\vec{\sigma})=Z^{-1}\exp(\sum h_i \sigma_i + \sum_{i,j} J_{i,j} \sigma_i \sigma_j)$, where $h_i$ define single-site fields that constrain the probability of observing the wildtype amino acid at a given site $i$, and $J_{ij}$ are interaction terms between amino acids (assuming a binary representation of the virus with $ \sigma_i=1$ denoting the consensus amino acid and $ \sigma_i=0$ a mutation at time $i$).
Such models have been inferred from multiple sequence alignment of the protein, and then used to evolve {\it in silico} viral proteins in a population of host individuals~\cite{Ferguson2013a,Barton2016}, using quasi-species equations introduced by Eigen~\cite{Eigen1971,  Leuthausser1998}, to make predictions about HIV evolution.
 
Instead of learning the fitness directly from the sequence data, another strategy is to derive the fitness of each strain  by decomposing it into two components
\beq
F_{\alpha}(t) = F^{\rm ag}_{\alpha}(t)+F^{\rm stability}_\alpha,
\eeq
where the $F^{\rm ag}_{\alpha}(t)$ is due to the antigenic component of the immune pressure exerted by the on the virus, and $F^{\rm stability}_\alpha$ encodes the fact that mutations that change the ability of the protein to fold should be detrimental to fitness. The fitness due to folding depends on the free energy of folding $G$, assuming a two state thermodynamic model of folding~\cite{Phillips_textbook}:
\beq
F_\alpha^{\rm stability} \sim (1+\exp[(G_\alpha-\bar{G}/G_0)])^{-1},
\eeq
where $\bar{G}$ and $G_0$ set the energy scale and are learned from data. Similarly, the antigenic component of the  fitness is taken to be a sigmoidal function of the binding affinity $H$:
\beq
 F^{\rm ag}_{\alpha} \sim (1+\exp[(H_\alpha-\bar{H}/H_0)])^{-1},
\eeq
where again $\bar{G}$ and $G_0$ set the energy scale and are learned from data. These constants are learned from data assuming that non-epitope mutations in the virus decrease protein stability, whereas epitope mutations decrease binding affinity. This approach was applied to influenza by \L uksza and L\"assig \cite{Luksza2014}.
The antigenic component $F^{\rm ag}$ was inferred from ferret blood titers, in which sera of blood from ferrets are challenged with different viral strains and their antibody response measured. To gain precision, the growth prediction of Eq.~\ref{fitnessgrowthflu} were considered at the level of clades rather than individual strains by summing over all strains in each clade. In the next section we will see how this model could be used to predict the upcoming dominant strain of influenza.

\subsection{Co-evolution between host and pathogen populations}
The adaptive immune system of hosts also evolves under the selective pressure of the antigenic environment, by expanding immune receptors that led to successful recognition. The antigens and the immune systems of individuals in a population engage on an arms race, where one forces the evolution of the other. The interaction between the immune system and different viruses takes on different forms. While HIV strains undergo very dynamic in-host evolution~\cite{Richman2003} experiencing many mutations and strong selection accompanied by clonal interference~\cite{Zanini2015}, the influenza virus evolves more slowly, and at scale of the whole population of hosts, with few mutations in the same host. We have discussed HIV evolutionary models in section~\ref{affinity_maturation}. Here we turn to the effects of population level co-evolution on immune repertoires, characteristic of viruses such as influenza. 

This is currently an emerging field. Most of the treatments of immune systems so far have been coarse-grained and reduced to the effective selective pressure they exert on pathogens. This pressure is seen on antigenic maps, which place pathogens in a common space according to the similarity of the immune response to them (measured in the sera of ferrets).
In practice, antigenic maps are produced using dimensionality reduction algorithms to reduce the response to a two dimensional manifold. These methods have been extremely useful in tracking influenza evolution~\cite{Smith2004} and show that single point mutations can result in a completely new response (new faraway cluster), while some multiple mutations do not change the type of response (same cluster).

However, the link between the molecular interaction between immune receptor (both BCR and TCR, as introduced in section~\ref{receptor_antigen_models}), and the these phenotypic maps remains to be explored. 
An outstanding question is what evolutionary constraints these molecular details impose. For example, is the order of mutations important, and are all the mutations independent? Deep mutational scanning experimental techniques exist now to both find the best binding proteins in multiple rounds of selection experiments, and to map out the spectrum of possible solutions.

Recently researchers have very successfully predicted short-term flu evolution~\cite{Luksza2014, Neher2014}. There are two successful types of models in this class. The first uses the diversity of existing strains and extrapolates the tree branches that have recently expanded to predict the dominant strains in the near future~\cite{Neher2014}. This approach makes use of the coalescence framework described in section~\ref{coalescence} and since it relies mainly on evolutionary properties, can easily be extended to other globally evolving viruses. The main idea is to statistically infer the characteristics of the evolutionary process from the existing influenza trees, with no reference to the immune system.
The second~\cite{Luksza2014} is based on molecular information about the influenza antigen~\cite{Smith2004, Bao2008} and identifying successful mutations. In practice these are stochastic models that assess the probability of future strains. This method incorporates an effective treatment of the immune system, based on the response of ferret serum to flu strains, as explained in the previous section. In practice this treatment ranks influenza strains by the strength of serum response, but has no information about the molecular basis of the response, potential overlap for similar strains and its evolution. It then builds a fitness model, where the future frequency of a given viral strain depends on its structural stability and ability to escape the immune system. The details of these interaction models are learned from data using advanced statistical inference techniques.  The second method relies much more on biophysical details and is therefore in principle more adapted to influenza, although the approach has been extended to predict the evolution of tumorous clones \cite{Luksza2017} as a function of their immunogenicity score (see Sec.~\ref{sec:immunogeneticity}).
Despite their differences both approaches are similar in style: they rely on recent evolutionary traces (from the last couple of months) to predict the dominant strain up to a year in advance. From the evolutionary perspective both models encode the idea of clonal interference between viral strains (see section~\ref{clonal_interference}).

Another theoretical approach to the viral-immune co-evolution problem is based on Susceptible-Infected-Recovered (SIR) Models. These models have been used for a long time to look at epidemic spreading within populations~\cite{Grenfell2008, Perelson2002}. Many of these approaches are based on simulating sets of nonlinear equations. Originally these models were used to study slowly evolving viruses, such as measles~\cite{Grenfell2008}. However recently, they have been tied with viral fitness models that account for the evolution of the virus. We already discussed one type of these models, for in-host evolution of HIV in section~\ref{dynamics}. Influenza, unlike HIV that evolves within the host organism, evolves mainly at the level of a population. Recent models \cite{Rouzine2018,Yan2018} of influenza evolution combine traditional SIR approaches with  data-derived knowledge about flu strain evolution. Specifically, in Yan et al.~\cite{Yan2018}, the model is of the form given by Eqs.~\ref{SIR_simple1}-\ref{SIR_simple3} for each antigenic strain $a$, with the Susceptible equation solved explicitly, $S_\alpha \approx \exp({-\sum_{\alpha'} K_{\alpha\alpha'} R_{\alpha'}})$, where $S_\alpha$ and $R_\alpha$ are the fractions of individuals who are susceptible and recovered from (and thus immune to) strain $\alpha$, respectively.
$K_{\alpha \alpha'}=e^{-\abs{\alpha-\alpha'}/d} $ defines a cross-reactivity Kernel of range $d$, where distances correspond to the number of mutations in an infinite-genome model.
This equation means that individual who are outside of the cross-reactivity range of the recovered individuals are susceptible to be infected by a given strain.

Within this setup, a mutation in the virus introduces a new strain $\alpha'$ that increases its distance to all existing strains by one. In a perturbative limit, this mutation increases fitness by a fixed and constant amount $s\propto d^{-1}$. This situation can be mapped onto a fitness wave model where many beneficial mutations compete against each other~\cite{Desai2007a}  (see section~\ref{clonal_interference}). In the fitness wave description, the viral population has a distribution of fitnesses and reproduces according to that fitness. The dynamics are dominated by events happening at the stochastic nose of the distribution, i.e. strains with the the largest fitness $f_m$.
The beneficial mutations that govern the future fate of the population all occur at that stochastic nose, with rate $r\sim f_m/\ln(f_m/\mu)$, where $\mu$ is the mutation rate.

The cumulative effect of these stochastic beneficial mutations at the nose can be described by an effective Langevin equation:
\beq
\frac{df_{\rm m}}{dt} = \frac{sf_m}{\ln(f_m/\mu)}- I_{\rm tot} +s\xi(t),
\eeq
where $\xi(t)$ is a Gaussian white noise of amplitude $\av{\xi(t) \xi(t')}= r \delta(t-t')$, and where the negative fitness term $-I_{\rm tot}$ comes from the population of immune systems catching up with the viral strains in the bulk of the distribution, decreasing the fitnesses of all existing strains. This Langevin description makes it possible to estimate the rate of ``speciation'', whereby two mutant strains at the nose escape the influence of the current immune systems in different way, creating two independently evolving antigenic niches which separate once their genetic distance reaches at least $\sim d$, $r_{\rm sp}\sim r e^{-d/q}$, where $q=f_n/s$ is the typical number of beneficial mutations that strains at the fitness nose accumulate relative to the rest of the population.

These arguments can be used to explain the deep splits in flu strain topologies that have been observed in data and simulated using similar types of equations~\cite{Bedford2012,Marchi2019}. In other words, the traveling wave picture also explains how influenza can constantly escape the immune system without continuous accumulation of genetic diversity.

Rouzine and Rozhnova \cite{Rouzine2018} applied a similar mapping onto a fitness wave description, but in a one-dimensional antigenic space. They
used the predictions of their continuous traveling wave framework to estimate an antigenic mutation rate of $\mu\sim 3\cdot 10^{-5}$ per transmission event and predict the cross-immunity distance of $d\sim 15$ nucleotide substitutions that agrees well with independent estimates.

While these models give general insights and scaling laws to understand viral-immune co-evolution, they ignore the specific molecular details and mechanisms of immunity. As more data becomes available about the specific interactions between immune cells and viral strains, it may be possible to combine these approaches with data-driven models of immune repertoires and viral genomes, and to use them to make specific predictions about the fate of particular viral strains and their relation to immune repertoires of hosts.

\section{Discussion}
Despite the length of this review, there are many subjects we did not touch upon that definitely fall into the global topic of quantitative immunology. Some of these, like the whole area of transcriptomics~\cite{Regev2017} applied to immunology, are new and while the methodology is quantitative, the current experiments are only just starting to give quantitative models of the immune system that can be linked with a physical understanding (although things are moving so fast that by the time this review is published, this sentence may be obsolete). We also did not go into the methodology of certain analysis or theoretical approaches that are widely used,  such as machine learning~\cite{Mehta2018a} or stochastic gene expression and biochemical regulation~\cite{Walczak2012,Tkacik2011}, because detailed reviews for the physics audience exist for these topics. We refer the curious reader to these papers for the necessary background. Lastly, we only briefly mention some amazing experimental advances, such as imaging~\cite{Victora:2012gx} since they are currently in the process of being used to verify quantitative models. We hope we have managed to give the idea of a vibrant and multi-direction field. We also note that we made presentation choices, which were not easy, because many of the presented topics are linked to other ones. These links will surely become better explored in the coming years as solid experimental quantification and validation of theoretical models becomes the norm.

\section{Glossary}~\label{glossary}

Immunology can be painful at time for physicists because of its ``jargon''. To help with accessibility, we try not to focus on specific molecules, but sometimes we need to. So here we give a brief overview of the main players  and rules of the game for the immune system, and also other molecules that are mentioned in the text. For an introduction to immunology, we strongly recommend reading ``Your Amazing Immune System'' courtesy of the European Federation Immunological Society \url{http://www.oegai.org/oegai/2-PDF/AmazingImmuneSystem.pdf}~\cite{AmazingImm_book}. After this first glance, we recommend the short but illuminating book ``How the immune system works'' by physicist Lauren Sompayrac~\cite{Sompayrac1999}.

\begin{itemize}
\item Adaptive Immune Response -- a response of the organism to specific pathogens that changes in the lifetime of each individual. It is based on lymphocytes (T and B-cells) recognizing pathogens, proliferating and then keeping a subset of cells called memory cells that are adapted to previously encountered pathogens. Recognition involves interactions with many components of the innate immune system (e. g. antigen presenting cells, cytokines) and well as with different cells of the adaptive immune system. Adaptive immune systems first appeared in jawed vertebrates. 
\item Innate Immune Response -- the innate immune response involves non-specific types of defense, from physical and chemical barriers (e.g. skin, clotting, scratching) to cells and molecules that recognize non-specific pathogenic patterns, such as proteins of the bacterial cell envelope (LPS - lipopolysaccharides). The innate immune system is evolutionary older than the adaptive one and is found in plants, insects, funghi, as well as vertebrates.
\item Cytokines - small protein secreted by leukocytes to enforce cell-to-cell communications in the immune system. Examples of specific cytokines mentioned in the text: IL-2 is a key anti-apoptotic cytokine for T cells as well as an activation cytokine for regulatory T cells; IL-7 is an hematopoietic growth (anti-apoptosis, pro-proliferation) factor for lymphocytes.
\item Interleukins (IL): cytokines historically thought to be produced by leukocytes.
\item Leukocytes: the white blood cells of the immune system. These include lymphocytes (B, NK or T cells), granulocytes, monocytes and macrophages (see Fig.~\ref{fig:hematopoiesis}).
\item TCR: T cell receptor, the main receptor on the surface of T cells that recognizes pMHC as a ligand for T cell activation.
\item BCR: B cell receptor,  the main receptor on the surface of Bcells that recognizes membrane-bound molecules as ligands for B cell activation.
\item Antigen: biomolecules that trigger an adaptive immune responses. Antigens can be proteins that are recognized by antibodies (B-cell mediated responses) or short peptides that are loaded onto MHC (T-cell mediated responses).
\item Epitope: part or whole of the antigen that the immune receptor bind to. For B cells, part of the protein that the B cell receptor bind to. For T cells, this corresponds to the peptide loaded by the MHC.
\item APC (antigen presenting cells) -- dendritic cells and macrophages. Surveilling cells that internalize molecules and cells from tissues and present them on their MHC type II to T-cells.
\item MHC: major histocompatibility complex; its main function is to ''present'' short peptides, as antigen for T cells. It comes in three types, but two of them are more relevant for the purposes of this review: type I and type II.  Type I MHC are expressed on most cells in the body, presenting random bits of protein fragments found in that cell -- this informs the surveilling cells if the MHC presenting cell is healthy or not. Type II MHC are expressed only on specialized cells (antigen presenting cells) and carry information about the cellular environment (whether there is an infection in a given tissue).
\item CD8(+) (killer) T-cell -- a type of T-cell identified by its CD8 marker involved in interactions with MHC type I presenting cells. These T-cells trigger apoptosis (kill by forcing the cells to kill themselves) the infected cells. 
\item CD4(+) (helper) T-cell -- a type of T-cell identified by its CD4 marker involved in interactions with MHC type II presenting cells that present peptides from antigens they engulfed. These T-cells produce cytokines and help orchestrate the response of other immune cells (e.g. B-cells, killer T-cells).  
\item pMHC: a complex of a short peptide (p) and MHC that constitutes a ligand for TCR. Depending on the nature of the embedded peptide, a pMHC can be agonistic (triggering an immune response), antagonistic (extinguishing an immune response) or null. Such hierarchy often lines up with self/non-self discrimination. 
\item Somatic (hyper)mutation: genetic alteration acquired by a cell that occurs in body cells (somaplasm) and that can be passed to the progeny of the mutated cell in the course of cell division; one of the driver for antibody maturation in B cells, and for repertoire generation in T cells.
\item Germline mutation: genetic alteration that occur in the gamete-producing cells (sperm and eggs)
\item VDJ recombination -- a DNA editing process that creates T- and B-cell receptors (see section~\ref{pgen} for details).
\item Ig class -- B-cells can express different types of constant regions of their receptors coupled to the same variable region. Throughout their lifetime, they can change (albeit in a specific order) which gene they express -- this is called class switching. The expressed constant region gene determines the so-called class of the B-cell and antibody. The different genes are aligned in the immunoglobulin locus, and after the B-cell moves on the next class, previous classes get deleted. For example the $\mu$ and $\delta$ genes are first in the heavy chain locus and they lead to the expression of IgM and IgD chains, both of which are expressed on naive cells. The order of the remaining classes for the heavy locus is IgG3, IgG1, IgA1, IgG2, IgG4, IgE, IgA2. Class switching is regulated by cytokines, through regulation of gene regulation (see sections~\ref{sec:cytokine_comm} and \ref{sec:celldiff}). 
\end{itemize}

\section{Methods}\label{sec:methods}

This section is devoted to introducing or expanding the more technical details of  methods relevant to quantitative immunology. These methods and concepts may be of different kinds. Some pertain directly to the biophysics of immunology and the molecular details of binding between receptors and cognate ligands. Some may be classical tools from other fields, from statistical learning to population genetics, but which carry some conceptual similarity with approaches from statistical physics. The presented methods are almost always cited in the context of recent work in physical or quantitative immunology. While they are often not established concepts of immunology, we believe that they might play an increasing role in its future development, with more and more imports from other fields.

\subsection{Physical kinetics}
\subsubsection{Diffusion-limited reaction rate}~\label{diff_lim}

To compute the association rates of \ref{eq:association_rate} of two molecules (Ligand and Receptor), let's place ourselves in the reference frame of the receptor, so that the receptor is immobile, but the ligand diffuses with coefficient $D=D_{\rm Ligand}+D_{\rm Receptor}$. The spatial distribution of the ligand concentration $C_{\rm Ligand}(\vec{r},t)$  at steady state obeys Fick's law:
\begin{equation}
\nabla^2 C(\vec{r})=0,
\end{equation}
with boundary conditons $C(\vec{r},t) \xrightarrow{r \rightarrow \infty} C_{\rm Ligand}$ and $C(r=R)=0$, where $R=R_{\rm Receptor}+R_{\rm Ligand}$ is the sum of the receptor and ligand radii (each modeled by a sphere). Because of spherical symmetry, $C(\vec{r})=C(r)$,  we can consider only the radial part of the Laplacian in spherical coordinates 
$$
\frac{\partial}{\partial r}\left(r^2\frac{\partial C}{\partial r}\right)=0,
$$
and the steady-state solution is:
$$
C(r)=C_{\rm Ligand}\left(1-\frac{R}{r}\right).
$$
The flux of ligands $j_{\rm Ligand}$ onto the receptor sphere is
$$
j_{\rm Ligand}(r=R)=D \nabla C=\frac{D C_{\infty} R}{r^2}=\frac{D C_{\infty}}{R},
$$
and integrating over the sphere surface one can compute total flux:
\begin{eqnarray}
\Phi_{\rm collision}
&=&\oint \vec{j}_{\rm Ligand}(r=R) d\vec{S}\\ \nonumber
&=&4\pi R^2 \frac{D C_{\rm Ligand}}{R}=4\pi DC_{\rm Ligand}R.
\end{eqnarray}
The total number of collisions per unit time is $4\pi\left(R_{\rm Ligand}+R_{\rm Receptor}\right)(D_{\rm Ligand}+D_{\rm Receptor})C_{\rm Ligand}$ and the rate of diffusion-limited interactions between receptors and ligands $k_{\rm collision}=\Phi_{\rm collision}/C_{\rm ligand}$ is:
\beq
k_{\rm collision}=4\pi\left(R_{\rm Ligand}+R_{\rm Receptor}\right)\left(D_{\rm Ligand}+D_{\rm Receptor}\right).
\label{eq:association_rate2}
\eeq

\subsubsection{The rates of dissociation between two biomolecules}~\label{dissociations_rates}

The physical chemistry of dissociation of two biomolecules can be modelled in two steps: a slow one that corresponds the breakage of chemical bonds between molecules, and a fast one that corresponds to the rushing in of the solvent. Estimating the rate of dissociation $k_{\rm off}$ thus corresponds to estimating the slow step, which is truly driven by the quantum physics of bond breakage. The rate is then encapsulated by the frequency of bond vibration ($k_B T/\hbar \approx 10^{12} s^{-1}$) multiplied by the probability of of successfully dissociating the molecular pair ($\exp \left(-\Delta G_{\rm off}/(k_BT)\right)$). This bond-breakage free energy sums up the contribution  of all the bonds (electrostatic interactions, hydrophobic bonds, Van der Waals forces, hydrogen-bonds etc.) holding a molecular complex together: it could be computed from purely quantum considerations, independently of the context in which the pair is considered (solvated in the intercellular medium, embedded in the plasma membrane, in vacuum etc.). Yet, this remains a tricky proposition as small errors in estimating the bond energies will yield to exponentially-inaccurate estimates of $k_{\rm off}$.

\subsubsection{The formation of ligand-receptor pairs: equilibrium and kinetics}~\label{eq_kinetics}

Once association and dissociation rates are available, estimating the kinetics of formation of the ligand-receptor pairs is done by considering the reaction:
\beq
{\rm Ligand} + {\rm Receptor} \xrightleftharpoons[k_{\rm off}]{\,k_{\rm on}\,} {\rm Complex},
\eeq
and integrating the equation:
\begin{eqnarray}
\frac{d[{\rm Complex}]}{dt}&=&k_{\rm on}[{\rm Ligand}][{\rm Receptor}]\\ \nonumber
&&-k_{\rm off}[{\rm Complex}].
\end{eqnarray}

We have three species whose kinetics we are estimating, and two conservation laws (for the ligand and for the receptor), hence the reaction coordinate for this reaction is one-dimensional and can be solved analytically. We call the complex concentration $x=[{\rm Complex}]$"
\beq
\frac{dx}{dt}=k_{\rm on}(L_{\rm total}-x)(R_{\rm total}-x)-k_{\rm off}x,
\eeq
where $L_{\rm total}$ is the total ligand concetration, free and complexed, and $R_{\rm total}$ is the total receptor concentration. In steady state (for $t\rightarrow \infty$) we have $dx/dt=0$ and
\beq
 \frac{(L_{\rm total}-x)(R_{\rm total}-x)}{x}=\frac{k_{\rm off}}{k_{\rm on}}=K_{\rm D},
\eeq
which has two formal solutions
\begin{eqnarray}
x_{\pm}& = &\frac{1}{2}\Big(L_{\rm total}+R_{\rm total}+K_{\rm D} \\ \nonumber
&&\pm \sqrt{(L_{\rm total}+R_{\rm total}+K_{\rm D})^2-4 L_{\rm total} R_{\rm total}}\Big) 
\end{eqnarray}
of which only $x_-$ is physical and satisfies $x<L_{\rm total},R_{\rm total}$.

The kinetics of relaxation 
\beq
\frac{dx}{dt}=k_{\rm on}(x-x_-)(x-x_+),
\eeq
can also be integrated:
\beq
x(t)=\frac{x_{\rm -}-ax_+e^{-k_{\rm reaction}t}}{1-ae^{-k_{\rm reaction}t}},
\eeq
where the characteristic rate for the reaction is: 
$k_{\rm reaction}=k_{\rm on}\sqrt{(L_{\rm total}+R_{\rm total}+K_{\rm D})^2-4L_{\rm total}R_{\rm total}}$, and $a$ is a constant depending on the initial condition.

In the limit of a few receptors $R_{\rm total}\ll K_D,L_{\rm total}$, the expression simplifies to $x=L_{\rm total}R_{\rm total}/(L_{\rm total}+K_{\rm D})$ and $k_{\rm reaction}=k_{\rm on}L_{\rm tot}+k_{\rm off}$.
Three regimes can be considered:
\begin{itemize}
\item When the ligand exists in high concentration, $L \gg K_{\rm D}$ and $x=R_{\rm total}$, the system relaxes and fluctuates with a characteristic rate $k_{\rm reaction}=k_{\rm on} L_{\rm total}$.

\item When the ligand is sparse and in limited amount, $L_{\rm total} \ll K_{\rm D}$, few complexes form, $x\approx L_{\rm total} R_{\rm total}/K_{\rm D}$, and the system relaxes and fluctuates with a characteristic rate $k_{\rm off}$.

\item When $[{\rm Ligand}]\approx K_{\rm D}$, half of the receptors are occupied and half are free of ligands, $x=R_{\rm total}/2$; this is the mid-point of the dose response when one assess receptor occupancy against increasing concentrations of ligands.
\end{itemize}

\subsection{Gene regulation}
\subsubsection{Basic model}~\label{genereg}

In the simplest quantitative model of gene regulation,
we can write the deterministic dynamics of gene transcription and translation as:
\begin{eqnarray}
\frac{d}{dt}[{\rm mRNA}] & = & p_{\rm mRNA}-\gamma_{\rm mRNA}[{\rm mRNA}]\\
\nonumber
\frac{d}{dt}[{\rm protein}] & = &  p_{\rm protein}[{\rm mRNA}]-\gamma_{\rm protein}[{\rm protein}] ,
\end{eqnarray}
where $p_{i}$ are the production rates and $\gamma_i$ the degradation rates of the molecules, for $i ={\rm protein}, {\rm mRNA}$. Since  transcription and degradation rates for mRNA are often larger than translation and degradation rates for the protein, timescale separation leads to
\beq
\frac{d}{dt}[{\rm protein}] =  p_{\rm protein}\frac{p_{\rm mRNA}}{\gamma_{\rm mRNA}}-\gamma_{\rm protein}[{\rm protein}].
\eeq
At steady state:
\beq
[{\rm protein}]=\frac{p_{\rm mRNA} p_{\rm protein}}{\gamma_{\rm mRNA}\gamma_{\rm protein}}.
\eeq

This straightforward expression can become arbitrarily complicated. Both the production and degradation rates of mRNA and proteins of interest can be complex functions of the signalling response, and post-translational modifications and regulated degradation can also happen. 

In simplest case, these rates can be approximated as constants, when adiabatic conditions apply such that signalling responses and cytokine production and consumption occur on timescales much shorter or much longer then gene regulation. Including positive feedback loops produces more interesting dynamical behaviour, especially when the reinforcement is through mutltimerized transcription factors that introduce additional nonlinearities.

\subsubsection{Auto-amplification with a single transcription factor}

A simple feedback loop corresponds to the case when the rate of transcription of a protein species present in the cell at concentration ${\rm [protein]}=X$ is a dose response of the $X$ itself (e.g. when $X$ is a transcription factor that binds to the promoter region of the gene for $X$). Then the dynamic equation for $X$ is:
\beq
\frac{d[X]}{dt}=k\frac{[X]}{[X]+K}-\gamma[X],
\eeq
where $k$ is the maximum production rate, $K$ the concentration of protein $X$ at half-maximum expression, and $\gamma$ is a degradation rate (we have used the separation of time scales to eliminate the mRNA stage, as in the previous paragraph).  
This equation has one fixed point $X=0$ for ${k}/{\gamma}<0$, and
$X_{\rm equilibrium}={k}/{\gamma}-K$,  for ${k}/{\gamma}>K$. A simple stability analysis around that non trivial fixed point, taking  $[X]=X_{\rm equilibrium}+\epsilon$,
\beq
\frac{d\epsilon}{dt}=\frac{\gamma^2}{k}\left(K-\frac{k}{\gamma}\right)\epsilon,
\eeq
shows negative restoring force ($K-{k}/{\gamma}<0$), meaning that all fluctuations get quenched and the fixed point is stable.

\subsubsection{Auto-amplification with multiple transcription factors}

The complexity and relevance of the auto-amplification gene regulatory circuit becomes more relevant when the expression of gene $X$ is regulated by two transcription factors $TF_1$ and $TF_2$, i.e. both transcription factors need to bind to the promoter region of gene $X$ to elicit its transcription. We can compute the state of the promoter using classical  tools of statistical mechanics.
The promoter for gene $X$ can exist in four possible states with the related probabilities: unoccupied promoter ($p_0$), $TF_1$ only ($p_1$), $TF_2$ only ($p_2$), and both ($p_{12}$), with
\begin{equation}
p_0=\frac{1}{Z},\ p_1=\frac{[TF_1]}{K_1},\ p_2=\frac{[TF_2]}{K_2},\ p_{12}=\frac{[TF_1][TF_2]}{K'_1K'_2},
\end{equation}
with the partition function:
\beq
Z=1+\frac{[TF_1]}{K_1}+\frac{[TF_2]}{K_2}+\frac{[TF_1][TF_2]}{K'_1K'_2},
\eeq
where $[TF_1]$ and $[TF_2]$ denote concentrations of the two transcription factors and $K_1= c_0\exp(\Delta G_1/RT)$, $K_2=c_0\exp(\Delta G_2/RT)$, $K'_1=c_0\exp(\Delta G'_1/RT)$ and $K'_2=\exp(\Delta G'_2/RT)$ their equilibrium binding constants related to the free energies of binding individually (non-primed) and cooperatively (primed) to the promoter binding sites. 

When the cooperative binding of transcription factors binding to the promoter region is usually thermodynamically favored, $K'_1\ll =K_1$ and $K'_1 \ll K_2$, the partition function simplifies to $Z\approx 1+[TF_1][TF_2]/(K'_1 K'_2)$.

A common case if when $[TF_1]=[TF_2]=[TF]$ and gene $X$ is transcribed upon homodimerization of a transcription factor in its promoter region (then $K'_i=K'$). The probabilities of the promoter to be unoccupied (transcribed at a basal level for auto-activation), $p_{\rm off}=p_0$, and  to be occupied (transcribed at an enhanced  level for auto-activation), $p_{\rm on}=p_{12}$ are:
\begin{equation}
p_{\rm on}=\frac{[TF]^2}{[TF]^2+K'^2},\quad p_{\rm off}=1-p_{\rm on}
\end{equation}
This form of regulation, which follows a Hill function (with Hill coefficient $h=2$), introduces a nonlinearity in the production of $X$ that is of critical relevance in cell differentiation. 
For auto-activation by homodimers, the dynamic equation for the regulation of $X$ is:
\beq\label{eq:hill2}
\frac{d[X]}{dt}=k\frac{[X]^2}{[X]^2+K^2}-\gamma[X],
\eeq
where $k$ and $\gamma$ are production and degradation rates for $X$ respectively. 

\begin{figure}
\includegraphics[width=\linewidth]{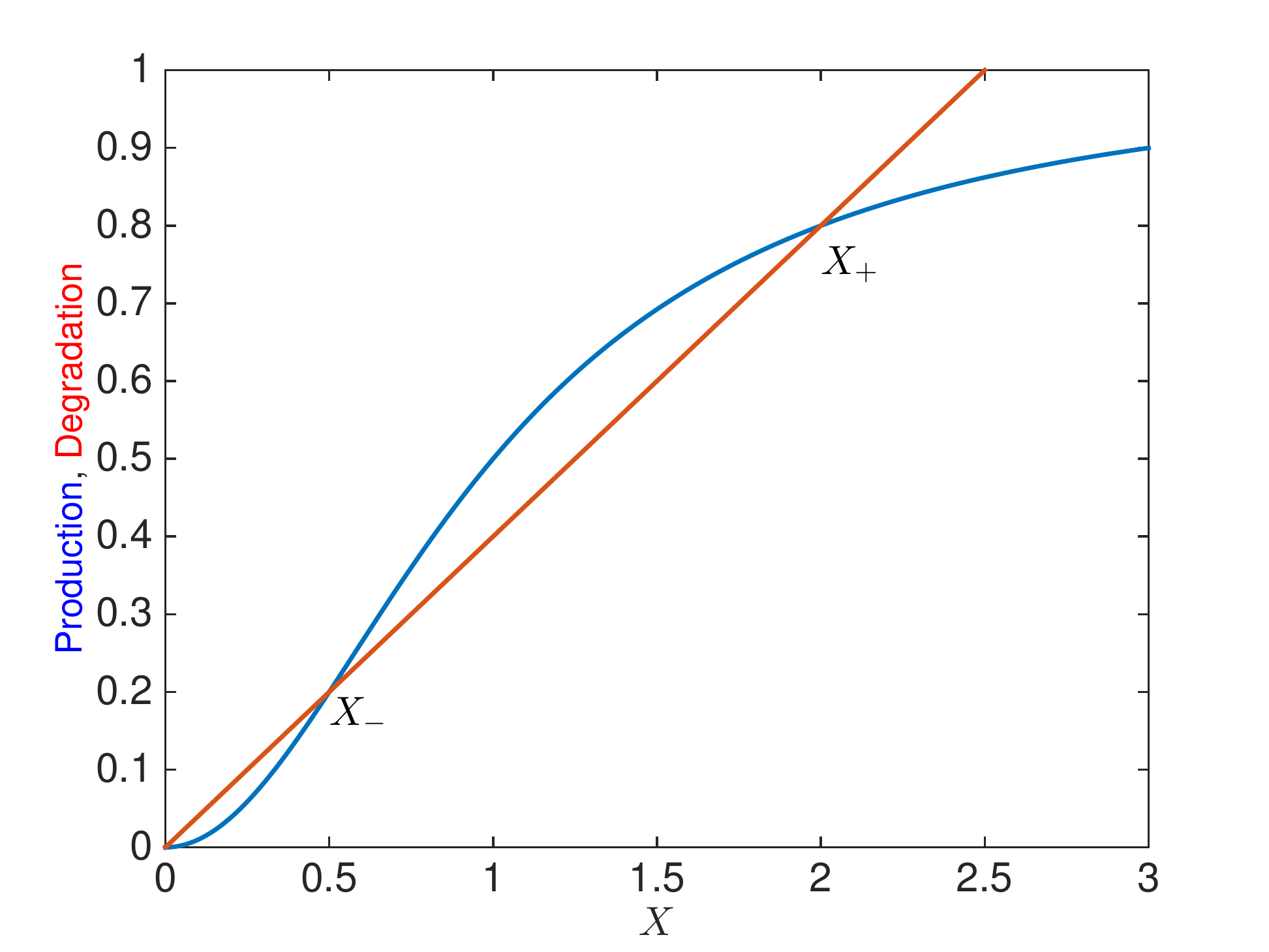}
\caption{\textbf{Bistable auto regulation of a gene.} Shown are the production and degradation rates of as a function of the protein concentration $X$, Eq.~\ref{eq:hill2}. The intersections of the curves define three fixed points: two stable ones ($0$ and $X_+$), and one unstable one ($X_-$).
}
\label{fig:hill}
\end{figure}

As in the non cooperative case, this dynamical system has a single fixed point if $k/\gamma<K$: $[X]=0$. But for $k/\gamma>K$, it has three fixed points at $0$, $X_-$ and $X_+$. The stability of these solutions can be deduced graphically (see section~\ref{sec:celldiff} for details): $X_-$ is an unstable fixed point and $0$ and $X_+$ are stable fixed points. Hence, this simple system of gene regulation will generate two types of cells: cells that do not express $X$, and cells that express a high level of $X$.  We see many examples of such bimodal distributions in expression of transcription factors, cytokines and surface markers in cells of the immune system (see section~\ref{sec:celldiff}).

\subsection{Population dynamics, genetics} 

To study the somatic evolution of immune clones and cell types it is useful to summarize some basic results from population genetics. The material in this section can be found in many evolution and population textbooks  \cite{Hartl_Genetics, Gillespie_book}. To describe the evolution of a mutating population, one typically considers a background of genetically identical individuals (or cells, in the context of lymphocyte population dynamics). Each one of these individuals can acquire a mutation with rate $\mu$, which starts a new mutant subpopulation. Each individual in both the ancestral and the mutant subpopulations can die or reproduce in each generation, and mutants carrying non-neutral amino acid substitutions have a different (higher for beneficial and lower for deleterious mutations) growth rates. Since it is more likely that a mutation will confer a disadvantage than an advantage, most mutants are deleterious. But as we shall see, even beneficial mutations are not necessarily destined to succeed and ultimately the fate of most mutants is to die. Out of the lucky few that establish subpopulations of significant frequency, some will completely take over the population --- they will {\it{fix}}. In this section we calculate the probabilities for fates of mutant subpopulations in the simplified case, compared to somatic evolution of immune repertoires, when there are no sources of new clones other than mutations. 

\subsubsection{Deterministic mutation-selection balance}\label{Ass:detmutsel}
Let us start with a situation where two subpopulations exist, the ancestral clone of size $n_2$ individuals and a mutant clone of  size $n_1$ individuals. The subpopulations grow (which accounts for both reproduction and death) with rates $\gamma_1$ and $\gamma_2$, respectively, where $g_i=r_i - f(n_1, n_2)$ is the balance of growth rate $r_i$ and a death rate $f(n_1,n_2$. Individuals also mutate from the ancestral population to the mutant one with rate $\mu_1$ and mutate back with rate $\mu_2$:
\begin{eqnarray}
\frac{d n_1}{dt}&=&\mu_2 n_2-\mu_1 n_1 +g_1(n_1, n_2) n_1 \\
\frac{d n_2}{dt}&=&\mu_1 n_1-\mu_2 n_2 +g_2(n_1, n_2) n_2.
\end{eqnarray}
We also assume, as is often the case in population genetics models that the population size is constant, $N=n_1+n_2= \rm{const}$, which sets $f(n_1, n_2) = (r_1 n_1 +r_2 n_2)/N$. The constant population size constraint means that knowing $n$ the size of both subpopulations is completely determined by the fraction of individuals in the ancestral population $x=n_1/n$, which follows:
\beq
\label{det_mut_sel}
\frac{d x}{dt}=\mu_2 -(\mu_1+\mu_2) x+sx (1-x),
\eeq
where the selection coefficient $s=r_1-r_2$ describes how much faster (or slower if $s<0$) the ancestral population grows compared to the mutant subpopulation. 

At steady state, in the absence of mutations $\mu_1=\mu_2=0$, either the ancestral subpopulation fixes, $x*=1$ for $s>0$, or the mutant one $x*=0$ for $s<0$. In the absence of selection but with mutations, the relative ratio of the mutation rates determines the fraction of each subpopulation, $x^*=\mu_1/(\mu_1+\mu_2)$. With both mutation and selection a mutation-selection balance is established (for $s<0$) at $x^*=(1/2)(1+{2\mu}/{|s|} - \sqrt{4({\mu}/{s})^2+1})$ for $\mu_1=\mu_2=\mu$. When mutations are rare compared to the fitness advantage, $|s|\gg \mu$, this balance simplifies to $x^*\approx \mu/|s|$.

\subsubsection{Genetic drift}\label{gendrift}
In order to observe a situation described by the deterministic equations in Eq.~\ref{det_mut_sel}, the mutant population needs to grow to a sizeable fraction of the population. However, every mutant appears first in only one individual and undergoes a subsequent random walk of reproduction and death, which means that the most likely fate is for it go immediately extinct. This makes the effect of small number noise coming from finite population sizes, called genetic drift in population genetics, relevant. We can explore this effect considering a population that produces only neutral mutants (meaning all mutants grow at the same rate as the ancestor), and keeping the population size fixed to $N$ individuals. If we focus on one individual at some initial time, and follow its offspring, at very long times only two outcome are possible. Either its offspring have taken over the whole population, or the lineage has gone completely extinct. Since we start with $N$ individuals, the probability of taking over is $1/N$.
The argument generalizes to a subpopulation of size $n$: the probability that one of its members has taken over the population is $n/N$.

\subsubsection{Wright-Fisher model}\label{popgen:WF}
While we cannot calculate deterministically the fate of any particular mutant individual, we can calculate the probability of the evolution of the ancestral and mutant fractions, assuming two subpopulations and a constant population size as we did in subsection~\ref{Ass:detmutsel}. Two models, the Wright-Fisher and the Moran models, each with slightly different setups, describe the neutral evolution of populations. The Wright-Fisher model assumes discrete, non-overlapping generations: at each generation, individuals from the previous generation are cleared, a new sample of $N$ individuals is drawn, each new individual picking an ancestor from the previous generation with probability $1/N$.
Following a subpopulation of size $n$ in the parent generation, the probability that their offspring comprise $m$ individuals is given by the transition probability:
\beq
\label{WFmodel}
P_{n \rightarrow m}= {{N}\choose{m} }\left(\frac{n}{N}\right)^m \left(1-\frac{n}{N}\right)^{N-m}.
\eeq
The probability distribution for the size of that subpopulation evolves according to the recursion: $P_{t+1}(m)=\sum_{n} P_{n\rightarrow m} P_t(n)$, where $t$ labels generations.

Within this model the mean frequency of the subpopulation does not change with the time $\sum_n n P_n(t)=\av{n(t)}=\av{n(t+1)}$. Since the binomial distribution for large N becomes Gaussian, the evolutionary trajectory of the number of individuals in each subpopulation size is well described by a random walk: $n(t+1)\approx n(t)+ \eta(t)\sqrt{n(t)(1-n(t)/N)}$, with $\eta(t)$ normally distributed.

The Wright-Fisher model can be generalized to include selection. The probability to pick a given member $i$ of the parent generation depends on its selective advantage or disadvantage $s_i$, as $(1/N) (1+s_i)/(1+\bar s)$, with $\bar s=(1/N)\sum_i s_i$.

In the case of two subpopulations, wildtype (of size $n_2=N-n$) and mutant (of size $n_1=n$), with selective (dis-)advantage $s$ for the mutant, Eq.~\label{WFmodel} is modified to: $P_{n \rightarrow m}= {{N}\choose{m} }p^m(1-p)^{N-m}$, with $p=(n/N)(1+s)/(1+\bar s)$ and $\bar s=1+sn/N$. For large populations, this leads to a biased random walk:
\beq\label{eq:WFRW}
n(t+1)\approx n(t)+ sn(t)\left(1-\frac{n(t)}{N}\right)+ \eta(t)\sqrt{n(t)\left(1-\frac{n(t)}{N}\right)},
\eeq
whose determistic part reproduces Eq.~\ref{det_mut_sel} in the absence of mutations.

\subsubsection{Probability of extinction}\label{extinctionprob}
We can calculate the probability of extinction of the mutant subpopulation in the setup of two populations we have described until now. This limit corresponds to a small mutation rate: each subpopulation can fix or go extinct before a new mutation appears. While many immunological situations (e.g. affinity maturation) may not be in that regime, it is an important result to know. The probability that a subpopulation of size $n$ goes extinct, $q(n)$, requires knowing the probability that all their offspring, calculated within the Wright-Fisher model, will go extinct:
\beq
q(n)= \sum_{m=0}^N {{N}\choose{m}} p^m(1-p)^{N-m}q(i),
\eeq
given a selective advantage $s$ for the population. For large $N$ the binomial distribution is peaked, and we can Taylor expand $q(i)$ around $n$, $q(i)\approx q(n) +(i-n) q'(n) + \frac{1}{2}(i-n)^2 q''(n)$, which results in:
\beq
0=sn\left(1-\frac{n}{N}\right)q'(n)+\frac{1}{2}n\left(1-\frac{n}{N}\right)q''(n).
\eeq
Solving for the extinction probability with boundary conditions $q(n=0)=1$ (an extinct subpopulation remains extinct)  and  $q(n=N)=0$ (a fixed population cannot go extinct) yields $q(n)=1-\frac{1-e^{-sn}}{1-e^{-sN}}$. The probability of fixation is then:
\beq\label{probfix}
P_{\rm fix}=1 - q(n)=\frac{1-e^{-sn}}{1-e^{-sN}}.
\eeq
If $s=0$ we recover the result in subsection~\ref{gendrift}.

In the limit of strong positive selection ($Ns\gg 1$, but $s\ll 1$) the probability that a founder mutant ($n=1$) fixes is proportional to the selection strength $P_{\rm fix}\approx s$, if the initial population is one individual, or more generally  $P_{\rm fix}\approx ks$, if the initial population is small $k\ll 1/s$ individuals. On the other hand $P_{\rm fix}\approx 1$ for large founder populations $k\gg 1/s$. This defines a threshold of $1/s$ individuals to ensure the survival of the subpopulation. For weak selection pressures $\abs{Ns}\ll1$, the probability of fixation is equal to the frequency of individuals in the population and is independent of $s$, $P_{\rm fix}\approx n/N$, making this regime effectively neutral. Lastly, for strongly deleterious mutations, $sN\ll -1$, the probability of fixation is non-zero $P_{\rm fix}\approx e^{s(N-n)}$, but is exponentially small.

\subsubsection{Moran model, continuous limit, and time varying selection}\label{varrying_sel_Moran}
The above results can also be obtained considering a diffusion process in the number of individuals in the subpopulation. To do this, we use the Moran model, in which at each time step we choose one individual to die and one to reproduce (it can be the same individual), ensuring the population size is kept fixed. This implies overlapping generations, and the typical generation time is of order $N$ time steps.

Assume a wildtype population of size $N-n$ and a mutant population of size $n$. The probability of having $n$ at time step $i$ is given by:
\begin{eqnarray}\label{eq:Moran}
p_n(i) &=& p_{n+1}(i-1) \frac{n+1}{N}\frac{N-(n+1)}{N}\\ \nonumber
&&+p_{n-1}(i-1) \frac{n-1}{N}\frac{N-(n-1)}{N} \\ \nonumber
&& +p_{n}(i-1) \left[ {\left (\frac{n}{N}\right)}^2+{\left (\frac{N-n}{N}\right)}^2 \right],
\end{eqnarray}
where the first term describes killing one dominant allele and reproducing a mutant, the second one reproducing a dominant allele and killing a mutant, and the last term describes the two possibilities of killing and reproducing the same allele. In the limit of large population sizes, rescaling time by the typical generation $N$, $t=i/N$, so that $p_n(i) - p_n(i-1)= (1/N)\partial_t p_n(t)$, and Taylor expanding in  $x=n/N$:
\beq
\frac{\partial p(x,t)}{\partial t} = \frac{1}{2N} \frac{\partial^2}{\partial x^2} \left[ x(1-x) p(x,t)\right].
\eeq
Mutation and selection can be added following a similar derivation and contribute in a mean ``drift'':
\beq\label{ForwardDiffusion}
\begin{split}
\frac{\partial p(x,t)}{\partial t} &=\frac{1}{2N} \frac{\partial^2}{\partial x^2} \left[ x(1-x) p(x,t)\right]\\
&-\frac{\partial}{\partial x} \left[\left(sx(1-x)+\mu_2-(\mu_1+\mu_2)x \right)p(x,t)\right].
\end{split}
\eeq
This Fokker-Planck equation is the stochastic version of \ref{det_mut_sel}, with effective diffusion coefficient $x(1-x)/2N$. This expression is also consistent with the random walk approximation of the Wright-Fisher model (\ref{eq:WFRW}), meaning that the two models are equivalent in that continuous limit. 

Eq.~\ref{ForwardDiffusion} has a general solution in terms of Gegenbauer polynomials. At steady state, assuming $\mu_1=\mu_2=\mu$ for simplicity it takes the form $p(x)=Z^{-1}[ x (1-x)]^{N\mu-1} e^{Nsx}$, where $Z$ is a normalization factor. We can reinterpret it as a Boltzmann distribution, $p(x)=Z^{-1} p_0(x) e^{N s x}$, where each mutant individual gives an energy gain $-s$, and $p_0(x)= \left[ x (1-x)\right]^{N\mu-1} $ is the neutral distribution in the absence of selection.

Since Eq.~\ref{ForwardDiffusion} is a one dimensional diffusion equation, the probability of fixation is calculated from the backward equation for reaching the absorbing barrier at $x=1$. This calculation gives the same result as Eq.~\ref{probfix}.

To model the fluctuations of the environment, we can consider a time varying selection pressure $s(t)=s_0+\sigma(t)$ with mean $s_0$ and random white noise fluctuations $\av{\sigma(t)\sigma(t')}=\delta(t-t')\Omega$ around this mean~\cite{Takahata:1975ug}. The probability $p(y, x,t)$ that the wildtype, starting at initial frequency $y$ at time $t=0$, reaches frequency at least $x$ by time $t$ is given by the backwards equation:
\beq
\partial_t p(y, x,t) =  v(y) \partial_y p(y, x,t) + D(y) \partial^2_y p(y, x,t),
\eeq
where 
\beq
v(y) = s_0 y (1-y) + \Omega y(1-y) (1-2y),
\eeq
and where the diffusion term cumulates genetic drift and fitness fluctuations:
\beq
D(y) = \frac{1}{2N} y (1-y) + \Omega y^2(1-y^2).
\eeq
With the boundary conditions $p(y\geq x, x,t)=1$ and $p(y=0, x,t)=0$, the probability of ever reaching frequency $x$ is:
\beq
p(y, x,t\to \infty) = \frac{1-\abs{\frac{1-y/\alpha_+}{1-y/\alpha_-}}^{\lambda}}{1-\abs{\frac{1-x/\alpha_+}{1-x/\alpha_-}}^{\lambda}},
\eeq 
where $\alpha_{\pm} = \frac{1}{2}\left[ 1 \pm \sqrt{1+2/N\Omega}\right]$ and $\lambda = s_0/(\Omega\sqrt{1+2/N\Omega})$. For $x=1$ we recover the fixation probability reported in Takahata et al~\cite{Takahata:1975ug}, and for $\Omega \to 0 $ we recover the result for constant fitness (\ref{probfix}), $P_{\rm fix}=(1-e^{-s_0Ny})/(1-e^{-s_0N})$.

For more than two genotypes \cite{Nourmohammad2013}, Eq.~\ref{ForwardDiffusion} generalizes to an equation for a vector $\vec{x}=(x^1,...,x^{K-1})$ of linearly independent genotype frequencies ($x^K=1-\sum_{\alpha}^{K-1} x^{\alpha}$):
\begin{eqnarray}\label{ManyDimDiffusion}
\frac{\partial p(\vec{x},t)}{\partial t} &=&\sum_{\alpha, \beta} \Big\{-\frac{\partial}{\partial x^{\alpha}} \left[ m^{\alpha}(\vec{x})+C^{\alpha \beta}(\vec{x}) s_{\beta}(\vec{x})  \right]+\nonumber \\ 
&&+\frac{1}{2N} \frac{\partial^2}{\partial x^{\alpha} x^{\beta}}C^{\alpha \beta}(\vec{x}) \Big\}p(\vec{x},t),
\end{eqnarray}
where the covariance matrices are:
\begin{eqnarray}\label{covariance}
C^{\alpha \beta}(\vec{x})=\Big \{ \begin{matrix} - x^{\alpha} x^{\beta} & {\rm if }\  \alpha \neq \beta \\x^{\alpha} (1-x^{\alpha}) &  {\rm if } \  \alpha = \beta \end{matrix},
\end{eqnarray}
 the mutation  coefficients $m^{a}(\vec{x})= \sum_{\beta} (\mu_{\beta \rightarrow \alpha} x^{\beta}-\mu_{\alpha \rightarrow \beta} x^{\alpha})$ and the selection coefficient is the relative growth rate of genotype $\beta$ compared to a reference genotype $s_{\alpha}= f^{\beta}- f^{{\rm ref}}$.

\subsubsection{Branching processes}\label{branching}
Branching processes are useful for tracking the fate of the offspring of a individual through time. We will first introduce it in the context of the Moran model, and then present its more standard applications. Assuming birth (with rate $1+s$) and death (with rate $1$) as the only possibly processes, we can track the evolution of the probability of having 1 individual at time $0$ and $n$ at time $t$. This probability satisfies a recursion that can be obtained by considering the possible events occuring in the first time step, between times $0$ and $dt=1/N$~\cite{Allen2003, Desai2007a}:
\beq
\begin{split}
&p(1, n, t) = \frac{1}{N} \left(1- \frac{1}{N}\right) \delta_{n,0}\\
&\quad+ \frac{1}{N} \left(1- \frac{1}{N}\right)(1+s) p(2, n, t-dt)+\\ 
&\quad +\left[1- \frac{1}{N} \left (1- \frac{1}{N}\right)(2+s)\right] p(1, n, t-dt),
\end{split}
\eeq
The first term corresponds to the lineage going extinct between times $0$ and $dt$ --- and remaining extinct until $t$. The second term corresponds to a division of the initial individual between times $t=0$ and $dt$, and the lineage then reaching size $n$ from size $2$ in the remaining time, $t-dt$. The last term corresponds to no change at all.

The next step, which characterizes the branching process approach, is to assume that the outcome of a lineage can be deduced from the outcome of each of the descendants of the first division, taken independently: $p(2, n, t) \approx \sum_{m=0}^n p(1, m, t) p(1, n-m, t)$, which assumes that once one individual gives birth to two, the births in these lineages happen independently. This approximation is valid as long as $n\ll N$. Then the recursion becomes in the continuous time limit:
\beq
\frac{\partial p(1,n)}{\partial t} = \delta_{n,0}+(1+s) \sum_{m=0}^n p(1, m) p(1, n-m)-(2+s) p(1,n),
\eeq
Defining the generation function $G(z,t) = \sum_{n=0}^{\infty} p(1,n,t) z^n$ we obtain:
\beq
\frac{\partial G(z)}{\partial t} = 1+(1+s)G^2(z)-(2+s) G(z),
\eeq
which is solved with boundary conditions $G(z=1)=1$ and $G(z, t_0)=z$:
\beq
G(z,t) =\frac{(z-1)(1-e^{st})+zs}{(z-1)(1-(1+s)e^{st})+zs)}.
\eeq
For $s=0$ (no selection), we obtain $G(z)=1-\frac{1}{1+(1-z)^{-1}}$, whose series expansion yields $p(1,n=0,t)=t/(1+t)$ and $p(1,n,t)=t^{n-1}/(1+t)^{n+1}$ for $n\geq 1$. 

For arbitrary $s$ the mean number of individuals at time t is $\av{n}=\partial_z G(z)|_{z=1}=e^{st}$.The probability of going extinct at time t is
\beq
G(z=0)=p(1,n=0,t)= \frac{e^{st}-1}{(1+s)e^{st}-1},
\eeq
which goes to $1-s$ for $s>0$ at $t\to \infty$. We thus recover the result that the fixation probability goes to $s$, as obtained previously in the strong positive selection limit. In that limit, by the time the mutant escapes genetic drift, for $n\gg 1/s$, its population size is still small compared to the total population size $n\ll N$, making the branching process approximation appropriate.

While we have illustrated branching processes in the context of a Moran model of evolution, branching processes are ubiquitous and used in a variety of contexts. One of the simplest 
 branching process is defined in discrete time, where at each step a individual can divide with probability $p$, or not divide with probability $1-p$~\cite{Allen2003, Yates2007, Zapperi1995}. A recursion relation for the generating function of the total number of individuals $n$ in a lineage at time $t$ (past and present), $G(z,t)=\sum_n p(n,t) z^n$, can be written following similar arguments as above:
\beq
G(z,t) = zpG^2(z,t-1)+(1-p)z.
\eeq 
This recursion equation is solved for $t\to\infty$ by:
\beq\label{branchproc_genfun}
G(z,t\to \infty)=\frac{1-\sqrt{1-z^24p(1-p)}}{2pz},
\eeq
which has a critical point as $p\rightarrow 1/2$~\cite{Zapperi1995}, when the average number of offspring equals 1. Rewriting the generating function as a series we recover the probability of the total number of individuals (which must be even) 
\beq
P_{2 k} = \frac{1}{2}\frac{\Gamma(k-1/2)}{\Gamma(1/2)\Gamma(k+1)} \frac{\left(4 p (1-p)\right)^{k}}{2p},
\eeq
For $p<1/2$, $P_{k}$ is a decaying exponential as expected, and for $p=1/2$ it decays as a power-law $P_{k}\sim k^{-3/2}$. Note that this class of critical branching processes has been used to explain power laws in the distribution of activity in neural networks \cite{Beggs2003}.

\subsubsection{Coalescence process}~\label{coalescence}
As seen when calculating fixation probabilities, it is often useful to think about evolutionary processes backwards in time. This is the basic idea behind a coalescence approach, which can be formalised using branching processes in the context of the Wright-Fisher model. Here we give just some basic intuition about how thinking back in time about the history of coalescing sequences in lineages can be useful when studying affinity maturation processes, or tracing phenotypic lineages. W will present the coalescence process in a neutral evolutionary framework. As before the neutral framework provides us with a null model in the case affinity maturation where selection is important, but it may also be useful in immunological phenotyping.
The coalescence approach does not concern itself with mutations, but simply tracks genealogies. Mutations can later be added to an existing genealogy (tree).

We consider two individuals and ask how long ago they shared a common ancestor. If one individual has a given parent, the probability that the second cell has the same parent, given there are  $N$ cells, is $1/N$. The probability that they do not have the same parent is $1-1/N$. Following this reasoning, the probability that they have the same parent $t$ generations ago, but not during the $t-1$ generations is 
\beq\label{t2}
P \left[ T_{2} =t \right]= \left[ 1-1/N\right]^{t-1} 1/N \approx 1/N e^{-t/N},
\eeq
where we have expanded for large $N$ and $T_{2}$ stands for time to mean recent common ancestor (MRCA). The mean time for two cells to coalesce is simply the mean expectation time of this distribution
\beq
\av{T_2}=\int_0^{\infty} dt t/N e^{-t N} =N.
\eeq
The mean time for two cells to coalescence is equal to the population size, in units of generation time. More generally, the probability that $k$ cells do {\it{not}} share the same parent is $P_{\rm diff} =\left[ 1-1/N\right] \left[ 1-2/N\right]...\left[ 1-(k-1)/N\right]\approx 1- {{k}\choose{2}} N^{-1}$. The probability that {\it at least}  two cells have the same parent  (or coalescence in our backwards picture) is then $P_c = 1-P_{\rm diff}  = {{k}\choose{2}} N^{-1}$. If $N\gg k$ then the probability that more than two cells share the same parent  in a single generation can be neglected and we will assume that in each generation only two cells will share a parent. The distribution of times until the first coalescence is:
\begin{eqnarray}\label{Pcoal}
& &P\left[ \textrm{1st coalescence at time }t \right]= \left[ 1-P_c\right]^{t-1} P_c \\ \nonumber
&&\approx P_c e^{-P_c t} = {{k}\choose{2}} \frac{1}{N}e^{-{{k}\choose{2}} t/N}.
\end{eqnarray}
After a coalescent event, there are $k-1$ individuals left, and the process can be repeated until the whole genealogical tree is reconstructed. The coalescence probability completely determines the statistics of the topology of the tree and the branch lengths in a neutral process. Once the branch lengths (times between each coalescence events) have been determined, neutral mutations are distributed randomly along the branch with some rate, so that their number on each branch follows a Poisson distribution. Specificially, the number of mutations $\pi$ between two individuals, also called pairwise heterozygocity, is given by:
\beq
P(\pi|t) = e^{-2 \mu_n t} \frac{\left[2 \mu t\right]^{\pi}}{\pi!}
\eeq
for coalescence time $t$ and per-generation mutation rate $\mu$ (the total time is $2t$ because it adds the two branches of length $t$ from the common ancestor).
Of course, all of this breaks down in the presence of selection.

Thus, the coalescence probability determines the genetic diversity within a population. 
The distribution of $\pi$ in the population can be obtained by integrating over the coalescence time \eqref{t2},
\beq
P(\pi)=\int_0^{\infty} dt \left[ \frac{1}{N} e^{-t/N} \right] \left[ e^{-2 \mu t} \frac{\left[2 \mu t\right]^{\pi}}{\pi!} \right] = \frac{1}{1+\theta} \left[\frac{\theta}{1+\theta}\right]^{\pi}, 
\eeq
where $\theta=2 \mu N =\av{\pi} $. Thus, in principle, in the regime of neutrality, characterizing the distribution of the mean mutational distance should allow us to read off the mutation rate. 

Departures from the Poisson distribution is one of the signatures of selection. In the presence of selection, the distribution of branch lengths depends on the details of the type of selection we are studying. If we do think about affinity maturation (where selection plays an important role), we see that cells undergo bursts of selection in the germinal centers, followed by periods outside of the germinal centers. Two cells sampled at the same time from the blood may therefore have very different recent histories and using their mutation distance to infer a mean mutation rate would be misleading.

\subsubsection{Site frequency spectra and tree balancing}\label{SFS}

Another statistics that is easy to calculate within the neutral model, and can thus be used as a null model to compare to data, is the site frequency spectrum (SFS). The SFS is the number of individuals in the population (in our case cells) that have a mutation at a given position in the aligned sequence with respect to the dominant base pair in the most recent common ancestor, and presents it as a histogram: ${(n_1, n_2, ...., n_{N-1})}$,  where $n_1$ is the number of mutations present in a single cell, $n_2$ is the number number of mutations present in two cells, etc. Sites that are not polymorphic (all have the same base pair) do not contribute to the spectrum as they coincide with the most recent common ancestor of the whole population. However a mutation at a site that is close to fixation will contribute to the spectrum, although this mutation now dominates the population. In absence of information about the most recent common ancestor, in population genetics the mutation is often called with respect to the outgroup. In the case of B-cell receptors, a good estimate can be the infered from the best alignment to the V, D, and J germline sequences. 

We can estimate the SFS for a neutrally evolving population of constant size $N$ within the Moran model. In the continuous time limit, the mean number of mutations shared by $k$ individuals evolves according to (for $k>1$):
\begin{equation}
\begin{split}
\frac{d n_k}{d t}=& \frac{(k+1)(N-k-1)}{N} n_{k+1} + \frac{(k-1)(N-k+1)}{N} n_{k-1} \\ 
 & -2 \frac{k(N-k)}{N} n_{k}=J_{k-1}-J_{k},
\end{split}
\end{equation}
with 
\begin{equation}
J_k=[{k(N-k)}/{N}] n_{k} - [{(k+1)(N-k-1)}/{N}] n_{k+1}
\end{equation}
the ``current'' of mutations across subpopulation sizes.
The first term corresponds to death event occuring in the subpopulation of size $k+1$ carrying the mutation of interest, and the second term to birth events in the subpopulation of size $k-1$.
New mutation always starts with one individual, so that
\begin{eqnarray}\label{BCSFS}
\frac{d n_1}{d t}& =& \frac{2 (N-2)}{N} n_{2}  -2 \frac{(N-1)}{N} n_{1}+\mu,
\end{eqnarray}
where $\mu$ is the mutation rate. At steady state the current is constant and equal to $J_k=J_1=\mu/N$,
because each new mutation has a probability $1/N$ to fix, resulting in a current $\mu/N$ of mutations traveling from from size $1$ to $N$.
This implies:
\beq
n_k = \frac{(k-1)(N-k+1)}{k(N-k)}n_{k-1} - \frac{\mu}{k(N-k)},
\eeq
which is solved by 
\beq\label{eq:neutralSFS}
n_k=\mu/k.
\eeq
The SFS can also be calculated approximately in models with selection, but still with fixed population sizes, so we do not recall these results here \cite{Neher2013}.

Another feature that can be used to identify selection through departure from the neutral model is how balanced lineage trees are. Intuitively, a neutral process does not favor adding a new mutation to any of the branches, hence the resulting trees should be symmetric and balanced. A process with selection will preferentially grow the favoured parts of the tree, resulting in some long branches, and other short ones. In the neutral model, the number of leaves $n$ in a sublineage at generation $t$, follows the distribution $p_t(n)$, which satisfies the recursion:
\beq
p_t(n) = \left(1-\frac{n}{t-1} \right)p_{t-1}(n)+\frac{n-1}{t-1} p_{t-1}(n-1).
\eeq
The sublineage of interest can add one leave by reproducing, with probability proportional to its size $n$, (second term), or not (first term).
Solving by recursion gives a uniform branch length distribution $p_t(n)={1}/({n-1})$.

\subsubsection{Clonal interference}\label{clonal_interference}
Another concept from population genetics that has been shown to be relevant in BCR affinity maturation is clonal interference, which we discussed in more detail in section \ref{affinity_maturation}. Here we recall the back-of-the envelope arguments of Desai and Fisher~\cite{Desai2007a} to show on a more classical example why clonal interference slows down the rate of adaptation. 

Given a population size $N$ and mutation rate $\mu$, a mutation occurs in any individual with rate $N\mu$, so the time between mutations is $(N\mu)^{-1}$. Most of these mutations go extinct due to genetic drift. In the strong selection regime, a mutation survives genetic drift with probability $s$. Thus, when mutations are rare, the average time for a new mutation to occur and fix is  $(N\mu s)^{-1}$, and the rate of adaptation is given by $v_{\rm rare}=N \mu s^2$.

However fixation itself may take time, and the above picture breaks down when mutations are no longer rare, i.e. when the typical time between succesful mutation becomes smaller or of the same magnitude as the time for it to fix.
After the subpopulation carrying the mutation overcomes genetic drift which results in an initial population size of $1/s$, we can consider that the subpopulation grows deterministically with rate $s$, $n(t)\sim (1/s)e^{st}$. The mutation will fix when $n(t)=N$, which gives the fixation time $t_{\rm fix} = (1/s) \ln{N s}$. Mutation are no longer rare when $(N\mu s)^{-1}\sim 1/s \ln{N s}$, or $N\mu \sim (\ln{N s})^{-1}$.

In that regime, many mutants can co-exist at the same time, and new mutations can appear in existing mutants before these have time to fix. Mutants existing at the same time will have different growth rates, depending on the number of mutations they have acquired, even if we assume for simplicity that all mutations give the same fitness advantage. The whole population can be described by a fitness distribution and we can notice that the fate of a mutation in the bulk of this distribution (a subpopulation that was created some time ago) is different that at its nose (a recently created subpopulation). The nose subpopulations are small and susceptible to genetic drift. The bulk subpopulations have more individuals and grow deterministically, but are subject to nonlinear effects of competition between individuals. We can consider these two subpopulations separately and then ``stitch'' the two solutions. 

The individuals at the nose of the fitness distribution have a fitness advantage, lets call it $qs$, meaning that they have $q$ mutations, each confering an advtantage $s$, compared to the bulk of the distribution. The time needed for an even fitter individual (with fitness $(q+1)s$) to appear is given by $\tau$ that satifies $\int_0^{\tau} dt \mu n_{\rm nose}(t) q s=1$, where $n_{\rm nose}(t) = 1/(qs) e^{qs t}$ is the number of individuals at the nose of the fitness distribution, following the same arguments as for the rare mutation case. In the limit of $e^{qs \tau}\gg 1$, one gets: 
\beq\label{taunose} 
\tau = 1/(qs) \ln{(qs/\mu)}.
\eeq 

Every time that a new fitter class is added to the distribution, the mean of the distribution also increases. After $q \tau$, the old nose of the distribution will become the mean (note that the individuals that are less fit than the mean go extinct deterministically). The populations in fitness class $q$ grow as $qs$, so in class $q-1$ they grow as  $(q-1) s$, etc. A lineage originating at the nose grows to dominate the population $q\tau$ later, while progressively losing its fitness advantage. By the time it dominates the population,
its relative fitness advantage is zero. On average, it will have grown at rate $qs/2$. This gives the consistency equation $(1/qs) e^{qst/2} =N$, with $t=q\tau$. This gives a second estimate of the establishment time:
\beq\label{taubulk}
\tau=\frac{2}{q^2s} \ln{Nqs}.
\eeq
Equating Eq.~\ref{taunose} and Eq.~\ref{taubulk}, we find the average rate at which the population grows:
\beq
q=\frac{2 \ln{Nqs}}{ \ln{(Nq/\mu)}}.
\eeq
We can solve this implicit equation approximately assuming that $q$ is not too big and neglecting the $\ln{q}$ terms:
\beq
q=\frac{2 \ln{Ns}}{ \ln{(N/\mu)}}.
\eeq
The rate of adaptation in the regime of multiple competing mutations is then:
\beq
v_{\rm CI} = \frac{s}{\tau} = \frac{s^2 \ln{Ns}}{\ln{s/\mu}}.
\eeq
This rate scales with the logarithm of the population size, $v_{\rm CI}\sim \ln{N}$, much slower than in the rare mutation regime $v_{\rm rare}\sim {N}$. The clonal interference regime is not only a regime in which there are many competing mutations, but a regime where new mutations arise on the background of still relatively low frequency mutations, forming competing lineages. The distinction between competition of different clones and clonal interference is especially important in affinity maturation, where the competition in germinal centers could be between very different clones (which is not clonal interference), or between clones with similar histories, in which case it is clonal interference. 

\subsubsection{Quantitative traits}\label{quantitative_traits}
While models of population genetics are often defined in the space of genotypes or alleles, what is often measured in the resulting phenotype, which is a (possibly nonlinear) function of the genotype. For instance, for lymphocyte receptors, the relevant phenotype may be defined as the binding affinity to epitopes of interest.

A projection of a description in phenotypic space, where mutations occur independently at all loci $i$ along the genome, results in an effective description in phenotypic space, which is useful because selection occurs at the phenotypic level and we often do not have access to the genotype--phenotype map. Given a fraction $x^{\alpha}$ of the population carrying allele ${\alpha}$, the population mean and variance of a phenotype $E$ is:
\begin{eqnarray}\label{QuantT_defs}
\Gamma&=&\av{E}=\frac{1}{N}\sum_{\alpha=1}^N E^{\alpha} x^{\alpha} \approx \int dE E w(E) \\ \nonumber
\Delta&=&\av{E^2} - \av{E}^2
\approx \int dE (E - \Gamma)^2 w(E),
\end{eqnarray}
where $N$ is the population size, and $w(E)$ is the distribution of this trait in the population.

One can derive an effective equation for the joint evolution of the mean phenotype and its variance \cite{Nourmohammad2013}:
\begin{eqnarray}\label{phenotype1}
\frac{\partial}{\partial t} Q(\Gamma, \Delta, t)& =& \Big[ - \frac{\partial}{\partial \Gamma} \left( \frac{d \Gamma^{\rm mut}}{dt} +\frac{d \Gamma^{\rm sel}}{dt}  \right)  \\ \nonumber
&&- \frac{\partial}{\partial \Delta} \left( \frac{d \Delta^{\rm mut}}{dt} +\frac{d \Delta^{\rm sel}}{dt}  \right)
 \\ \nonumber
 &&+\frac{1}{2N} \left( \frac{\partial^2}{\partial \Gamma^2}C^{\Gamma \Gamma} + \frac{\partial^2}{\partial \Delta^2}C^{ \Delta  \Delta}\right) \Big] Q(\Gamma, \Delta, t),
\end{eqnarray}
where we calculate the drift and diffusion terms below.

First lets focus on the diffusion terms, which correspond to genetic drift are are independent of selection:
\begin{eqnarray}\label{Cgammagamma}
C^{\Gamma \Gamma} &= & \sum_{\alpha,\beta}  \frac{\partial  \Gamma}{\partial x^{\alpha}}\frac{\partial  \Gamma}{\partial x^{\beta}} C^{\alpha \beta} \\ \nonumber
&=&  \sum_{\alpha}  E^2_{\alpha} x^{\alpha} (1-x^{\alpha})  -\sum_{\alpha \neq \beta} E_{\alpha}E_{\beta}x^{\alpha} x^{\beta}  \\ \nonumber
&=& \av{E^2} - \av{E}^2 = \Delta,
\end{eqnarray}
where we have used the definitions of the covariance matrices in high dimensional space from Eq.~\ref{covariance}. Similarly
\beq
\label{Cdeltadelta}
C^{\Delta \Delta} =  \sum_{\alpha,\beta}  \frac{\partial \Delta}{\partial x^{\alpha}}\frac{\partial  \Delta}{\partial x^{\beta}} C^{\alpha \beta}
=\av{(E-\Gamma)^4}- \Delta^2 \approx 2 \Delta^2,
\eeq
where in the last step we have assumed $w(E)$ is approximately Gaussian. A similar calculation shows that $C^{\Delta \Gamma}  = \av{(E-\Gamma)^3} \approx 0$ with the Gaussian assumption.

To compute the drift terms, which arise from selective effects, we need to specificy how fitness and phenotype are related.
For simplicity, let us assume that fitness of each individual is a quadratic function of the phenotype (or expand fitness close to its peak at $E^*$), $f(E)=-c_0 (E-E^*)^2$, where $c_0$ is a prefactor that measures the width of the fitness peak.  The mean fitness is then:
\begin{eqnarray}
F(\Gamma,\Delta)&=&\sum_{\alpha}^N f^{\alpha} x^{\alpha} = -c_0 \sum_{\alpha}^N (E^{\alpha} -E^*)^2 x^{\alpha} \\ \nonumber
&=& -c_0\left[ \Delta +(\Gamma - E^*)^2\right].
\end{eqnarray}
The change in the mean phenotype $\Gamma$ due to selection is:
\begin{eqnarray}
\frac{d \Gamma^{\rm sel}}{dt} &=& \frac{d }{dt} \int dE E w(E) = \int dE E (\frac{d w(E)}{dt})  \\ \nonumber
 &=& \int dE E \left( f(E)- F\right) w(E)  \\ \nonumber
 &=& -c_0\left( (E -E^*)^2 - \Delta - (\Gamma - E^*)^2\right) w(E)  \\ \nonumber 
 &=& - 2 c_0 \Delta (\Gamma - E^*).
\end{eqnarray}
Since  (by analogy with Eq.~\ref{ManyDimDiffusion}) $\frac{d \Gamma^{\rm sel}}{dt}  = C^{\Gamma \Gamma} s_{\Gamma}$, using Eq.~\ref{Cgammagamma} we can verify that the selection coefficient is $s_{\Gamma} =\frac{\partial}{\partial \Gamma} F(\Gamma, \Delta) = - 2 c_0 (\Gamma - E^*) $.

In principle we can calculate the  change in the mean variance due to selection in the same way $\frac{d \Delta^{\rm sel}}{dt} = \frac{d }{dt} \int dE (E - \Gamma)^2  w(E)$ assuming a peaked phenotype distribution,  but in practice its easier to use the analogy with Eq.~\ref{ManyDimDiffusion} and Eq.~\ref{Cdeltadelta}:
\begin{eqnarray}
\frac{d \Delta^{\rm sel}}{dt} &=& \frac{\partial}{\partial \Gamma} F(\Gamma, \Delta) \approx -2 c_0 \Delta^2.
\end{eqnarray}

To consider the changes of $\Gamma$ and $\Delta$ due to mutations we have to write down a more detailed genotype model since mutations act on base pairs. If each locus can take a WT value ($\sigma_i=0$) or mutant value ($\sigma_i=1$), each linearly additive phenotype (for simplicity) $E$ (for example binding energy)  can be written as $E^{\alpha}=\sum_i^L E_i^\alpha \sigma_i$, where $E_i^\alpha$ is a given sites contribution to the overall phenotype. We have $\Gamma = (1/N) \sum_{i,\alpha} E_i^\alpha \sigma^{\alpha}_i x^{\alpha}$ and:
\begin{eqnarray}
\frac{d \Gamma^{\rm mut}}{dt} &=&\frac{1}{N}\sum_{i,\alpha} \frac{\partial \Gamma}{\partial \sigma^{\alpha}_i} \frac{d  \sigma^{\alpha}_i}{d t} = \\ \nonumber
 &=&-\sum_{\alpha,i} x^{\alpha} E^{\alpha}_i \mu (2  \sigma^{\alpha}_i-1) = -2\mu (\Gamma - \Gamma_0),
\end{eqnarray}
where $\Gamma_0 = (1/2)\sum_{i} E^{i}$ is a phenotype average. 

Similarly, the variance defined in Eq.~\ref{QuantT_defs} is also a sum over sites, and  the change in the variance due to mutations is:
\beq
\frac{d \Delta^{\rm mut}}{dt} =\sum_{i, \alpha} \frac{\partial\Delta}{\partial \sigma^{\alpha}_i} \frac{d \sigma^{\alpha}_i}{d t}
\approx  -4\mu (\Delta - E^2_0)
\eeq
where $E^2_0=(1/4) \sum_{i} E^2_{i}$. 

Now we have all the elements, the effective diffusion equation (\ref{phenotype1}) in phenotypic space reads:
\beq
\begin{split}
&\frac{\partial Q(\Gamma, \Delta, t)}{\partial t}  = 
\frac{1}{2N} \Big[ \frac{\partial^2}{\partial \Gamma^2}\Delta + \frac{\partial^2}{\partial \Delta^2} \left( 2 \Delta^2 \right) \Big] Q(\Gamma, \Delta, t)\\
&\ +\Big[ 2 c_0 \Delta  \frac{\partial}{\partial \Gamma} \left( \Gamma-E^* \right)+2c_0 \frac{\partial}{\partial \Delta} \Delta^2 \Big]  Q(\Gamma, \Delta, t) \\ 
&\ +\Big[ 2 \mu  \frac{\partial }{\partial \Gamma} \left( \Gamma-\Gamma_0 \right)+4\mu \frac{\partial } {\partial \Delta} \left(\Delta^2-E^2_0 \right) \Big ] Q(\Gamma, \Delta, t).
\end{split}
\eeq

The stochastic equation for the evolution of the mean phenotype can then be read off as:
\begin{eqnarray}
\frac{\Gamma}{\partial t} & =&  - 2 \mu \left( \Gamma-\Gamma_0 \right) + \Delta \partial_{\Gamma} F + \sqrt{\frac{\Delta}{N}} \xi_{\Gamma},
\end{eqnarray}
where the first term describes mutations, the second term selection and
$\partial_{\Gamma} F =-2{c_0}  (\Gamma-E^*(t))$
and $\xi_\Gamma$ is a normalized Gaussian white noise. $E^*(t)$ can be a time dependent moving fitness maximum. An analogous equation holds for $\Delta$. Assuming that $\Delta$ changes on much faster timescales than $ \Gamma$, it can be replaced by its mean, and the steady state solution has a Boltzmann form $Q_{\rm eq} (t) = Z^{-1} Q_0 (\Gamma) e^{-2 N F(\Gamma)} $, where as before $Q_0 (\Gamma) \sim \exp\Big[-\frac{1}{2} \frac{\left( \Gamma-\Gamma_0 \right)^2}{\av{\Delta}/(4N)} \Big])$ is the distribution with no selection. 

In a time dependent environment the peak of the distribution also changes with time according to a prescribed model. The population tries to track the fitness peak, without really ever reaching it.
From the physical point of view, the system is maintained out of equilibrium.
In a changing fitness landscape, where the population history is described by a sequence of phenotypic trait measurement $(\Gamma_0, .... \Gamma_M)$ over time $(t_0, ..., t_M)$ the fitness flux of a population history $\Phi$~\cite{Mustonen2010b} describes a cumulative selective effect of phenotypic trait changes:
\beq
\Phi = \sum_{i=1}^M \delta \Gamma_i \partial_{\Gamma_i} F(\Gamma_i, t_i) \neq  F(\Gamma_M, t_M) -  F(\Gamma_0, t_0).
\eeq
The lack of equality holds in general because $F$ also has an explicit time dependency.
One can show that this discrepancy is formally equivalent to dissipation in thermal systems, and can be related to the entropy production:
\beq
\av{2 N \Phi} = {\rm KL}({\cal P}| {\cal P^T}) + {\rm boundary \ terms},
\eeq
where $ {\cal P^T}$ is the probability of the forward trajectory and  ${\cal P^T}$ of the backward trajectory and the Kullback-Leibler divergence is defined in Eq.~\ref{DKL}. This dissipation has been called ``fitness flux'' in the context of population genetics.

These kinds of phenotypic trait models have been used as starting points for studying co-evolution as we describe in section~\ref{affinity_maturation}. The main difference is that the fitness objective $E^*$ is itself a function of the composition of the population with which the initial population evolves. That other population is also subject to selection and drift and evolves stochastically, giving rise to coupled equations \cite{Nourmohammad2016}.

\subsubsection{Lineage reconstruction}\label{lineage_reconstruction}
Lineage reconstruction is a necessary first step in both BCR evolutionary analysis and phenotypic tracking. While many software methods exist to reconstruct lineages for evolutionary problems, they are not always well adapted for immunological data. Nevertheless they are often used, and more adapted methods are usually built upon classical ones, so here we provide a general overview of these existing approaches. 

The first problem in reconstructing lineages requires taking the sequence data and identifying clusters of sequences that share a common ancestor, and therefore belong to the same lineage. A classical strategy to cluster datapoints is single-linkage clustering, which builds hierarchical clusters by iteratively merging pairs together  \cite{Everitt2011}.
The Partis software~\cite{Ralph2016a} uses a likelihood ratio test to decide if a give set of sequences $\sigma_1,\sigma_2,...,\sigma_N$ can be grouped as descending from one ancestor or not. Specifically Partis uses a Hidden Markov Model (HMM) based method to annotate each nucleotide in a specific BCR sequence as coming from a given a set of hidden states corresponding to V, D or J genomic template, or from a set of exponentially distributed non-templated insertions.  The sum over paths determines the probability $P(\sigma_i)$ for each sequence. The same procedure repeated for a  set of $N$ sequences, results in the total probability of a common scenario for these sequences $P(\sigma_1,..,\sigma_N)$. If the likelihood ratio $P(\sigma_1,..,\sigma_N)/\Pi_{i=1}^N P(\sigma_i)>{\rm threshold}$, one concludes that the sequences originate from a common recombination.  Partis starts with pairs of sequences, keeping the cluster with the largest likelihood ratio, and then adding new members to the cluster based on the same test. In practice, to speed up the algorithm Partis does not test all possible pairs but creates initial subsets of data with low Hamming distances. Joining a cluster is irreversible, which can lead to errors in clustering.

Once a lineage is defined, annotation softwares such as Partis~\cite{Ralph2016} or IGoR~\cite{Marcou2018} can be used to propose the naive root of the tree.

This problem alone is simpler than finding the whole tree genealogy, known as the topology of the tree or the branching pattern, because part of the sequence is templated by the V, D, and J genes (see Sec.~\ref{pgen}). There are two main classes of methods for tree topology reconstruction: maximum parsimony and maximum likelihood. Let us first explain why it is practically impossible to exhaustively sample all tree topologies, and then explain the differences in the two approaches.

We will use the example of binary trees --- trees in which each node gives rise to only two branches --- since essentially all loopless trees can be cast into a binary form by adding branches of zero length. Starting from the simplest unrooted tree of two leaves connected by one branch, and adding new leaves one at a time, it is simple to convince oneself that in each step we add $1$ leave, $1$ internal node and $2$ edges, such that a tree with $N$ leaves has $N-2$ internal nodes and $2N-3$ edges (or branches).  Adding the $N$th leave ($N-1\rightarrow N$) adds $2N-5$ topologies, such that the number of unrooted tree topologies with $N$ leaves $T_N$, is $T_N=(2N-5)T_{N-1}$, which can be recursively solved to give $T_N=(2N-5)!!$. For rooted trees, the number of rooted topologies with $N$ leaves is given by $T_{N+1}$ in terms of the number of non-rooted tree topologies, since it just requires adding a root to the existing unrooted tree topologies.

The goal of tree reconstruction is to find the topology that is consistent with the data. For a tree with $N=10$ leaves, that means exploring $T_N\sim 10^6$ trees and for $N=50$,  $T_N\sim 10^{76}$ trees, which is prohibitively large. BCR clusters often have hundreds of leaves. 

There are two ways of determining distance between two node sequences along an edge of a tree: one is to calculate the Hamming distance by simply counting the number of single nucleotide differences between two sequences $h_j$, the other is to use the estimated time between two mutations $t_j$. The number of mutations in a given time $t_j$ is distributed according to a Poisson distribution of mean $\mu t_j$: $P(h_j)=\frac{\mu t_j}{h_j!}e^{-\mu t_j}$. The branch length refers to the time between two nodes. The tree length is defined as the total number of substitutions, $h_{\rm tot} = \sum_{\rm{branches}} h_j$, and is not equal to the sum of branch lengths. Maximum parsimony methods use Hamming distance to define distances on trees, which requires knowing the identity of all the internal nodes and the tree topology, whereas maximum likelihood sums over possibilities for the sequences on the internal nodes, but requires a mutational model. 

The maximum parsimony approach is easy to formulate: for a given topology, it assigns internal nodes so as to minimize $h_{\rm tot}$. It then selects topologies with minimal $h_{\rm tot}$.
The parsimony scores differ from $0$ (when all the leaves are the same) to $NL$ (when all the leaves are different, which essentially means this position carries no information). There are $(2N-5)!!$ topologies to scan so many trees have the same $h_{\rm tot}$, including at its minimum. Therefore there is often a large family of most parsimonious trees. Apart from the practical problems, the assumptions behind the evolutionary model in this approach are not clear~\cite{ZYang2006}. In practice, the reconstructed trees match the true structure if there is little variation in the internal nodes.  The main advantage is that it embodies the basic intuition for reconstructing the phylogeny.

In the maximum likelihood method, given an overall lineage age $T$ and tree topology, the likelihood is given by a specific mutation model by varying the branch lengths $t_j$. Within the independent site assumption, the likelihood factorizes over sites and the log-likelihood $\ell=\sum_{r=1}^{L}\ln{\ell_r}$ independently traces the evolution of each site $r$. Thus the following equations apply to a single site, but generalize readily to many sites. The probability of mutating the base pair from $x_i$  into $x_j$ between nodes $i$ and $j$, is defined as $P_{x_i,x_j}(t_j)$. For instance, this probability can take the following form, although many others are possible:
\beq
P_{x_i,x_j}(t_j) = e^{-\mu t_j}\delta_{x_i,x_j} + (1-e^{-\mu t_j}) \pi_{x_j},
\eeq
where $\pi_{x}$ (with  $\sum_x \pi_{x}=1$) describes the probability that a mutation results in base pair $x$, and is equal to $1/4$ for the unbiased 4-base pair model, and $t_j$ is the branch length (time between nodes $i$ and $j$). The first term corresponds to no mutation, and the second  to a mutation (including one into the same base pair). This formulation guarantees detailed balance: $\pi_x P_{xy}(t)=\pi_y P_{yx}(t)$ for all $t$.

We can now build the likelihood recursively, assuming a fixed root identity $0$, with sequences denotes as $x_0$ that are not fixed. First consider a tree made of a root and two branches of length $t_1$ and $t_2$ leading to leaves $1$ and $2$, whose identity $x_1$ and $x_2$ is known. The likelihood of this tree is:
\beq
L=\sum_{x_0} \pi_{x_0} P_{x_0x_1}(t_1)P_{x_0x_2}(t_2).
\eeq
We can now add an internal node $4$ to the tree, descending from $0$ and ancestor of $1$ and $2$, and a leave $3$ descending from 0:
\begin{eqnarray}
L&=&\sum_{x_0} \sum_{x_4} \pi_{x_0} P_{x_0x_4}(t_4) P_{x_0x_3}(t_3)P_{x_4x_1}(t_1) P_{x_4x_2}(t_2)= \nonumber \\
&=&  \sum_{x_0}  \pi_{x_0}P_{x_0x_3}(t_3)  \sum_{x_4} P_{x_0x_4}(t_4)L_4(x_4),
\end{eqnarray}
where $L_4(x_0)= P_{x_4x_1}(t_1) P_{x_4x_2}$. This recursive form allows for efficient calculation of likelihood by successively joining roots of trees into a new root. Formally, joining root $j$ and $k$ into common parent $i$:
\beq
L_i(x_i)=\left[  \sum_{x_j} P_{x_ix_j}(t_j) L_j(x_j)\right] \left[  \sum_{x_k} P_{x_ix_k}(t_k) L_k(x_k)\right],
\eeq
Leaves are initialized to $L_j(x_j)=1$, and the final results is given by:
\beq
L=\sum_{x_0}\pi_{x_0}L_0(x_0).
\eeq

Given a fixed topology, we can now maximize the likelihood $L$ over the set of intermediate branch lengths $\{ t_j\}$, which results in a score for each topology. The topology ${\cal T}$ with the best best ranking score, is the most likely topology (similarly to the maximum parsimony approach). 

Existing maximum likelihood methods use the independent site assumption to decrease computational time. Many methods also assume homogeneous mutation rates across sites and the
reversibility of the evolutionary process. The GTRGAMMA substitution model of the RAxML software \cite{Stamatakis2014} uses gamma-distributed mutation rates for different base pairs, and IgPhyML~\cite{Hoehn2017} encodes a non-reversible mutation model that effectively accounts for the context dependence of mutations (although in a site-independent setup). 

Davidsen and Matsen~\cite{Davidsen2018} compared maximum likelihood and maximum parsimony methods for BCR lineage reconstruction. They concluded that an improved and informed maximum parsimony method outperforms classic maximum likelihood.

\subsubsection{Population growth rates}\label{maxLAMBDA}
Another way to study the evolution of populations is to study dynamics that maximize the long term growth rate of the population given a fixed environment~\cite{Kussell2005, Rivoire2011}:
\beq
\Lambda ({\rm dynamics, environment}) =\lim_{T \rightarrow \infty} \frac{1}{T} \sum_{t=0} \ln(N_t),
\eeq
where $N_t$ is the population size at a given time, whose evolution is driven by the dynamics. One needs to specify classes of evolutionary dynamics to consider, and the class of interactions it has with the environment. The framework is clearly very general, and reduces to a constrained optimization problem. The main conceptual difference with population genetics models studied in this section is that populations cannot go extinct, and all possible variants are represented. Populations are in a regime where fast growth governs evolution. Alternative approaches could consider optimal strategies for avoiding extinction~\cite{Bradde2019}, by minimizing the probability of extinction, $P_{\rm ext}$, whose form depends on the problem considered.

\subsection{Ecological models}

Population of lymphocytes interact with other through signaling, competition for resources, cytokines, or antigens. For this reason, concepts and mathematical models from ecology are often useful to describe their dynamics.

\subsubsection{Generalized Lotka--Volterra models}\label{GenLV}
Ecological generalized Lotka-Volterra models ~\cite{Lotka20, volterra56, Keshet2005} describe the co-habitation of multiple species in the same environment and account for their interactions, both direct (one species needs another to reproduce) and indirect (through competition for external resources). In general these models have the form:
\beq\label{LotkaVolterra}
\frac{d x_i}{ dt} = x_i \left[ \alpha( \vec{x}, \vec{c}) - \beta( \vec{x}, \vec{c}) \right],
\eeq
where $x_i$ is the frequency of a given species, $c_i$ are external factors such as nutrients $\alpha( \vec{x}, \vec{c})$ is the growth rate and $\beta( \vec{x}, \vec{c})$ the death rate with $\vec x=(x_1,x_2,\ldots)$ and $\vec c=(c_1,c_2,\ldots)$. The form of the dependence of the growth rate on the different species and external factors determines the non-linearity of the problem. This is usually encoded by an interaction matrix between species. Models where the growth and death rates are the same for all species are called neutral. Often one introduces a carrying capacity which is a form of non-linearity that describes competition for resources.

This formulation of the model is very general and the solution depends on the specific assumptions. In general these models are solved numerically, and the results depend on the numerical values of the parameters that are often not known with great certainty. These models are often used to ask questions about the coexistence of different species, as well as speciation itself --- why are ecological environments with different species stable? Since we often do not have experimental information to parametrize the interaction matrix, random matrix models have been succesfully used to effectively describe the interactions~\cite{May2001}. In this approach one uses the fact that within a family of random matrices (i.e. matrices whose elements are chosen from the same distribution) the eigenvalues of these matrices are the same.  Near the fixed point, the stability of the system is explored by linearizing the system of equations in Eq.~\ref{LotkaVolterra}, and the eigenvalues of this linearized matrix determine the stability fo the ecosystem:
\beq\label{LotkaVolterra_lin}
\frac{d x_i}{ dt} = -x_i +\epsilon \sum_j K_{ij} x_j(t),
\eeq
where $\epsilon$ is the interaction strength and $\bf{K}$ is a random interaction matrix.  As a result the stability of the system does not depend on the realization of the interaction matrix, just on its statistics. For the matrix to be stable, all the eigenvalues of $\bf{K}$, $\lambda_i$ have to satisfy $\epsilon \lambda_i \leq 1$, which is fulfilled in the largest eigenvalue satisfies $\lambda_{\rm max}<1/\epsilon$.  In the case of Gaussian random matrices, the properties are determined by their first two moments and a strong transition occurs for $N\rightarrow \infty$ where the system is stable for $\epsilon<1/\sqrt{2}$ and otherwise unstable. This result holds only if the connectivity scales with the size of the system. 

Incidentlly, random matrix theory also has been used when looking at covariances in sequence variation (see section~\ref{inference_maxentropy}). 
The general idea is that beyond the first couple of eigenvalues of the covariance matrix of amino acid variability (which often stems from phylogenetic bias), eigenvalues are well approximated by a random matrix. For $R\times C$ Gaussian random matrices, where the elements $K_{ij}$ are chosen from a Gaussian distribution with mean $0$ and variance $\sigma$,  the density of eigenvalues of the covariance matrix $Y=\av{KK^{T}}/R$ is given by the Marchenko-Pastur distribution~\cite{MarchenkoPastur1967}:
\beq
\rho(x)=\frac{1}{2\pi \sigma^2} \frac{\sqrt{(\theta_+-x)(x-\theta_-)}}{\theta x},
\eeq
where $\theta=C/R$ and $\theta_{\pm}=\sigma^2(1{\pm}\sqrt{\theta})$.

\subsubsection{Susceptible-Infected-Recovered (SIR) Models}\label{SIR}
Susceptible-Infected-Recovered (SIR) models are used to study the spread of epidemics at the population level. An SIR model considers three possible kinds of individuals: ones that are susceptible to the infection (S), infected (I) and recovered (R) and therefore usually immune. The simplest SIR model assumes that the birth-death of the host individuals happens on slower timescales than the spread of the epidemic itself. This model is a good description of measles and other infections by slowly evolving pathogens.
\begin{eqnarray}
\frac{d S}{d t} &=& - \beta I S\label{SIR_simple1}, \\ 
\frac{d I}{d t} &=& \beta I S - \nu I  = x I\label{SIR_simple2}, \\
\frac{d R}{d t} &=&\nu R,\label{SIR_simple3}
\end{eqnarray}
where $\beta$ is an effective infection rate, $\nu$ is the rate of recovery, and $\nu (\beta S/\nu -1)=x$ defines an effective growth rate, or fitness of the host population. Within this formulation no one can die. Since $\partial_t S +\partial_t I+\partial_t R=0$, the total population size is constant, $N=S+I+R=$const. The typical recovery time is $T_{\rm r}=\nu^{-1}$ and the typical timescale for a infected-susceptible individual interaction is $T_{\rm inf}=\beta^{-1}$, meaning that an infected individual manages to infect $R_0=T_{\rm r}/T_{\rm inf}=\beta/\nu$ individuals before recovering. $R_0$ is also called the infection radius or basic reproduction number. 

Using Eqs.~\ref{SIR_simple1}-\ref{SIR_simple3}, we find $S=S(t=0) e^{-R_0 (R-R(t=0)}$. At infinite time ($t \rightarrow \infty$, $I=0$), the number of recovered individuals reads:
\beq
R_{\infty}=N-S(t=0) e^{-R_0 (R_{\infty}-R(t=0))},
\eeq
which means that at an end of an epidemic there are still susceptible individuals. From Eq.~\ref{SIR_simple2} we find that $S(t=0) R_0>1$, leads to $\partial_t I>0$ and an exponentially growing infection (i.e. epidemic outbreak), whereas $S(t=0)  R_0<1$ leads to $\partial_t I<0$ and quenching of the infection. Since these equations are deterministic, the initial condition determines the infection spread for all times. The effective number of susceptible individuals that can propagate the infection determines its future. 

SIR equations can be extended to account for host birth (with rate $\alpha$) and death (with rate $\mu$), that lead to equations of the form:
\begin{eqnarray}\label{SIR_vital}
\frac{d S}{d t} &=&\alpha-\mu S - \beta I S, \\
\frac{d I}{d t} &=& \beta I S - \nu I - \mu I  = x I, \\ 
\frac{d R}{d t} &=&\nu R- \mu R,
\end{eqnarray}
with two stable steady state solution: pathogen free, $(S, I, R)=(\alpha/\mu, 0, 0) $; and permanent infection, $(S, I, R)=\left((\nu+\mu)/\beta, (\mu/\beta) (R_0-1), (\gamma/\beta) (R_0-1)\right)$, where $R_0=\beta\alpha/ \left(\mu(\mu+\nu)\right)$ is the infection radius. For $R_0\leq1$, the system evolves towards the pathogen-free solution, while for $R_0 \geq 1$ the permanently infected solution is reached.

The basic SIR model can be made more complicated in many different ways. One other variation called the SIS (Susceptible -- Infected -- Susceptible) model describes a situation where an infected individual does not have long lasting immunity. This model is inspired by influenza, where the virus mutates fast enough that a past infection does not necessarily guarantee immunity (although current models of influenza account for more detailed descriptions of cross-reactivity -- see section~\ref{popleveldyn}). The model just has two states:
\begin{eqnarray}\label{SIS}
\frac{d S}{d t} &=& - \beta I S +\nu I,\\ 
\frac{d I}{d t} &=& \beta I S - \nu I  = x I.
\end{eqnarray}
Using $S=N-I$, we obtain a single differential equation:
\beq
\frac{d I}{d t} = (\beta N   - \nu) I -\beta I^2,
\eeq
which has two fixed points $I=0$ for $R_0=(\beta/\nu) N<1$ and $I=(\beta N-\nu)/\beta$ for $R_0>1$. The second fixed results from the solution:
\beq
I(t)= \frac{(\beta N-\nu)}{\beta} \frac{1}{1+(\frac{(\beta N-\nu)}{\beta I_0}-1)e^{(\nu-\beta N)(t-t_0)}}.
\eeq  

We note that the SIR equations (Eqs.~\ref{SIR_simple1}-\ref{SIR_simple3}) differ in its assumptions from the two species Lotka-Volterra equations (see section~\ref{GenLV}) between and a host and a pathogen. Specifically, SIR equations  do not assume that everyone can get infected, while Lotka-Volterra equations do. However, generalized Lotka-Volterra equations can easily be modified to account for different subclasses of hosts. Lotka-Volterra equations on the other hand do explicitly model the viral population, and allow for host-pathogen oscillations that SIR models do not, and they do do not assume a constant population size. Notably, the SIS equations (Eq.~\ref{SIS}) are a two-species realization of Lotka-Volterra dynamics, with the additional assumption for constant population size. 

In section~\ref{dynamics} we describe equations that model the in-host evolution during HIV infections. They are very similar in spirit to SIS equations which explicitly consider viral dynamics.

\subsubsection{Solution of stochastic population dynamics with a source}\label{neutralthymicderivation}
Here we present the calculations leading the results presented in Sec.~\ref{dynamics}, some of which are original to this review.
To solve Eq.~\ref{eq:neutralthymic}, we define the generating functions:
\bea
G(x)&=&\sum_{C=1}^{+\infty} N_C x^C,\\
\Theta(x)&=&\sum_{C=1}^{+\infty}\theta_C x^C.
\eea
The time evolution of the population is then governed by:
\beq\label{eq:Gx}
\frac{dG(x)}{dt}=\Theta(x)-\mu N_1 + (\nu x^2+\mu - (\mu+\nu) x)G'(x).
\eeq
At steady state, we have:
\beq
G'(x)=\frac{\Theta(1)-\Theta(x)}{(\nu-\mu x)(1-x)},
\eeq
where we have used the balance between birth and death of clones $dG(1)/dt=\Theta(1)-\nu N_1=0$. Using $(\Theta(1)-\Theta(x))/(1-x)=\sum_{C>0}\theta_c\sum_{i=0}^{C-1}  x^i$, and expaning $1/(\nu-\mu x)=(1/\nu)\sum_{j=0}^{\infty} (\mu/\nu)^j x^j$, we obtain:
\beq
\begin{split}
N_C&=\frac{1}{\mu-\nu}\frac{1}{C}\left\{{\left(\frac{\nu}{\mu}\right)}^C\sum_{k=1}^{C-1}\theta_k \left[{\left(\frac{\mu}{\nu}\right)}^k-1\right]
\right.\\ &\left.
+\sum_{k=C}^\infty \theta_k \left[1-{\left(\frac{\nu}{\mu}\right)}^C\right] \right\},
\end{split}
\eeq
which reduces to 
\beq
N_C=\frac{\theta}{\nu}\frac{1}{C}{\left(\frac{\nu}{\mu}\right)}^C
\eeq
 in the simple case $\theta_C=\theta\delta_{C,1}$. The total number of clones reads:
\beq
\begin{split}
N_{\rm tot}&=G(1)=\frac1{\mu-\nu}\sum_{k}\theta_k\left\{ \sum_{C=1}^k \frac{1-{\left(\frac{\nu}{\mu}\right)}^C}{C}
\right.\\ &\left.
+\left[{\left(\frac{\mu}{\nu}\right)}^k-1\right]\sum_{C=k+1}^\infty {\left(\frac{\nu}{\mu}\right)}^C\frac{1}{C}\right\},
\end{split}
\eeq
and 
\beq
N_{\rm tot}=(\theta/\nu)\ln[1/(1-\nu/\mu)]
\eeq
for $\theta_C=\theta\delta_{C,1}$.
In the limit of rare divisions $\nu\ll \mu$, only clone sizes below the initial clone size, $C\leq k$, contribute, so that 
\beq
N_{\rm tot}=\frac{\sum_k \theta_k H_k}{\mu}=C_{\rm tot}\frac{\sum_k  H_k\theta_k}{\sum_k k\theta_k},
\eeq
where $H_k=\sum_{i=1}^k 1/i$ are harmonic numbers. In the opposite limit of division balancing death, $\mu-\nu\ll \mu$, the long tail of the second sum dominates:
\beq
N_{\rm tot}=\frac{\ln(1/\epsilon)\sum_k k\theta_k }{\mu}=C_{\rm tot}\epsilon\ln(1/\epsilon),
\eeq
with $\epsilon=(\mu-\nu)/\mu$.

Eq.~\ref{eq:Gx} can be solved out of steady state with the method of characteristics.
Defining $F(x)=G(x)-G(1)$ and making the change of variable $y=\ln[(x-\mu/\nu)/(x-1)]/(\mu-\nu)$, this equation reduces to:
\beq\label{eq:Fequation}
\left(\frac{\partial}{\partial t}-\frac{\partial}{\partial y}\right)F=\Theta(x)-\Theta(1).
\eeq
This equation can be solved in absence of a source term ($\Theta=0$): $F_0(x,t)=A(t+y)$, with $A(y)=F_0(x,0)$, yielding:
\beq\label{eq:F}
F_0(x,t)=F_0\left(\frac{\mu/\nu - (x-\mu/\nu)/(x-1) e^{t(\mu-\nu)}}{1 - (x-\mu/\nu)/(x-1) e^{t(\mu-\nu)}},0\right)
\eeq
Starting with a single clone of size $s$ at $t=0$, $G_0(x,0)=x^s$ and $F_0(x,0)=x^s-1$, the coefficients of the solution $G_0(x,t)$ to the homogeneous equation (Eq.~\ref{eq:Fequation} with $\Theta=0$) give Green's function, defined as the probability that the clone starting at size $s$ at $t=0$ has size $C$ at a later time $t$, $G_0(x,t)=\sum_C P(C,t|s,0)x^C$. Expanding \eqref{eq:F} in $x$, one obtains:
\beq
\begin{split}
&P(C,t|s,0)=g(C,s,t)=\frac{r^C}{(1-rz)^{C+s}}\\
&\times \sum_{n=1}^{\min(C,s)}\binom{s}{n}\binom{C-1}{n-1} r^n(1-r^{-1})^{2n}z^n(1-z)^{s+C-2n},
\end{split}
\eeq
with the shorthands $z=e^{-t(\mu-\nu)}$, $r=\nu/\mu$. The combinatorial factor in the sum comes from counting the number of ways there are to choose $s-n$ factors of order $0$ in $x$, $\binom{s}{n}$, and $n$ factors of order $(j_1,\ldots,j_n)$ so that $\sum_{i=1}^nj_i=C$, $\binom{C-1}{n-1}$, in a series expansion of the form $[a+b((cx)+(cx)^2+(cx)^3+\ldots)]^s=[a+bcx/(1-cx)]^s$ to which \eqref{eq:F} can be reduced.

We can now use Green's function to calculate the dynamics of the distribution with an abitrary initial condition:
\beq\label{eq:full}
N_C(t)=\sum_k N_k(0)g(C,k,t)+\int_{0}^t dt' \sum_{k}\theta_k(t') g(C,k,t-t').
\eeq
In the case of an empty immune system at $t=0$, $N_C(0)=0$, and constant thymic output $\theta_k$, we get:
\beq
\begin{split}
&N_C(t)=\frac{r^C}{\mu-\nu}\sum_k\theta_k \sum_{n=1}^{\min(C,k)}\binom{k}{n}\binom{C-1}{n-1}\frac{r^n(1-r^{-1})^{2n}}{n}\\
&\times \left[{z^n}F_1(n,2n-k-C,C+k,1+n,z,rz)\right]_{e^{-t(\mu-\nu)}}^1,
\end{split}
\eeq
where $F_1$ is the Appell hypergeometric function of two variables, and where we made the change of variable $z=e^{-(t-t')(\mu-\nu)}$ to carry out the integral in \eqref{eq:full}. This equation simplifies greatly for $\theta_k=\theta \delta_{k,1}$:
\beq
N_C(t)=\frac{1}{\nu}\frac{r^C}{C}{\left(\frac{1-e^{-t(\mu-\nu)}}{1-re^{-t(\mu-\nu)}}\right)}^C.
\eeq

\subsubsection{Solution to foward jump process with opposing drift and source}\label{jump}
Consider  the process described by \eqref{eq:jump} generalized to an arbitrary distribution of jumps:
\beq\label{eq:jumpgen}
\frac{\partial \rho}{\partial t}=\mu \frac{\partial \rho}{\partial x}+\int_0^\infty dy\,J(y) \left[\rho(x-y)-\rho(x)\right]+ \tilde\theta(x),
\eeq
where absorbing boundary condition at $x=0$.
where $J(y)dy$ is the rate (per unit time) of jumps of size between $y$ and $y+dy$. This equation can be solved exactly by considering the Laplace transform of $\rho$:
\beq
\hat \rho(k,t)=\int_{0}^{\infty} dx\, e^{-kx}\rho(x,t),
\eeq
which satisfies:
\beq
\frac{\partial \hat \rho}{\partial t}=\mu(\rho(0,t)+k \hat\rho(k,t))+(\hat J(k)-\hat J(0))\hat \rho(k,t) + \hat \theta(k),
\eeq
where $\hat J(k)$ and $\hat\rho(k)$ are the Laplace transforms for $J$ and $\tilde \theta$. The steady state solution is:
\beq
\hat\rho(k)=\frac{\hat\theta(k)-\mu \rho(0)}{\hat J(k)-\hat J(0)+\mu k}.
\eeq
To get a physical solution, a pole at $k=0$ cannot exist, so that the total number of clones $\hat\rho(0)$ is well defined. This is satisfied for $\mu\rho(0)=\hat \theta(0)$, which is simply the condition that the rate of birth of new clones, $\hat\theta(0)$ be balanced by the rate of death, $\mu\rho(0)$. Then solution becomes
\beq
\hat\rho(k)=\frac{\hat\theta(k)-\hat\theta(0)}{\hat J(k)-\hat J(0)+\mu k}.
\eeq
To understand the behaviour of $\rho(x)$ at large $x$, we have to examine poles for negative $k$, $k=-\alpha$, which satisfy the condition:
\beq\label{eq:alphagen}
\int_0^\infty dx\, J(x)(e^{\alpha x}-1)=\mu \alpha.
\eeq
The left-hand side has derivative $\bar\nu\equiv \int dx\, J(x)x$ with respect to $k_0$, which is the average growth rate of a clone. To guarantee that all clones eventually go extinct, that number must be smaller than $\mu$. Therefore \eqref{eq:alphagen} has only one solution $\alpha>0$. At large $x$ the behaviour is dominated by that pole, $\hat\rho(k)\sim 1/(k+\alpha)$, yielding $\rho(x)\sim e^{-\alpha x}$ and:
\beq
N_C\sim\frac{1}{C^{1+\alpha}}.
\eeq
{\rev The total numbers of clones and cells read:
\bea
N_{\rm tot}&=&\hat\rho(0)=\frac{\int_0^\infty dx\,\theta(x) x}{\int_0^\infty dx\, J(x)x - \mu},\label{Ntotsel}\\
C_{\rm tot}&=&\hat\rho(-1)=\frac{\int_0^\infty dx\, \theta(x) (e^x-1)}{\int_0^\infty dx\,J(x) (e^x-1)-\mu}.\label{Ctotsel}
\eea
}

In the specific case of fixed-size jump, $J(x)=sp\delta(x-\Delta x)$, we get back \eqref{eq:alpha} and \eqref{CtotNtotsel}.
Eq.~\ref{eq:alphagen} simplifies in the limit of many very small jumps, corresponding to $J_{\rm tot}\equiv \int_0^\infty dx\,J(x)\to \infty$ while keeping the average effect of jumps, $\bar\nu<\mu$, as well as its second moment, $\Gamma=\int_0^\infty dx\,J(x)x^2$, finite, meaning that the average jump size $\<\Delta x\>=\nu/J_{\rm tot}$ goes to zero. Expanding \eqref{eq:alphagen} at small $x$, we get:
\beq
\alpha=2\frac{\mu-\bar\nu}{\Gamma}.
\eeq
which gives back \eqref{eq:alphajonathan} in the case of fixed-size jumps.

\subsubsection{The Yule process}\label{Yule}
The Yule-Simon~\cite{Yule1925, Simon1955} process is another type of ``neutral'' process that is characterized by a distribution with power long tails, which in its tail is reminiscent of Zipf's law (see section~\ref{neutral_discussion}). The model was initially formulated to describe the distribution of biological genera. It was later adapted in network science and is known under the name of preferential attachement, or the rich-get-richer model. In general it is a realization of a birth-death process, where the birth coefficient is proportional to the number of individuals. 
In the model, new individuals are added at each time step. With probability $\alpha$, the new individual gets a mutation (or undergoes a speciation event if we think about species), and with probability $1-\alpha$ the new individual's type is chosen among all the possible ancestors, i.e. proportionaly to the abundance of each type.

The probability that the rank of an individual is $k$ results in a power law distribution at large $C$:
\beq\label{eq:yule}
\frac{N_C}{C_{\rm tot}} = \frac{\alpha}{2-\alpha}\frac{\Gamma(C) \Gamma(\rho+1)}{\Gamma(C+\rho+1)}\sim \frac{1}{C^{\rho+1}},
\eeq
where $\Gamma(x)$ denotes a gamma function, and $\rho=1/(1-\alpha)$.

A precise derivation of \eqref{eq:yule} can be found in \cite{Simon1955},  but the power-law behaviour can be understood intuitively in a continous approximation. Assume that the rate of division is $\mu$, so that there are $C_{\rm tot}=e^{\mu t}$ individual at time $t$, starting from a single individual at time $t=0$. The rate of new emergence of new mutants is $\alpha \mu e^{\mu t}$, so that at time $t$ there are $n(t)=\int_0^t dt' \alpha\mu e^{\mu t'}=\alpha (e^{\mu t}-1)$ new types. The key point is that the abundance of a mutant is an exponentially increasing function of its age. A mutant that arose at $t'$ has size $C=e^{\mu(1-\alpha)(t-t')}$ at time $t$. $n(t')$ can thus be viewed as the rank of the mutant, ordered by increasing frequency. The abundance of the mutant arising at time $t'$, of rank $n=n(t')$ is, at time $t$: 
\beq
C=\frac{e^{\mu(1-\alpha)t}}{(1+n/\alpha)^{(1-\alpha)}}\sim \frac{1}{n^{1-\alpha}},
\eeq
where the approximation is only valid at large $n$. We recognize Zipf's law \eqref{eq:zipf}, which implies a power law in the clone size distribution according to \eqref{eq:powerlawfrequencydistribution}:
\beq
N_C\sim \frac{1}{C^{1+1/(1-\alpha)}}=\frac{1}{C^{\rho+1}}.
\eeq

\subsection{Inference} 
Here we introduce basic concepts and methods for constructing models and inferring their parameters from data. These approaches are becoming increasingly important as large datasets are being produced by high-throughput methods, from imaging to sequencing. Some of these methods have been used in only a few applications in computational immunology, but we anticipate that their usage will become widespread in future studies. 

\subsubsection{Probabilistic inference, maximum likelihood and Bayesian statistics}\label{inference}

Often, empirical data originates from a stochastic process. The first source of variability, which is present in almost any measurement, is experimental noise, which is random. In many cases related to biological systems, from gene expression to cell signaling and random recombination of DNA, the underlying processes are intrinsically stochastic, and so are the models that must describe them.

Let us assume that a model can be encapsulated into a probability distribution for observations $P(\bx|\theta)$, where $\bx$ is the empirical data, and $\theta$ the set of parameters defining the model. In many cases, the data can be decomposed into a set of independent datapoints, $\bx=(x_1,\ldots,x_N)$, so that the probability distribution factorizes over these observations: 
\beq
P(\bx|\theta)=\prod_{i=1}^N P(x_i|\theta).
\eeq
This probability is called the likelihood (of the data given the model). Because of its often multiplicative form, it is common to consider instead its logarithm, the log-likelihood $\mathcal{L}(x|\theta)$, which is additive, or its normalized variant $\ell(x|\theta)=(1/N)\sum_{i=1}^N\ln P(x_i|\theta)$. One can think of $x_i$ as a sequence, for instance of an immune receptor or of a pathogen; a fluorescence signal coming from flow cytometry or single-cell microscopy imaging; an abundance of RNA transcripts for gene expression, or of sequencing reads of immune receptors when estimating the abundance of lymphocyte clones. The models may be statistical models for the occurence of a particular data point, or stochastic dynamical models of the process.

A popular way to infer the parameters $\theta=(\theta_1,\ldots,\theta_K)$ of a model is to find those that maximize the likehood of the data:
\beq
\theta^*(\bx)=\argmax_{\theta} P(\bx|\theta).
\eeq
This estimate of the parameters is called the maximum likelihood estimator, and it can be shown to be unbiased and optimal in the limit of large numbers of observation, in the sense that it saturates the Cramer-Rao bound. The Cramer-Rao bound gives a lower bound on the variance of unbiased estimators of a fixed and deterministic parameter. In practice, this implies that the fluctuations of the estimator around its true value, $\theta^*(\bx)-\theta$ are given by a multivariate Gaussian distribution of mean 0 and co-variance $I^{-1}/N$, with:
\beq
I_{ab}(\theta)=\sum_{x} P(x|\theta) \frac{\partial \ln P(x|\theta)}{\partial \theta_a\partial \theta_b}
\eeq
is called the Fisher information matrix.

When one has information about the process and its parameters independently of the data, through a prior belief in the distribution of $\theta$, $P_{\rm prior}(\theta)$, this information can be combined with the data using Bayes rules:
\beq\label{eq:bayes}
P(\theta|\bx)=\frac{P(\bx|\theta)P_{\rm prior}(\theta)}{P(\bx)}.
\eeq
This expression defines the posterior distribution of parameters, given the data. For instance, some parameters may have been measured with some uncertainty, which can be modelled using a Gaussian distribution $P_{\rm prior}(\theta_i)=(1/\sqrt{2\pi\sigma_i^2})\exp[-(\theta_i-\theta_{i,\rm meas})^2/\sigma_i^2]$, or there exists natural physiological range for them. A uniform prior, $P_{\rm prior}=$const, amounts to considering the likelihood alone. Also note that in the limit of large datasets, $N\gg 1$, the log-likelihood dominates over the logarithm of the prior, which becomes negligible. In that case, the posterior over possible parameters, $P(\theta|\bx)$, is given by a Gaussian distribution over $\theta$ with mean $\theta^*$ and covariance given by the same Fisher information matrix $I^{-1}/N$. This means that the Bayesian fluctuations, which correspond to our uncertainty about the true values of the parameters, match the fluctuations of the error we make when picking the maximum-likelihood estimate.

Using Eq.\,\ref{eq:bayes}, one can either consider the full range of acceptable parameter values from the posterior, by e.g. sampling from it. Alternatively, one can take its maximum, $\theta^*=\argmax_\theta P(\theta|\bx)$, called the maximum a posteriori estimator. In the limit of large $N$, that estimator is equivalent to the maximum likelihood estimator.

We now turn to two simple examples of inference problems for which the maximum likelihood estimator recovers an intuitive answer. First consider linear regression. Suppose we have data points $(x_1,\ldots,x_N)$ taken at times $(t_1,\ldots,t_N)$, which we want to fit with a linear model, $x=\theta_1+\theta_2 t$. Implicitly, we must assume that the data is noisy, otherwise a fit would not be necessary. Assuming Gaussian noise, the stochastic model reads:
\beq
\begin{split}
P(\bx|\theta)&=\prod_{i=1}^N \frac{1}{\sqrt{2\pi\sigma^2}}e^{-(x_i-\theta_1-\theta_2 t_i)^2/2\sigma^2}\\
&=\frac{1}{(2\pi\sigma^2)^{N/2}}\exp\left[-\frac{1}{2\sigma^2}\sum_{i=1}^N(x_i-\theta_1-\theta_2 t_i)^2\right]
\end{split}
\eeq
where $\sigma$ is the amplitude of the noise. Examining this expression, one can see maximizing the likelihood corresponds to minimizing the mean squared error, $\sum_{i=1}^N(x_i-\theta_1-\theta_2 t_i)^2$, which is the standard way to do linear fit. This argument generalizes to any fit of a parametrized function, assuming Gaussian distributed noise.

The second example is estimating frequencies from a finite set of outcomes. For concreteness, let us assume that we want to describe the probability of finding a particular nucleotide, $x=$A,C,G,T, at a particular location in a genomic sequence. The parameters of the model are the frequencies of A, C, and G, denoted $\theta_1,\theta_2,\theta_3$, while the frequency of T is $1-\theta_1-\theta_2-\theta_3$. The probability of observing a certain set of observations $\bx=(x_1,\ldots,x_N)$ is given by:
\beq
P(\bx|\theta)=\theta_1^{n_A}\theta_2^{n_C}\theta_3^{n_G}(1-\theta_1-\theta_2-\theta_3)^{n_T}
\eeq
where $n_A=\sum_{i=1}\delta_{x_i,A}$ and similarly for the other nucleotides. Maximizing the likelihood with respect to the parameters yields the intuitive counting estimator: $\theta^*_1=n_A/(n_A+n_C+n_G+n_T)$, and similarly for the other nucleotides. Note that this argument holds for sequences of nucleotides rather than single nucleotides, provided that the underlying probabilistic model assumes that the choices of nucletoides at each position along the sequence are independent of each other.

\subsubsection{Model selection}\label{modelselection}
The maximum likelihood and Bayesian rules outlined above assume that the class of models, if not its parameters, are known, but this is not always the case. For instance, we may have competing hypotheses between different models, or we may consider models of increasing complexity. For instance, in the example of linear regression, we may want to know whether the data is not better explained by a quadratic model $x=\theta_1+\theta_2t+\theta_3t^2$. Comparing the likelihood directly always favors models with more parameters, which can in turn lead to overly complex models with the risk of overfitting the data. We need a way to compare models with different structures and numbers of parameters.

The general approach to compare two model A and B, described by probabilities $P(\bx|\theta_A,A)$ and $P(\bs|\theta_B,B)$, is to evaluate the overall probability of the data given each model, by integrating the over the parameters:
\bea\label{eq:modelsel}
P(\bx|A)&=&\int d^{K_A}\theta_A P_A(\bx|\theta_A)P_{\rm prior}(\theta_A|A)\\
P(\bx|B)&=&\int d^{K_B}\theta_B P_B(\bx|\theta_B)P_{\rm prior}(\theta_B|B).
\eea
The relative probabilities of each model is then evaluated by Bayes's rule:
\beq\label{eq:bayesmodelsel}
P(A|\bx)=\frac{P(\bx|A)P_{\rm prior}(A)}{P(\bx|A)P_{\rm prior}(A)+P(\bx|B)P_{\rm prior}(B)}.
\eeq
In the limit of large $N$, $P_A(\bx|\theta_A)$ (and the same for B) is very peaked around its maximum likelihood value, with Gaussian fluctuations around that peak given by the Fisher Information matrix $I_A$:
\beq
P_A(\bx|\theta_A)\approx P(\bx|\theta^*_A)e^{-N(\theta_A-\theta_A^*)I_A(\theta_A-\theta^*_A)/2}.
\eeq
By the saddle-point approximation we obtain:
\beq\label{eq:saddle}
P(\bx|A)\approx \sqrt{(2\pi/ N)^{K_A}/\mathrm{det}(I_A)} P_A(\bx|\theta^*_A)P_{\rm prior}(\theta^*_A|A).
\eeq
If we replace this expression into Eq.~\ref{eq:modelsel}, and focus on two terms that do not depend on our prior assumptions, we see two terms: the maximized likelihood for each model $P_A(\bx|\theta^*_A)$, which quantifies the quality of the fit, and ``parsimony'' term $\sqrt{(2\pi/ N)^{K_A}/\mathrm{det}(I_A)}$, which quantifies the volume of parameters that are consistent with each model. As evident for this expression of the parsimony, the more parameters, the smaller that volume will be.

The ratio of probabilities in \eqref{eq:modelsel} between A and B is sometimes called Bayes's factor, while the ratio of the parsimony terms is called Occam's factor. If we expand the logarithm of \eqref{eq:modelsel} in $o(\ln(N))$ using \eqref{eq:saddle}, we obtain, up to a $-2$ factor, the so called Bayesian Information Criterion (BIC):
\beq
 -2\ln P(A|\bx) \approx K_A\ln(N)-2N\ell(\bx|\theta_A^*)\equiv \mathrm{BIC}(A).
\eeq
The BIC is a popular score for comparing models, and can be intuitively understood as a correction to the likelihood by a term that penalizes large numbers of parameters. In the large $N$ limit, Eq.~\ref{eq:bayesmodelsel} tells us that the model with the smallest BIC is more likely to be correct. Another popular score is the Akaike information criterion (AIC), defined as $2K-2N\ell(\bx|\theta^*)$, which puts a smaller penalty on the number of parameters.

Applications of model selection criterion have been numerous in computational immunology, see e.g. \cite{Mccoy2015,Marcou2018c,Cotari2013,Laydon2015,Kepler14}.

\subsubsection{Expectation-Maximization}\label{EM}

Some models are better defined in terms of variables that are not accessible to the observer. If we call those hidden variables $\bh$, and $\bx$ the visible variables, then the likelihood of the data can be expressed as the sum over the hidden variables:
\beq
P(\bx|\theta)=\sum_{\bh}P(\bx,\bh|\theta)=\prod_{i=1}^N\sum_{h_i}P(x_i,h_i|\theta).
\eeq
Expectation-Maximization (EM) consists of maximizing $P(\bx|\theta)$ iteratively. Starting with a guess $\theta^t$, one can write a pseudo log-likelihood:
\beq
\mathcal{\tilde L}(\bx|\theta)=\sum_{i=1}^N \sum_{h_i}P(h_i|x_i,\theta^t)\ln P(x_i,h_i|\theta),
\eeq
which is essentially an average of the log-likelihood of the full model (i.e. including the hidden variables), weighted by the posterior distribution of the hidden variable $h_i$ under the current model $\theta_t$. In an iteration step, one sets $\theta_{t+1}=\argmax_\theta \mathcal{\tilde L}(\bx|\theta)$. As this scheme is iterated, $\theta_t$ converges to a maximum of the true likelihood $P(\bx|\theta)$.

Expectation-maximization has notably been used in the inference of generation model for antigen-specific lymphocyte receptors (TCR and BCR), in Refs.~\cite{Murugan2012,Elhanati2014,Elhanati2015,Elhanati2016}.

\subsubsection{Hidden Markov models}
Hidden Markov models (HMM) are models with a hidden variable $h$ which is the trajectory of a Markov process, $h=(h_x)_{x=1,\ldots,L}$:
\beq
P(h)=p_1(h_1)\prod_{j=2}^{L}w(h_{j}|h_{j-1}).
\eeq
The visible variable $x$, is also a vector of size $L$, whose elements $x_j$ are called 'emissions' and are drawn according to the value of the hidden variable according to the emission probabilities $P(x|h)=\prod_{j=1}^L e_j(x_j|h_j)$.
A key feature of HMM is that high-dimensional sums over $h=(h_1,\ldots,h_L)$ can be performed recursively using technique which is equivalent to the transfer matrix technique in statistical physics. For instance, the forward algorithm allows one to calculate $P(x)$, which requires to sum over all possible trajectories of the hidden variable. This is done by defining $z_j(h_j)=P(x_
1,\ldots,x_j,h_j)$ and using the recursive equation:
\beq
z_j(h_j)=\sum_{h_{j-1}}e_j(x_{j}|h_{j})w_j(h_{j}|h_{j-1})z_{j-1}(h_{j-1}).
\eeq
The probability $P(x)=P(x_1,\ldots,x_L)$ is then obtained as $\sum_{h_L}z_L(h_L)$.
A similar trick allows one to calculate any marginal of the model, e.g. $P(x_j)$ or correlation functions. Combining these recursions with EM defines a powerful algorithm for learning the parameters of the model, $\theta=\{w_j(\cdot|\cdot),e_j(\cdot|\cdot)\}$, called the Baum-Welch algorithm.

HMM have been applied to model BCR and TCR annotation to the germline in various software, e.g. SODA \cite{Munshaw2010}, Partis~\cite{Ralph2016}, and repgenhmm \cite{Elhanati2016b}.

\subsubsection{Information theory}\label{infotheory}
The {\bf entropy} of a probability distribution $P(x)$ is defined as:
\beq
S[P]=-\sum_x P(x)\ln P(x).
\eeq
It quantifies the randomness of the distribution, and is maximum for a uniform distribution (and zero for perfectly peaked one). For this reason is is often used as a measure of diversity. Unlike all other diversity measures, it is additive, meaning that the entropy of a joint distribution of two independent variables $(P(x),Q(y))$ is given by $S[P]+S[Q]$.

The entropy is the building block for several information-theoretic measures. The {\bf Kullback-Leibler divergence} or relative entropy, defined as
\beq\label{DKL}
{\rm KL}(P\Vert Q)=\sum_x P(x)\ln \frac{P(x)}{Q(x)},
\eeq
is used a distance measure between probability distributions (although not a metric in the mathematical sense because of its asymmetry). Another popular measure is the {\bf mutual information} between two variables $x$ and $y$:
\beq
I(x,y)=S[P(x)]-S[P(x|y)]=\sum_{x,y}P(x,y)\ln\frac{P(x,y)}{P(x)P(y)}.
\eeq
It quantifies how much the knowledge of $y$ reduces the entropy of $x$, and vice-versa since it is symmetric in $x$ and $y$. It is often used as a non-parametric measure of correlations between two variables.

\subsubsection{Maximum entropy models}~\label{inference_maxentropy}
When the underlying mechanisms that give rise to the data are not known, it can be useful to define phenomenological models based on the observables that are deemed important. A convenient way to do this is to infer {\rm maximum entropy models} \cite{Presse2013a}, which are probabilistic models $P(x)$ of maximum entropy $S[P]$ subject to the constraint that they agree with the data on a choice of key mean observables $\mathcal{O}_a(x)$: $\sum_x P(x)\mathcal{O}_a(x)=(1/N)\sum_{i=1}^N \mathcal{O}_a(x_i)$. It can be shown that the distribution takes an exponential form:
\beq\label{eq:maxent}
P(x)=\frac{1}{Z}\exp\left(\sum_a \theta_a \mathcal{O}_a(x)\right),
\eeq
where $\theta_a$ are Lagrange multipliers that must be adjusted to satisfy the constraints, and $Z$ is a normalization constant. It turns out that maximizing the likelihood of a dataset $(x_1,\ldots,x_N)$ under Eq.~\ref{eq:maxent} over the model parameters $(\theta_a)$ is equivalent to satisfying the constraint over the mean observables. The inference procedure is computationally hard, and is usually performed using a combination of Monte Carlo sampling methods and gradient descent, or mean-field techniques \cite{Barton2016a}.
Often, the observables are taken to be marginals of variables of interest, as well as pairwise of higher-order correlations between them. The resulting models then fall into classes of inverse statistical physics models, such as disordered Ising or Potts models \cite{Nguyen2017}. 

\subsubsection{Machine learning and Neural networks}
Modern machine learning techniques are increasingly popular in computational immunology, and are anticipated to become even more popular in the near future. We refer the reader to Ref.~\cite{Mehta2018} for a review of machine learning methods aimed at physicists.

The best established use of deep neural networks in immunology is for predicting peptide-MHC binding through a set of tools call netMHC \cite{Buus2003,Andreatta2015}. More traditional machine learning approaches have been used to compare and characterize immune repertoires \cite{Thomas2014b,Cinelli2017}.

Current efforts aim at predicting TCR-antigen binding using machine learning techniques \cite{Jurtz2018,Sidhom2018,Jokinen2019}, although the amount of data necessary to obtain truly predictive models is probably still insufficient at this stage.

\subsection{High-throughput repertoire sequencing}

The variable region of an antigen receptor chain is about 400 bp long. The sequencing challenge is to capture the whole region in one read. Modern methods of high-throughput genomic or metagenomic sequencing use shotgun sequencing which breaks the genome of interest into short fragments and after sequencing pastes the reads together by putting together overlapping reads with the help of a reference guiding template. Since Repertoire Sequencing (RepSeq) focusses on highly variable regions which includes non-templated insertions and deletions, the sequence must be acquired in one read. Read lengths required for TCR sequences are typically shorter {($\sim  100-150$bp)} since short fragments of V and J genes are enough to distinguish different V genes with the help of known genomic templates. However, since B-cells carry a large number of hypermutations in the V gene, sequencing of longer fragments {($\sim600$bp)} that encompass explicitly most of the V gene is necessary.

Different methods have been developed for sequencing DNA and mRNA. In general, the protocols start with isolating the mRNA or DNA of the cells of interest (BCR, TCR, in subsets of cells of interest sorted  using FACS). The mRNA product is then reverse transcribed onto cDNA and barcoded, sometimes with unique molecular identifier (UMI), and amplified by Polymerase Chain Reaction (PCR) before sequencing. Different techniques exist for PCR amplification that either add primer specific sites or multiple primers. Rapid amplification of cDNA ends (RACE) is one of the most common techniques, which is based on adding a linker with a primer binding site to a conserved region, on the 5' end of the V gene where no constant template is available (unlike on the 3' end). In DNA sequencing multiple primers that target specific regions of the DNA, introducing a primer specific amplification bias that can be controlled for by spiking in known sequences. 

Protocols based on sequencing directly DNA have the advantage that the experiments is free of mRNA expression bias. However, currently mRNA based technologies are able to reliably report sequence counts thanks to the use of UMI barcoding techniques that are still being developed for DNA sequencing. The difficulty with using UMIs in the DNA protocol lies in introducing the barcode before the original sequence is amplified. In the mRNA protocole the barcode is introduced during the initial reverse transcription of mRNA into cDNA for every RNA molecule. The barcoded sequence is then purified and the whole product is amplified by PCR. For DNA an additional ligation step is needed, in which the DNA is cut close to the region of interest (keeping in mind that the sequenced lengths is short). Ligation is not an efficient reaction. This approach is simple when many gene copies are available in a sample, but this is not the case in immune repertoire sequencing. Alternatively, the barcode could be added in the first PCR cycle. The product then needs to be well purified making show no barcode carrying primers are left, at the same time making sure that the barcoded sequences is not lost due to its low fraction compared to non-barcoded sequences, which remains technically tricky.

The above procedures sequence the chains of TCR or BCR repertoire in bulk, leaving no possibility to figure out the pairing of alpha and beta, or light and heavy chains in the same cell. Naturally, this limits the discussions of repertoire diversity, and more importantly attempts to link sequence to a functional phenotype. Recently these limitations have been overcome by the development of single cell repseq sequencing, where the PCR reaction is performed either in wells~\cite{Dekosky2014} or in droplets~\cite{Klein2015,Mcdaniel2016,Grigaityte2017}. Alternatively, a computational approach that exploits the statistics of co-occurence of the rare events partitioning the chains separately into wells is used to identify pairs of TCR sequences that come from the same cell from bulk DNA sequencing experiments~\cite{Howie2015}. The results of these analysis show that the pairing of alpha-beta chains is largely independent for TCR~\cite{Dupic2018,Grigaityte2017}, so the results of bulk analysis for diversity hold. However the affinity and functional properties of the BCR and TCR need to be estimated at the level of the whole receptor.

Lastly, novel high-throughput single-cell barcoded mRNA technologies~\cite{Zheng2017} are promising to change the field of immune repertoire sequencing, providing paired chain reads in large numbers. These technologies are very similar to those used in 5'RACE with UMI. Droplet based platforms allow in principle for large numbers of cells (up to 80 000 per chip) to be analyzed at the same time, although these numbers are still much smaller than the $10^6$ cells in bulk experiments. Given that clones have very low frequencies in bulk experiments (possibly down to a single cell out of $10^{11}$), sampling of the order of $10^4$ does not guarantee reproducible experiments. Additionally the cost remains high, at about $1$ USD per cell. At the moment of writing, the first analyses based on this technology are still underway.

~\label{repseq}

~\label{repseq}

\section{Acknowledgments}
The authors thank all past and current collaborators for the many useful discussions. TM and AMW thanks Meriem Bensouda Koraichi, Barbara Bravi, Victor Chard\`es, Thomas Dupic, Cosimo Lupo, Carlos Olivares, Jacopo Marchi, Maria Ruiz Ortega and Natanael Spisak for their comments on the manuscript and Mikhail Pogorelyy and Anastasia Minervina for discussions. GAB thanks Sooraj Achar, Emanuel Salazar Cavazos and Van Truong for their comments on the manuscript.  This work was partially supported by the European Research Council Consolidator Grant n. 724208 [TM \& AMW] and the intramural research program of the National Cancer Institute [GAB].  We also thank the KITP, Les Houches EdP and IES Carg\`ese institutes for hospitality during workshops on the topics reported in this review.

\bibliographystyle{pnas}

\end{document}